\documentclass{aa}  
\usepackage[utf8]{inputenc}
\usepackage{natbib,epsfig,graphicx,color,psfrag,pdflscape}
\usepackage{amsmath}
\usepackage{calrsfs}
\usepackage[varg]{txfonts}
\bibpunct{(}{)}{,}{a}{}{,}

\usepackage[utf8]{inputenc}
\def\digitspace{\hphantom{0}}
\def\periodspace{\hphantom{.}}
\DeclareUnicodeCharacter{00B7}{\digitspace}    
\DeclareUnicodeCharacter{00AC}{\periodspace}   

\newcommand{\rsun}{\ensuremath{R_{\odot}}}

\newcommand{\Teff}{\ensuremath{T_{\rm eff}}}
\newcommand{\vinf}{\ensuremath{v_{\infty}}}
\newcommand{\mdot}{\ensuremath{\dot{M}}}

\newcommand{\msunyr}{\ensuremath{M_{\odot} {\rm yr}^{-1}}}

\newcommand{\mdu}{\ensuremath{10^{-6}\,M_{\odot} {\rm yr}^{-1}}}

\newcommand{\beq}{\begin{equation}}
\newcommand{\eeq}{\end{equation}}
\newcommand{\beqa}{\begin{eqnarray}}
\newcommand{\eeqa}{\end{eqnarray}}

\newcommand{\nbeq}{\begin{equation*}}
\newcommand{\neeq}{\end{equation*}}

\newcommand{\kms}{\ensuremath{{\rm km}\,{\rm s}^{-1}}}

\newcommand{\rarrow}{\rightarrow}
\newcommand{\dd}{{\rm d}}

\newcommand{\HeI} {He\,{\sc i}}
\newcommand{\HeII}{He\,{\sc ii}}

\newcommand{\CIII}{C\,{\sc iii}}
\newcommand{\CIV}{C\,{\sc iv}}

\newcommand{\NII}{N\,{\sc ii}}
\newcommand{\NIII}{N\,{\sc iii}}
\newcommand{\NIV}{N\,{\sc iv}}
\newcommand{\NV}{N\,{\sc v}}
\newcommand{\NVI}{N\,{\sc vi}}

\newcommand{\OII}{O\,{\sc ii}}
\newcommand{\OIII}{O\,{\sc iii}}

\newcommand{\OV}{O\,{\sc v}}
\newcommand{\OVI}{O\,{\sc vi}}

\newcommand{\SiIV}{Si\,{\sc iv}}

\newcommand{\SIV}{S\,{\sc iv}}
\newcommand{\SV}{S\,{\sc v}}
\newcommand{\SVI}{S\,{\sc vi}}

\newcommand{\PV}{P\,{\sc v}}

\newcommand{\FeII}{Fe\,{\sc ii}}
\newcommand{\FeIII}{Fe\,{\sc iii}}
\newcommand{\FeIV}{Fe\,{\sc iv}}
\newcommand{\FeV}{Fe\,{\sc v}}
\newcommand{\FeVI}{Fe\,{\sc vi}}

\newcommand{\Hd} {H$_{\rm \delta}$}

\newcommand{\Ha} {H$_{\rm \alpha}$}
\newcommand{\Lya}{Ly${\rm \alpha}$}
\newcommand{\Lyb}{Ly${\rm \beta}$}

\newcommand{\Rstar}{\ensuremath{R_{\ast}}}
\newcommand{\Rmax}{\ensuremath{R_{\rm max}}}
\newcommand{\zmax}{\ensuremath{z_{\rm max}}}
\newcommand{\xmax}{\ensuremath{x_{\rm max}}}
\newcommand{\Rmin}{\ensuremath{R_{\rm min}}}

\newcommand{\logg}{\ensuremath{\log g}}
\newcommand{\YHe}{\ensuremath{Y_{\rm He}}}

\newcommand{\Trad}{\ensuremath{T_{\rm rad}}}

\newcommand{\vsini}{\ensuremath{v{\thinspace}\sin{\thinspace}i}}

\newcommand{\vmic}{\ensuremath{v_{\rm mic}}}

\newcommand{\fcl}{\ensuremath{f_{\rm cl}}}
\newcommand{\fv}{\ensuremath{f_{\rm X}}}

\newcommand{\taur}{\ensuremath{\tau_{\rm Ross}}}

\newcommand{\trip}{\ensuremath{\lambda\lambda4634-4640-4642}}
\newcommand{\qua}{\ensuremath{\lambda\lambda4510-4514-4518}}

\newcommand\red{\textcolor{red}}

\begin{document}
\title{Atmospheric NLTE models for the spectroscopic analysis of blue 
stars with winds. V. Complete comoving frame transfer, and updated
modeling of X-ray emission}

\author{J. Puls\inst{1}, F. Najarro\inst{2}, J.O. Sundqvist\inst{3},
\and K. Sen\inst{4}}

\institute{LMU M\"unchen, Universit\"atssternwarte, Scheinerstr. 1, 
           81679 M\"unchen, Germany, \email{uh101aw@usm.uni-muenchen.de}
	   \and
           Departamento de Astrof\'isica, Centro de Astrobiolog\'ia 
	   (CSIC-INTA), Ctra. Torrej\'on a Ajalvir km 4, 28850 Torre\'on 
	   de Ardoz, Spain
           \and
           KU Leuven, Instituut voor Sterrenkunde, Celestijnenlaan 200D, 
	   3001 Leuven, Belgium
	   \and
	   Argelander Institut für Astronomie der Universit\"at Bonn, 
	   Auf dem H\"ugel 71, 53121 Bonn, Germany
}
\date{}

\abstract 
{Obtaining precise stellar and wind properties and abundance patterns of
massive stars is crucial to understanding their nature and
interactions with their environments, as well as to constrain their evolutionary
paths and end-products.} 
{To enable higher versatility and precision of the complete
ultraviolet (UV) to
optical range, we improve our high-performance, unified, NLTE
atmosphere and spectrum synthesis code {\sc fastwind}. Moreover, we
aim to obtain 
an advanced description of X-ray emission from wind-embedded shocks,
consistent with alternative modeling approaches.}
{We include a detailed comoving frame radiative transfer for the
essential frequency range, but still apply methods that
enable low turnaround times. We compare the results of
our updated computations with those from the alternative code {\sc
cmfgen}, and our previous {\sc fastwind} version, for a representative 
model grid.}
{In most cases, our new results agree excellently with those from {\sc
cmfgen}, both regarding the total radiative acceleration, strategic
optical lines, and the UV-range. Moderate differences concern
\HeII~$\lambda\lambda$4200-4541 and \NV~$\lambda\lambda$4603-4619. The
agreement regarding \NIII~\trip\ has improved,
though there are still certain discrepancies, mostly related to line
overlap effects in the extreme ultraviolet (EUV), depending on
abundances and micro-turbulence. In the UV range of our coolest
models, we find differences in the predicted depression of the
pseudo-continuum, which is most pronounced around \Lya. This
depression is larger in {\sc cmfgen}, and related to different \FeIV\
atomic data. The comparison between our new and previous {\sc
fastwind} version reveals an almost perfect agreement, except again
for \NV~$\lambda\lambda$4603-4619. Using an improved, depth-dependent description for
the filling factors of hot, X-ray emitting material, we confirm
previous analytic scaling relations with our numerical models.}
{We warn against uncritically relying on transitions, which
are strongly affected by direct or indirect line-overlap effects. The
predicted UV-continuum depression for the coolest grid-models needs to
be checked, both observationally, and regarding the underlying atomic
data. Wind lines from ``super-ionized'' ions such as \OVI\ can, in
principle, be used to constrain the distribution of wind-embedded
shocks. The new {\sc fastwind} version v11 is now ready to be used.}

\keywords{methods: numerical - stars: atmospheres - stars: early-type - stars: massive - X-rays: stars}

\titlerunning{Complete comoving frame transport with FASTWIND}
\authorrunning{J.~Puls et al.}

\maketitle
%
%
\section{Introduction} 
\label{intro}

The impact of massive stars on cosmic and galactic evolution (for
example, \citealt{Bresolin08}) has been widely appreciated within the
astronomical community. In particular, the observational detection of
merging black holes and neutron stars, via gravitational
waves \citep{Abbott16, Abbott17}, has renewed the interest
in massive stellar objects, and especially in the evolution of (binary)
black-hole progenitors (cf. \citealt{Marchant16, Langer19,
Petit17}). 
However, whether for single objects or objects in binary and multiple
systems, current models of massive stars suffer from a number of
uncertainties and simplified descriptions\footnote{This is also true
for low-mass stars.}, mostly related to the need for intrinsic multi-D
processes to be ``boiled down'' to 1-D. This is because of the much longer
evolutionary timescales as compared to timescales governing the
dynamics of specific processes. 
An important example is rotation (\citealt{Langer97, MM00}) 
and the induced mixing, which,
moreover, is often treated by a simple diffusion approach, even if
advective terms play an important role (exemplarily, for mixing due to
meridional circulations, \citealt{MaederZahn98, MM15}).

In order to test these evolutionary models and related predictions on
the one hand (for instance, regarding the surface composition, which
strongly depends on rotational mixing, and, in binary systems, also on
mass-overflow), and to calibrate various basically unknown
coefficients (such as convective overshoot length, and mixing
efficiencies) on the other, a careful comparison with observations
(that is, with real objects) is inevitable. 

Though ``comparison with observations'' sounds simple, it is not,
since one does not ``observe'' temperatures, luminosities, mass-loss
rates, rotational rates, surface abundances, etc., but rather infers
them from the observed spectral energy distribution, often by applying
a technique called ``quantitative spectroscopy.'' In brief, this
procedure also adopts a simplified model, here for the outer,
atmospheric layers of a star,
and derives, on top of this model, the emitted photonic energy
distribution.

Such synthetic spectra then depend on the specific combination of
atmospheric parameters and chemical composition, and, by varying these
quantities, one tries to simulate synthetic energy distributions that
are as close as possible to the observed ones. The variation itself is
obtained via comprehensive, pre-calculated model grids, or on the fly,
for example when genetic algorithms are used to minimize the deviation
between observed and synthetic spectra \citep{Mokiem05}. 

After an optimum fitting distribution has been found (for a large
number of proven-to-work, diagnostic features, and spectral ranges),
one then claims to have derived (or even observed) the atmospheric
parameters and surface abundances. 
Obviously, there is the immediate question of uniqueness (Can
different combinations of parameters and abundances yield a similar
agreement?), and the question about the influence of specific
approximations and data on the final results.

Irrespective of these questions, it is clear that for such a
procedure, a large number of theoretical spectra have to be
synthesized, because the number of parameters describing an atmosphere
is large, particularly for massive stars that display line-radiation
driven winds with densities that increase with stellar luminosity
(reviewed by \citealt{puls08b}). 

Consequently, a prime factor for efficient spectral analyses is
computational performance. Unfortunately, since massive stars are
often hot and/or have a low-density atmosphere, the most time-saving
assumption made for cooler stars with larger densities, namely that
the atomic and ionic occupation numbers can be approximated from local
thermodynamic equilibrium (LTE) conditions\footnote{Because of the
much larger impact of collisional than radiative processes.} (but see,
for example, \citealt{Bergemann12}), is no longer applicable. Instead, one
needs to set up and solve the equations of statistical (sometimes also
called kinetic) equilibrium, commonly denoted by non-LTE or NLTE. Such
calculations are computationally expensive, since the radiation field,
required to set up the radiative transition rates, and affected by
velocity-field induced Doppler-shifts, needs to be computed at many
frequency points, in an
iterative approach.
Moreover, NLTE requires the knowledge of numerous atomic and ionic
properties of contributing species, such as cross-sections and
collision strengths, which themselves can suffer from uncertainties of
different extent.

In recent decades, a variety of computational codes have been
released that can deal with the above problem (NLTE
atmospheres and synthetic spectra for massive stars including winds),
namely {\sc phoenix} \citep{haus92}, {\sc cmfgen}
\citep{hilliermiller98}, WM-{\sc basic} \citep{pauldrach01}, and PoWR
\citep{graefener02, Sander15}, where all of them (except for WM-{\sc basic},
which uses a Sobolev approximation to calculate the radiative
bound-bound rates) require considerable turnaround times, due to
their objective to deliver the highest-possible precision for any of
the considered processes. 

Already in 1995, within a collaboration between A. Herrero (Instituto
de Astrof\'isica de Canarias, La Laguna, Spain) and J.P., the idea was
developed to design an alternative approach where computational speed
should be of highest priority. The basic philosophy of the emergent
code, baptized as {\sc fastwind} (for previous versions, see
\citealt{santo97, Puls05}, and \citealt{rivero12}, 
\citealt{Carneiro16}, \citealt{SP18} for the most current versions in
use, v10.1 to v10.3) was to concentrate on the optical and
infrared (IR) spectroscopy of OB-stars and A supergiants, and to differentiate
between so-called ``explicit'' and ``background'' elements. The former
are those used as diagnostic tools (H and He always, and other
elements such as C,N,O, or Si, dependent on application). They are
treated with high precision, by detailed atomic models, following a
flexibe, {\sc detail-} \citep{ButlerGiddings85} like input-format, and
by means of a comoving frame transport for the line transitions.  In
this approach, the background elements (that is, the rest, particularly
the iron-group elements, with atomic data taken from the fixed-format
WM-{\sc basic} data base), are important ``only'' for the
line-blocking and blanketing calculations, and were treated, until now,
using various methods for setting-up the radiative bound-bound rates
(detailed in Sect.~\ref{philosophy}; see also
Table~\ref{calc_scheme}).  The targeted computational efficiency was
obtained by applying appropriate physical approximations to processes
where high accuracy was not needed (regarding the objective of the
analysis - optical and IR lines), in particular, concerning the treatment
of the metal-line background opacities. Here, the individual opacities
and source functions are added up to build continuum-like quantities
that subsequently determine the background radiation field over the
complete spectrum. Most importantly, all methods and approximations have
been carefully tested during the course of development, by comparing
with codes based on more ``exact'' methods, particularly {\sc cmfgen}
(and {\sc tlusty}, \citealt{hubeny98}, for models where the wind does
not play a role), but also with WM-{\sc basic}.

From the first line-blanketed version on, {\sc fastwind} has
significantly evolved during the last years, and meanwhile acts as a
working-horse for spectral analyses that require the computation of a
large number of atmospheric models and synthetic spectra (particularly
within the VLT-FLAMES survey of massive stars, \citealt{evans08},
within the LMC Tarantula survey -- VFTS, \citealt{evans11}, --, and
within the IACOB-project, \citealt{simon-diaz11, simon-diaz11b},
for Galactic objects). In addition, {\sc fastwind} has been used
in applications aiming at the analysis of non-spherical objects (for
example, binaries in a common envelope phase), by means of patching their
surfaces by numerous 1-D models with position-dependent parameters
\citep{AbdulMasih20}, and for the analysis of combined
\citep{simon-diaz15a} or disentangled \citep{AbdulMasih19} spectra of
multiple systems. Because of the specific way the multitude of
background lines is considered (to form a pseudo-continuum, which can
be described by relatively few frequency points), our
treatment might even allow us to develop multi-D NLTE models including
line-blocking and blanketing effects, operating on reasonable
computational time scales.

Of course, the downside of fast performance is the failure to achieve
precision in {\bf all} potentially interesting spectral regions, and
though {\sc fastwind} has been carefully tested and compared with
other codes, there are certain situations where specific
approximations might have a decisive impact. First of all, this might 
happen if individual line-overlap effects become influential, where
such overlaps are, to a major extent, neglected until now. This
downside has been emphasized from early on \citep{Puls05}, and one
such effect was identified within the formation of the diagnostic
\NIII~\trip\ triplet, at least for objects in a specific temperature
regime \citep{rivero11}. Certainly, there are more such effects, 
for example those participating in the formation of diagnostic carbon lines
\citep{MartinsHillier12}.

Moreover, due to our approach and (previous) philosophy, {\sc
fastwind} v10 cannot reliably synthesize spectral regions outside
individual lines from explicit elements. Consequently, the analysis of
a large, continuous portion of the spectrum, populated by lines from
dozens of elements different from the explicit ones, and required,
for example, when analyzing a UV-spectrum as a whole, is prohibitive.

To cure these problems, and to allow for applications that have not
been possible for {\sc fastwind} until to date, we have improved our
approach by performing a precise comoving-frame radiative transfer for
the complete spectrum (as done, for example, in {\sc cmfgen}, PoWR, and {\sc
phoenix}), but we still try to use methods that minimize the
computational effort. A brief announcement of the new version (without
providing any details) has already been published by \citet{Puls17},
and the new version itself has been used by \citet{Sundqvist19}, for
calculating the radiative acceleration in self-consistent models of
massive star winds.

In the current paper, we explain our improvements in fair
detail (Sect.~\ref{cmf_complete}), and extensively compare our new
results with those from {\sc cmfgen} (Sect.~\ref{comparison}), as
already done previously with respect to {\sc fastwind} v10. Of course, we
will also compare with results from the latter, previous version
itself, to
evaluate which diagnostics might be affected by our improved approach.
In Sect.~\ref{xrays}, we describe and discuss specific updates of our
treatment of X-ray emission from wind-embedded shocks. Such updates
are necessary to ``unify'' previous approaches based on ideas by
\citet{Hillier93} and \citet{Feldmeier97b} on the one side, and more
recent studies by \citet{OwockiSundqvist13} on the other, which, at
first glance, seem to be somewhat contradictory (see
\citealt{Carneiro16}). In Sect.~\ref{summary}, we finally summarize
our findings and conclusions, and the appendices provide some
additional technical details regarding the implementation of our code.

\section{The new Fastwind version v11}
\label{cmf_complete}
\subsection{General philosophy}
\label{philosophy}

To understand the changes and improvements in our new {\sc fastwind}
version (v11), it is necessary to briefly summarize the underlying,
general philosophy -- which remains untouched --, and, in particular,
the methods and approximations within our previous versions (v10, for
details, see \citealt{Puls05}, \citealt{rivero12},
\citealt{Carneiro16}, and \citealt{SP18}). Indeed, our new
version v11 has been set up in such a way that the new functionalities
(described below) are included in a separate module, and that by
changing one specific option inside the code all methods of v10 can be
recovered. Such a ``switch-back'' might be advantageous (because of
faster turnaround times) whenever the new features are not needed. As
an example, we mention already here the optical analysis of
photospheric and wind parameters by means of only H and He as explicit
elements, due to the only marginal differences between corresponding
results from v11 and v10 (for details, see Sect.~\ref{HHe_optical}).

\paragraph{Density, velocity, and temperature structure.} As 
detailed in \citet{santo97}, the deeper atmospheric layers are
approximated by hydrostatic equilibrium (in spherical symmetry), that
is, by neglecting the advection term in the equation of motion. In the
initial modeling phase, the flux-weighted opacities required to evaluate the
radiative acceleration are approximated by a Kramer's like formula
(with constants and exponents fitted to iterated opacity estimates). 
In a later phase, the structure is updated by using the actual opacities (see
also Appendix~\ref{convergence}). 

Densities, $\rho$, are obtained from the hydrostatic solution for
gas pressure $p$, via the equation of state, where the plasma is 
adopted as an ideal gas, 
\beq
\rho(r) = p(r) / v_{\rm sound}^2(r)
\eeq
with $v_{\rm sound}$ the isothermal sound speed.
Velocities in this deeper, photospheric part are derived from the
continuity equation (solving Eq.~\ref{conteq} provided below for
$v(r)$), using the density stratification from above.
The resulting photospheric structure is smoothly connected to the wind
outflow, at a pre-defined ``transition velocity'', with a default value of
10\% of $v_{\rm sound}$ (evaluated at \Teff\ for the
specific composition).
The wind-structure is specified by a typical $\beta$ velocity law, 
\beq
v(r)=\vinf(1-b/r)^\beta,
\eeq
and mass-loss rate \mdot\ (input),
\beq
\label{conteq}
\rho(r)=\frac{\mdot}{4\pi r^2 v(r)},
\eeq
with terminal wind speed, \vinf\ (input), $b$ a parameter calculated in
parallel with the location of the transition point and the transition
velocity, and $\beta$ the input parameter controlling the steepness of
the wind velocity field.
In parallel to densities and velocities, a consistent temperature
structure is determined, using a flux-correction method in the lower
atmosphere, and the electron thermal balance (cf. \citealt{Kubat99})
in the outer part.

\paragraph{Wind clumping.} In our current v11 version as described
here, we ``only'' allow for conventional, optically thin wind
clumping, consistent with many earlier versions of {\sc fastwind}. A
variety of stratifications for the clumping factor, $\fcl(r)$ (=
overdensities in clumps, if the interclump medium is assumed to be
void) can be chosen by the user (or, if desired, newly defined). These
include (i) spatially constant clumping factors from a pre-defined
velocity on, (ii) the default parameterization of {\sc cmfgen} \citep
{hilliermiller99, hillier03}, and (iii) generalizations of the latter
(see \citealt{Najarro11}). An implementation of optically thick wind
clumping and porosity in velocity space, as already included in v10.3 (see
\citealt{SP18}, and references therein) into our new v11 -- more
precisely, into the corresponding module --, is foreseen for the near
future. Because of its higher complexity (compared to optically
thin clumping), careful tests preceeding a final release are required
though.

\begin{table*}
\caption{Schematic comparison of {\sc fastwind} v10 and v11 (in red):
specific methods and data regarding the treatment of 
explicit and background elements. The default
values for $\lambda_{\rm min}$ and $\lambda_{\rm max}$ are 200 and
10,000 \AA, respectively. For further details, see text.}
\label{calc_scheme}
\begin{center}
\begin{tabular}{l|c|c|c}
\hline 
\hline
\multicolumn{1}{c|}{\rule[-3mm]{0mm}{8mm}}
&\multicolumn{1}{c|}{} 
&\multicolumn{2}{c}{background elements}\\ \cline{3-4}
\multicolumn{1}{c|}{}
&\multicolumn{1}{c|}{\raisebox{1.5ex}[-1.5ex]{explicit elements}}
&\multicolumn{1}{c|}{selected}
&\multicolumn{1}{c}{non-selected}\\
\multicolumn{1}{c|}{}
&\multicolumn{1}{c|}{\raisebox{1.5ex}[-1.5ex]
   {(e.g., H,He; H,He,N;}}
&\multicolumn{1}{c|}{(typically: C,N,O,Mg,Ne,Si,P,S,Ar,}
&\multicolumn{1}{c}{(remaining elements until}\\
\multicolumn{1}{c|}{}
&\multicolumn{1}{c|}{\raisebox{1.5ex}[-1.5ex]
   {H,He,C,N,O, {\ldots)}}}
&\multicolumn{1}{c|}{Fe,Ni, minus explicit elements)}
&\multicolumn{1}{c}{Zn, without Li,Be,B,Sc)}\\
\hline
\rule[-3mm]{0mm}{8mm}atomic data (v10 \red{\& v11})& user supplied\tablefootmark{a} 
& \multicolumn{2}{c}{fixed, from WM-{\sc basic}\tablefootmark{b} database} \\
\hline
\rule[-3mm]{0mm}{8mm}NLTE (v10 \red{\& v11})& exact & exact 
& approximate\tablefootmark{c} \\
\hline
\rule[0mm]{0mm}{5mm}v10: radiative transfer
&   
& strong lines: CMF; weaker lines:  
& Sobolev, irradiated by\\
\rule[-3mm]{0mm}{5mm}for individual lines 
&{\raisebox{1.5ex}[-1.5ex] {CMF}}  
&  wind: Sobolev; photosphere: static
& pseudo-continuum\\ \cline{2-4}
\rule[0mm]{0mm}{5mm}\red{v11: radiative transfer}
& \multicolumn{2}{c|}
{\red{$\lambda_{\rm min} <\lambda < \lambda_{\rm max}$: CMF (allowing
for multiple line overlap);}}
&\red{Sobolev, irradiated by re-} \\
\rule[-3mm]{0mm}{5mm}\red{-- lines and continuum} &
\multicolumn{2}{c|}{\red{else: as in v10}}
&\red{mapped\tablefootmark{d} CMF rad. field}\\
\hline
\rule[0mm]{0mm}{5mm}v10: mean intensity for 
& \multicolumn{2}{c|}
     {from pseudo-continuum with bound-free and free-free} 
& expressed in terms \\
photoionization rates 
& \multicolumn{2}{c|}
     {opacities/emissivities from {\bf all} elements, and
     combined\tablefootmark{c}} 
& of \Trad\ from \\ 
\rule[-3mm]{0mm}{5mm}and other quantities 
& \multicolumn{2}{c|}{line opacities/source-functions from 
     background elements} 
& pseudo-continuum\\ \cline{2-4}
\rule[0mm]{0mm}{5mm}\red{v11: mean intensity for} 
& \multicolumn{2}{c|}{}  
& \red{expressed in terms} \\
\red{photoionization rates} 
& \multicolumn{2}{c|}{\red{$\lambda_{\rm min} <\lambda < \lambda_{\rm max}$:
      from re-mapped\tablefootmark{d} CMF-radiation field; else: as in v10}}
&\red{of \Trad\ from re-mapped\tablefootmark{d}}\\
\rule[-3mm]{0mm}{5mm}\red{and other quantities} 
& \multicolumn{2}{c|}{} & \red{CMF radiation field}\\
\hline
\end{tabular}
\tablefoot{
\tablefoottext{a}{{\sc detail} \citep{ButlerGiddings85}-like input format}
\tablefoottext{b}{\citet{pauldrach01}}
\tablefoottext{c}{see \citet{Puls05}}
\tablefoottext{d}{see Appendix \ref{remap}}
}
\end{center}
\end{table*}

\paragraph{NLTE and radiative transfer in {\sc fastwind} v10.}
Within our {\bf previous} {\sc fastwind} version(s), the concept of
"explicit" and "background" elements is used, as outlined in the
introduction. The explicit elements are then treated straightforwardly
and with high precision, namely by solving the NLTE rate equations 
and performing the line transport in the comoving frame (CMF), though
neglecting explicit line overlap effects. Regarding the background
elements, the procedure is more complex (for a schematic representation,
see Table~\ref{calc_scheme}). 

At first, we divide them into two subgroups, and call the more important
ones -- essentially those with a higher abundance -- ``selected''
background elements. Currently, and if not included in the explicit
elements, these are C, N, O, Mg, Ne, Si, P (because of its important
UV-line), S, Ar, Fe, and Ni, but other elements of interest
can be included into this list as well. 

Whereas the occupation numbers of the remaining, ``non-selected'' 
elements\footnote{With Zn the heaviest element
considered, and Li, Be, B, and Sc discarded because of very low
abundances.} are estimated via an approximate NLTE approach
\citep{Puls05}, for the selected ones we solve the detailed NLTE rate
equations (as for the explicit elements). To save time, however, the
required radiative bound-bound rates are calculated in three different
ways: For the most important (strongest) transitions, we again solve
the CMF transfer; for the weaker lines, the bound-bound rates are
either approximated from a Sobolev approach (including the
pseudo-continuum radiation field, see below), or, in the photospheric
regime, derived from a static radiation transfer. All this only for
individual lines, again neglecting specific line overlaps, except
regarding transitions collected within the pseudo-continuum.

This pseudo-continuum, which is required to correctly describe 
line-blocking and blanketing effects, is obtained by sampling (almost)
all individual opacities and source functions to continuum-like 
quantities (accounting for Doppler-induced frequency shifts in an
approximate way), which are then used to solve the radiation transfer in
the observer's frame (cf. \citealt{Puls05}). The resulting radiation
field serves as input for a variety of calculations (such as
approximate NLTE -- see above --, pseudo-continuum background for the
Sobolev line rates, scattering continuum emissivity for the detailed
CMF transport, radiative bound-free rates for exact NLTE
calculations, bound-free heating and cooling rates, and photospheric
radiation force), and the circle is closed.

\subsection{Comoving frame transfer in Fastwind v11}
\label{basic_cmf}

The major change between the previous and the current {\sc
fastwind} version concerns the radiative
transfer. In a new module, almost {\bf all} lines and continua
(both from explicit and background elements) are now treated in the
CMF, thus allowing us to increase the precision, particularly by
``automatically'' accounting for line-overlap effects (both due to
coincidental identities or similarities of transition frequencies, and
wind-induced, cf. \citealt{puls87}). This complete CMF transport is
performed inside a wavelength range $\lambda_{\rm min}$ to
$\lambda_{\rm max}$, where, in the current version, the default values
are 200 and 10,000 \AA, respectively. Extending $\lambda_{\rm max}$ to
the near infrared (NIR) will be tested in future work. For late B-types and cooler
(with vanishing \HeII\ ionization edge), $\lambda_{\rm min}$ might be
set to 400 \AA, whereas for the hottest O-subtypes, it might be
extended to lower values, for example, 130~\AA, to include the \NV\ edge.
Test calculations have shown that such extensions do not change
current results using our default value of 200 \AA\ though.  Outside
the range $\lambda_{\rm min}$ to $\lambda_{\rm max}$, we follow our
previous, pseudo-continuum approach, but always check that the
transition between both regimes is monotonic, and that no jump occurs
(for an example regarding $\lambda_{\rm max}$, see
Fig.~\ref{app_xj_iminus2}. Black: detailed CMF transport; green:
pseudo-continuum approach). In these outer frequency domains (until
X-ray frequencies in the blue, and radio frequencies in the red), all
line-rates are calculated as in the previous versions described
above.

Subsequently, our current approach solves the NLTE rate equations, for
both explicit and selected elements, with radiative rates calculated
from the detailed CMF transport. Thus, in the new code the most
important difference between explicit and selected elements is now the
source of atomic data, either flexible (explicit elements) or fixed
(see Sect.~\ref{intro}). Once the atomic data are incorporated, the
method makes (almost) no distinction between explicit and selected elements
when solving the rate equations and the radiative transfer (again, see
Table~\ref{calc_scheme}). The only
additional difference refers to the degree of precision aimed at.
Selected elements are considered as converged by following the changes
within the ionization fractions, whilst for the explicit elements,
all levels have to fulfill the required convergence criterion (see also
Appendix~\ref{convergence}). Thus, the accuracy of specific excited
levels might be higher when a certain element is treated as an explicit one,
in the original spirit.

For the remaining (non-selected) background elements (also with
fixed-format atomic data), the approximate NLTE approach is still in
use, where the required radiation field quantities are taken from a
re-mapped CMF solution (see Appendix~\ref{remap}). Opacities and
emissivities inside the CMF transport comprise all elements.

The CMF transport itself can be solved in two ways. Either, we perform
a formal (``ray-by-ray'') solution for the Feautrier variables alone
(applying a fully implicit scheme, following \citealt{mihalas75}, in
the conventional $p$-$z$\ geometry, for example \citealt{Puls05}, 
\citealt{Puls20}, and references therein), or -- this
is the default -- we calculate the
corresponding Eddington factors and solve the moments equations
subsequently. To avoid numerical problems, and following \citet[their
Eq.~13, with $\epsilon=1$]{hilliermiller98}, we apply the ratios of
third to zeroth moment, $N_\nu/J_\nu$, instead of the more commonly
used ratios of third to first moment, $N_\nu/H_\nu$. In specific
cases, particularly for (almost) vanishing fluxes, the former approach
results in a more stable solution. 

The reason for considering the moments equations is twofold. First,
the solution for the angle-dependent Feautrier variables is affected
from certain approximations related to the (standard) discretization
on the $p$-$z$\ grid, such that specific intensities and corresponding
moments (particularly flux-like quantities) suffer from inaccuracies.
Since these inaccuracies mostly cancel within moment ratios, a
subsequent solution of the moments equations can provide a more exact
outcome. Moreover, by definition, such a solution needs to be
performed only on the radial grid, and thus is computationally
inexpensive. Comparing the results for a comprehensive model grid (see
Sect.~\ref{comparison}, and Table~\ref{tabgrid}) from both methods has
revealed that the differences in the emergent spectra in most cases
are marginal, and only the temperatures at large optical depths (which
are irrelevant for most applications) can be affected to a
non-negligible extent.

The second reason for additionally solving the moments equations
refers to computational performance. As long as the majority of
occupation has not stabilized close to our convergence criterion (on
the order of few per mille), which in our scheme is true as long as the
temperature structure of the atmospheric model has not converged (see
Appendix~\ref{convergence}), it is suitable to fix the Eddington
factors within one to three subsequent iterations. Then, we can solve
the moments equations alone, without any formal solution. (In
later iteration stages, such a fixing would be (slightly)
inconsistent, and would destroy the final convergence of sensitive
transitions). Fixing the Eddington factors (when possible) decreases
the computational time significantly, since the otherwise required
angle-dependent CMF transport is the most time-consuming part of the
total calculation. This, because it scales with the number of radial
grid points, the number of p-rays, and the large number of frequency
points, $N_f$, to be considered. For given (fixed) Eddington factors,
on the other hand, the moments equations scale ``only'' with the
number of radial grid points, and $N_f$.

The latter primarily depends on the assumed micro-turbulent velocity,
\vmic, and can be estimated\footnote{As long as the Doppler width of
the line-profile is dominated by \vmic, which is true in hot stars
when aiming at a reasonable resolution of lines also from the heaviest
elements.} by 
%
\beq 
\label{N_f}
N_f \approx \frac{\log \displaystyle \frac{\lambda_{\rm max}}{\lambda_{\rm
min}}}{\log(1+\displaystyle \frac{\vmic}{n_{\rm Dop}\, c})}, 
\;\mbox{roughly}\; \propto \frac{1}{\vmic}. 
\eeq 
For our default values, and $n_{\rm Dop}$ the number of frequency
points per Doppler-width ($=3$ in our simulations), this results in
$N_f \approx 230,000$ and $N_f \approx 710,000$ for $\vmic = 15$ and
$5$~\kms, respectively, which are prototypical values for O-supergiants
and B-dwarfs. 

As a last, more technical aspect, we approximate the incident
intensity at the outermost grid point, $I^-$ (and its frequency
derivative), as described in Appendix~\ref{app_iminus}, to keep
the computational effort as low as possible. This differs from the {\sc
cmfgen} approach as introduced by \citet{hilliermiller98}, namely to
``extend'' the atmosphere in the ray-by-ray solution toward larger
radii and optically thin conditions (using extrapolated opacities and
emissivites), and then set $I^-$ to zero at the new, extended
boundary.

\medskip
%
%

\subsection{Accelerated lambda iteration, and approximate
lambda operators}

To improve (or even enable) the convergence of our solution scheme, we
apply, as already done in {\sc fastwind} v10, an
accelerated lambda iteration (ALI), contrasted to {\sc cmfgen} that
uses a linearization method. The required, approximate lambda
operators (ALOs), are calculated in parallel with the ray-by-ray 
solution, following \citet{Puls91}.

We stress here that the actual ALO entering the
pre-conditioned (\citealt{RH91, Puls91}, denoted by ``reduced''
in the latter work) radiative bound-bound rates
needs to be weighted at each frequency (before integration
over the profile function), due to the manifold line-overlaps. The
weighting factor is given by the ratio between the line opacity of
the considered transition, $i$, and the total line opacity present in
the CMF transfer at frequency $\nu$. This is necessary since the ALO
constructed by our method refers to the total line source function
entering the radiative transfer (the contribution by continuum
processes is seperately accounted for), whereas the pre-conditioned
line rates refer to the individual ones, related via 
\beq
\label{sl_tot}
S_L^{\rm tot}(\nu) = \frac{\sum_{j \ne i} \bar \chi_j \phi_j(\nu)S_L^j}
 {\sum \bar \chi_j \phi_j(\nu)} + 
 \frac{\bar \chi_i \phi_i(\nu)} {\sum \bar \chi_j \phi_j(\nu)} S_L^i,
\eeq
Because the approximate lambda operator, within the bound-bound rates,
needs to act on the line-specific source function, $S_L^i$ (frequency
independent, when assuming complete redistribution), it must be
weighted by the fore-factor of the second term in Eq.~\ref{sl_tot}. In
this equation, $\bar \chi_j \phi_j(\nu)$ is the frequency dependent
line opacity for transition $j$ (with $1 {\ldots} i {\ldots} j$
overlapping components), and $\phi_j(\nu)$ the line profile
function\footnote{Adopted in the NLTE CMF transport as a pure Doppler
profile. As long as the final formal integral correctly accounts for
the actual broadening (for example, Stark- and pressure broadening), this has
a marginal effect on the resulting occupation numbers and
line-profiles \citep{Hamann81a, Lamersetal87}.}.

\begin{figure*}
\begin{minipage}{9cm}
\resizebox{\hsize}{!}
  {\includegraphics[angle=90]{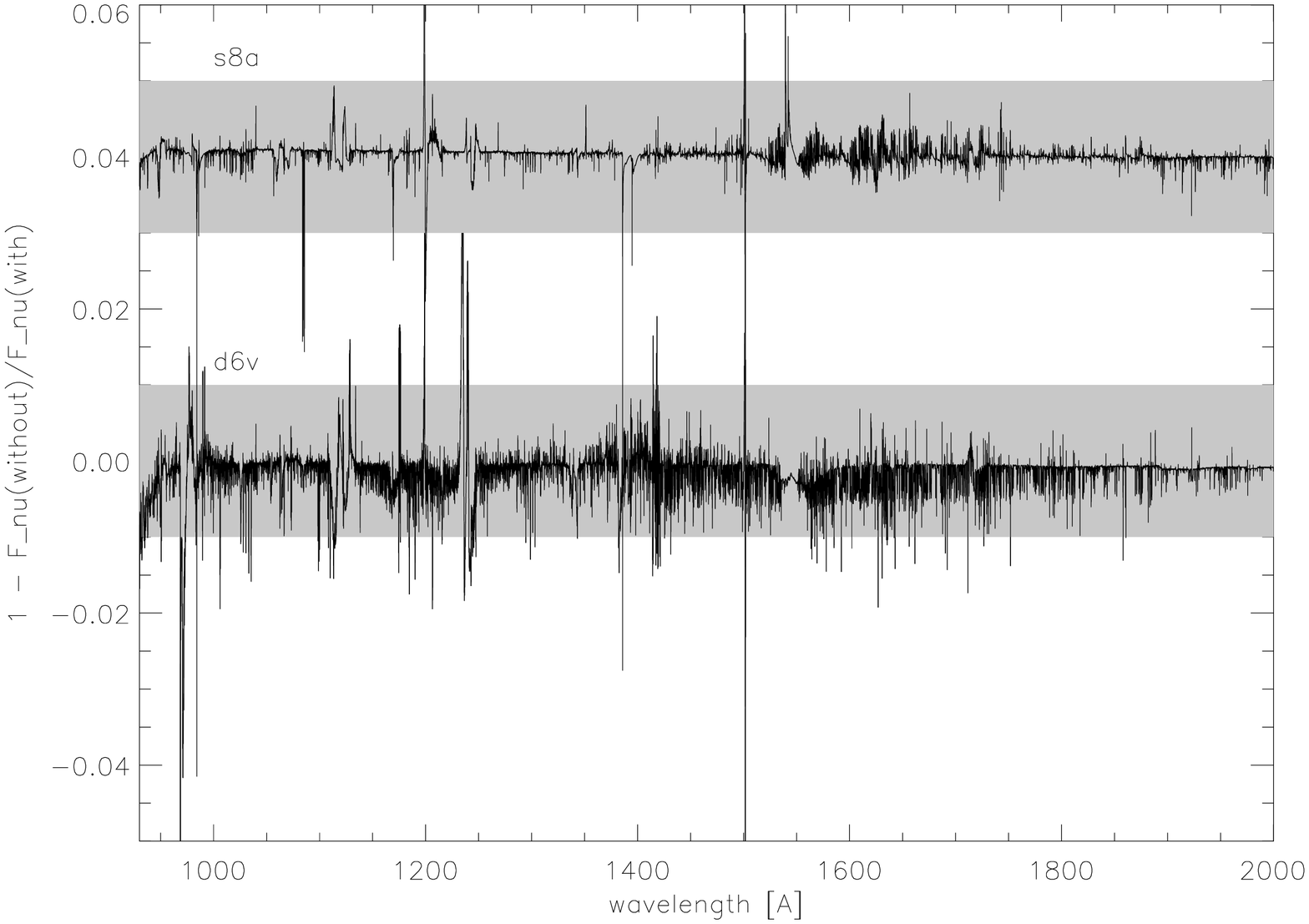}} 
\end{minipage}
\begin{minipage}{9cm}
\resizebox{\hsize}{!}
  {\includegraphics[angle=90]{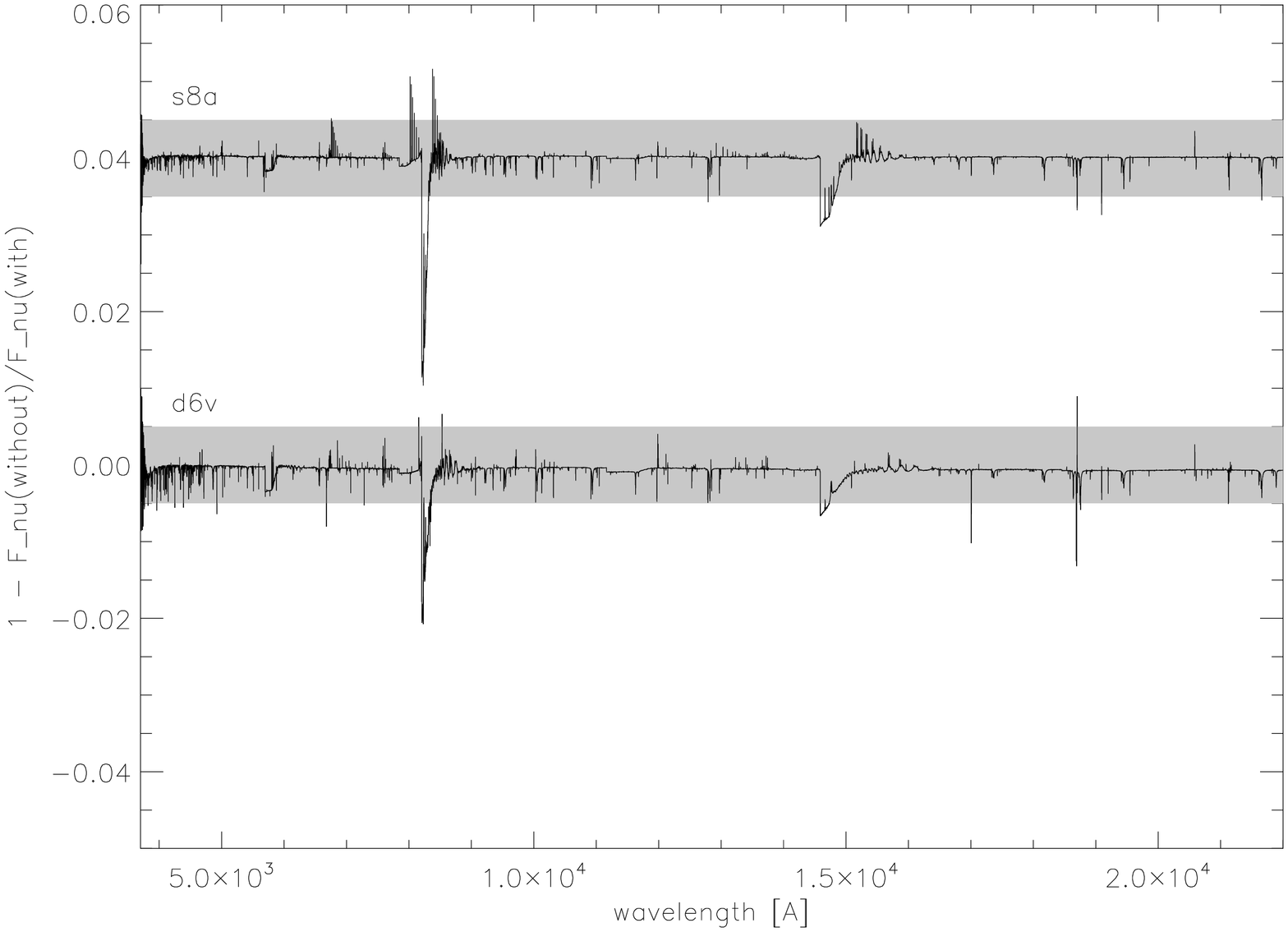}} 
\end{minipage}
\caption{Effects of level dissolution in the UV (left) and the
optical and NIR (right), for models s8a and d6v (see Table~\ref{tabgrid}).
Displayed is the deviation of the emergent fluxes, $1-F_\nu^{\rm
without}/F_\nu^{\rm with}$, without and with accounting for level
dissolution, as calculated by {\sc cmfgen}. The results for model s8a
have been vertically shifted by 0.04. To guide the eye, the
gray-shaded region refers to deviations of $\pm 1$\% for the UV range,
and of $\pm 0.5$\% for the optical and NIR range, respectively. For
clarity, wavelength ranges with $\lambda < 930$~\AA\ and $\lambda <
3700$~\AA\ (regions close to the Lyman and Balmer edges, where,
as expected, larger changes are found) are not displayed. All
wavelengths refer to vacuum.}
\label{dissol}
\end{figure*}

\medskip 
\noindent 
As outlined above, our solution scheme does not only use the
ray-by-ray solution, but also the corresponding moments equations.
Since, however, the latter yield slightly different mean intensities
than the former, in principle it might be necessary to calculate a
second set of approximate lambda operators (ALOs) which are consistent
with the solution of the moments equations, and thus can be used in
parallel with the corresponding scattering integrals, $\bar J_i = \int
J_\nu \phi_i(\nu) \dd \nu$, within the (pre-conditioned) radiative
bound-bound rates. Contrasted to the case of static
radiative transfer, however, the development of optimum ALOs within the CMF
transport is quite complex (due to the presence of the frequency
derivatives), and, thus far, has been performed only for the
angle-dependent, formal solution \citep{Puls91}. A corresponding ALO
to be calculated in parallel with the moments equations is, to our knowledge,
still not available (though certainly possible). 

Fortunately, and after many tests, it turned out that it is
sufficient to use the ALOs from the ray-by-ray solution, even if the
$\bar J$'s have been calculated from the moments equations. Only in
those cases when the ALOs are very close to unity ($>0.99$ in our
implementation using diagonal operators), we reset them to the latter
value, to account for potential inconsistencies. Such a reduction is
``allowed,'' since lower than optimum diagonal operators do not lead
to a divergence of the ALI, contrasted to overestimated ones (see
\citealt{OAB86}). On the other hand, whenever sub-cycles with fixed
Eddington-factors are performed (see Sect.~\ref{basic_cmf}), the ALOs
need to be set to zero, since otherwise the occupation numbers would
be strongly disturbed, due to the somewhat inconsistent approach.
Since the latter cycles appear only at earlier stages of the
calculation (before the temperature structure has converged), a
neglect of the ALO does not play a role though. This even more, since
each second to fourth iteration is still performed with both solutions
(ray-by-ray and moments), such that at this stage a consistent ALO
becomes known, and speeds up the convergence again. 
Details on the overall iteration cycle and convergence properties are
outlined in Appendix~\ref{convergence}.

\subsection{Additional issues}
\label{add_details} 

In the following subsection, and also Appendix~\ref{more_details}, we
discuss additional issues which are important for our specific {\sc
fastwind} implementation.

\paragraph{Formal integral.} 
After the model and all occupation numbers have converged (or
quasi-converged, when oscillating, see Appendix~\ref{convergence}), we
calculate, in a separate program package, the formal integral in the
observer's frame to obtain (normalized) synthetic spectra.
Basically, we follow our approach from \citet{santo97} of
interpolating opacities and emissivities onto a spatial micro-grid (in
$z$), with a typical resolution corresponding to Min$(\vmic(r))/3$,
but we abstain from a separation of continuum and line
processes\footnote{Because of the high line density, this approach
would not lead to any advantages in computational time.}, and
integrate over total opacities and source-functions. The latter are
calculated in the spirit of \citet{hilliermiller98}. Namely, before
any interpolation, the line list (for background and explicit
elements), the occupation numbers, and continuum
opacities and emissivities are re-stored from the last iteration of our
NLTE-model calculations. Subsequently, corresponding total 
opacities and emissivities are derived, still at comoving-frame 
frequencies (that is, without any velocity field induced Doppler-shifts),
by summing up line and continuum quantities. At this point, all
relevant broadening mechanisms (see below) are accounted for. This
calculation has to be performed only once, and is quite fast, since it
needs to be executed only on the coarse radial grid, and for the
spectral range defined by the user (see below). Only after the total
opacities and emissivities have been calculated (and stored), they are
interpolated onto the spatial micro-grid, and evaluated at the
corresponding {\bf local} comoving frame frequency (again by
interpolation), in dependence of rest-frame frequency and projected
velocity. Specifically, if the local CMF frequency is $\nu_{\rm CMF}(z)
\approx \nu_{\rm obs}(1-\mu v(z)/c)$, then the opacities and emissivities have 
to be interpolated from the pre-calculated values (see above) at
CMF frequencies $\nu_{\rm CMF}^i > \nu_{\rm CMF}(z) > \nu_{\rm
CMF}^{i+1}$, if the (unshifted) CMF-frequency grid has indices $i$, and
$\mu$ is the cosine of the angle between radial and radiation
direction.

After the formal solution for the specific intensity has been derived
(for all considered rest-frame frequencies, and all impact parameters,
with a default value of 80, inclusive 10 core rays), the
frequency-dependent emergent fluxes can be calculated by angular integration.
The normalization is finally obtained by calculating one additional formal
solution, accounting for the continuum opacities and emissivities alone (no
micro-grid required here), and dividing the total emergent fluxes
by the continuum ones. The range and resolution of the synthetic
spectra can be specified by the user (default: 900-2000~\AA\, and
3400-7000~\AA, with a resolution of 0.1~\AA), as well as the
micro-turbulence. For the latter, either a constant or
depth-dependent value can be chosen 
(to simulate the effects of the large velocity dispersion seen in
hot-star instability simulations, and observed through the so-called
black troughs in saturated UV P-Cygni lines, see 
\citealt{Hamann80, Lucy82, POF93, Sundqvist11a}).
To compare with previous models,
and to obtain an impression about the impact of blends from other
elements, our program package allows us to also calculate
individual line profiles for specified transitions within the explicit
elements, identical with the approach followed by {\sc fastwind} v10.
Finally, we note that line-broadening is accounted for in the standard
way, using Stark broadening for the hydrogen and helium lines, and
Voigt profiles (with damping parameters derived from data collected in
the LINES.dat file, see Appendix~\ref{no_upper}) for the metals. If no
broadening data are available (as for Fe and Ni in our current
data-base), simple thermal and micro-turbulent broadening is adopted.

\paragraph{Level dissolution.} Contrasted to {\sc cmfgen} and some 
other NLTE-codes dedicated to hot star atmospheres, the new {\sc
fastwind} version does {\bf not} account for level dissolution
\citep{HummerMihalas88, Hubeny94}. By switching off such processes in
{\sc cmfgen}, and comparing with its standard treatment including
level dissolution, we have convinced ourselves that in most cases the
corresponding emergent fluxes do not differ substantially, except
(very) close to the Lyman and Balmer edges. Indeed, larger variations
begin, for model d6v, only for Lyman (and corresponding \HeII)
transitions with an upper principal quantum number, $n_u \ge 8$
($\lambda \la 926$~\AA), and for Balmer transitions with $n_u \ge 11$
($\lambda \la 3770$~\AA), whereas the transitions close to the Paschen
threshold display only moderate effects. Generally, the dwarf models,
because of a higher density at typical line-formation depths,
display larger effects than supergiant models, as visible in 
Fig.~\ref{dissol}. In this figure, we have excluded the Lyman range
below 930~\AA, and the Balmer range below 3700~\AA, to allow for a
better vertical resolution of the diagnostic wavelength regimes.
Typical deviations in the line cores are, for a dwarf model, on the
order of less than 1\% in the UV, and on the order of 0.3 to 0.5\% in
the optical {\bf and} NIR. For supergiant models, the differences are
even lower. Thus, we conclude that except for a realistic
representation close to strong ionization edges (in particular, Lyman
and Balmer), and potentially for corresponding lines between
high-lying levels (in the far infrared and radio regime), our neglect of
level-dissolution effects does not lead to significant inaccuracies
(at least if we will not apply our code to white dwarfs). 
However, the above restrictions should also prevent the user from
``blindly'' applying our new {\sc fastwind} version, for example, for
estimating the Balmer-decrement, or for constraining the gravity from the very
high series members close to corresponding ionization edges.

\subsection{Computation time and memory requirements}
\label{comp_time} 

As already mentioned, the major fraction of computational time within
our implementation is typically spent for the ray-by-ray
solution,
which increases almost linearly with
$1/\vmic$. For our standard set-up (see Sect.~\ref{default}), with H,
He, and N as explicit elements, and 67 depth points, a program
run with 140 iterations requires 1.2 and 1.6 CPU hours on an Intel
Xeon processor with 3.7 and 2.7 GHz, respectively, if we use \vmic =
15~\kms, and 3.2 versus 4.4 CPU hours for \vmic = 5~\kms. For the same
specifications, 1.6 GB RAM needs to be allocated, which is
comparatively modest. We tried to keep the required RAM as low as
possible, to allow us to calculate as many models as possible in
parallel (comments on a future parallelization are given in
Sect.~\ref{summary}).

Since models with H and He as explicit elements converge much faster
(typically, after 80 to 100 iterations\footnote{Because of the somewhat
simpler atomic model for N when used as a background element, and
because of the corresponding, less rigorous convergence criterion, see
Sect.~\ref{basic_cmf}.}), they require ``only'' 1 CPU
hour on a 2.7 GHz machine, for \vmic = 15~\kms. Already such 
model types are well suited to calculate the total radiative
acceleration required for self-consistent massive star wind models
(\citealt{GraefenerHamann05, KK17, Sander17, Sundqvist19}, the
latter authors  already using the new {\sc fastwind} version), and also the
UV spectrum of hot stars. 

The formal integral, on the other hand, has much shorter turnaround
times, due to the frequency and spatial interpolation of the {\bf
total} CMF opacities and emissivities. For our default parameters for
spectral range and resolution (see above), the typical execution
times are on the order of 5 to 15 minutes, mostly depending on maximum
wind speed (and processor frequency). To calculate individual line
profiles for strategic lines from explicit elements, the turnaround times are
even shorter. If we resolve each line by 161 frequency points, the
most important strategic lines from H, He, and N are calculated in
less than one minute.

\begin{table}
\caption{Stellar and wind parameters of our model grid used to check
specific details of our new {\sc fastwind} version (v11), and to
compare with results from {\sc cmfgen} and {\sc fastwind}~v10.  All
models have been calculated with \vmic = 15~\kms, an unclumped wind,
no X-ray emission from wind-embedded shocks, and the ``older'' solar
abundances from \citet{grevesse98}, in particular a helium abundance,
\YHe = $N_{\rm He}/N_{\rm H}$ = 0.1, $\epsilon_{\rm N} = 7.92$, and
$\epsilon_{\rm Fe} = 7.50$, where $\epsilon_{\rm X} = \log_{10}(N_{\rm
X}/N_{\rm H}) +12$.}
\label{tabgrid}
\tabcolsep1.5mm
\begin{center}
\begin{tabular}{ccrcccc}
\hline 
\hline
\multicolumn{7}{c}{Luminosity class V}\\
\hline
\multicolumn{1}{c}{Model}
&\multicolumn{1}{c}{\Teff}
&\multicolumn{1}{c}{\Rstar}
&\multicolumn{1}{c}{\logg}
&\multicolumn{1}{c}{\mdot}
&\multicolumn{1}{c}{\vinf}
&\multicolumn{1}{c}{$\beta$}\\
\multicolumn{1}{c}{}
&\multicolumn{1}{c}{(K)}
&\multicolumn{1}{c}{(\rsun)}
&\multicolumn{1}{c}{(cgs)}
&\multicolumn{1}{c}{(\mdu)}
&\multicolumn{1}{c}{(\kms)}
&\multicolumn{1}{c}{}
\\
\hline
d2v & 46100 & 11.4 & 4.01 & 2.52   & 3140 & 0.8\\
d4v & 41010 & 10.0 & 4.01 & 0.847  & 2850 & 0.8\\
d6v & 35900 & 8.8  & 3.95 & 0.210  & 2570 & 0.8\\
d8v & 32000 & 8.0  & 3.90 & 0.056  & 2400 & 0.8\\
d10v& 28000 & 7.4  & 3.87 & 0.0122 & 2210 & 0.8\\
\hline
\multicolumn{7}{c}{Luminosity class I}\\
\hline
\multicolumn{1}{c}{Model}
&\multicolumn{1}{c}{\Teff}
&\multicolumn{1}{c}{\Rstar}
&\multicolumn{1}{c}{\logg}
&\multicolumn{1}{c}{\mdot}
&\multicolumn{1}{c}{\vinf}
&\multicolumn{1}{c}{$\beta$}\\
\multicolumn{1}{c}{}
&\multicolumn{1}{c}{(K)}
&\multicolumn{1}{c}{(\rsun)}
&\multicolumn{1}{c}{(cgs)}
&\multicolumn{1}{c}{(\mdu)}
&\multicolumn{1}{c}{(\kms)}
&\multicolumn{1}{c}{}
\\
\hline
s2a & 44700  & 19.6 & 3.79 & 12.0  & 2620 & 1.0\\
s4a & 38700  & 21.8 & 3.57 & 7.35  & 2190 & 1.0\\
s6a & 32740  & 24.6 & 3.33 & 3.10  & 1810 & 1.0\\ 
s8a & 29760  & 26.2 & 3.21 & 1.53  & 1690 & 1.0\\
s10a & 23780 & 30.5 & 2.98 & 3.90  & 740  & 1.0\\
\hline
\end{tabular}
\end{center}
\end{table}

\section{First results, including comparisons to \uppercase{cmfgen}
and \uppercase{fastwind} v10}
\label{comparison}

Most tests of our new {\sc fastwind} version have been performed for
stellar and wind parameters as defined by a model grid that covers early B
to hot O-type dwarfs and supergiants below \Teff = 47~kK. This grid
also served for comparing with analogous results from {\sc cmfgen}
models. The latter (for the same grid-parameters) have been calculated
by one us (F.N.), with a recent {\sc cmfgen} version that solves
the hydrostatic equation in the inner atmosphere, and which has also been
slightly modified to improve the transition between photosphere and
wind (Najarro et al., in prep.). The grid itself is a subset of the grid
introduced by \citet{lenorzer04}, and has already been used in
previous comparisons by \citet{Puls05} and \citet{rivero11,
rivero122}. For convenience, grid parameters and basic assumptions are
repeated in Table~\ref{tabgrid}.

\subsection{Default specifications used for the model grid} 
\label{default}

To enable a basic check, and to avoid the impact of additional
effects, in the following we concentrate on models with homogeneous
(unclumped) winds, and without X-ray emission from wind-embedded
shocks. Specific aspects of the latter will be discussed in
Sect.~\ref{xrays}. We use H, He, and N as explicit elements, unless
explicitly stated otherwise. The corresponding atomic models are
identical to those used in {\sc fastwind} v10. Details regarding H and
He have been presented in \citet{Puls05}, and our nitrogen model atom
has been discussed in \citet{rivero12}. For the model-atmosphere
calculations, we use a (default) grid of 67 carefully distributed
radial depth-points, and for the $p-z$ geometry, a set of 77 impact
parameters, including 10 core-rays (a similar\footnote{Though not
identical, to allow for a reasonable integration-error control.} set of
80 impact parameters is used in the formal integral). Tests have shown
that these numbers are sufficient to obtain a reasonable resolution,
and negligible errors in the angular integrations. For supergiant
atmospheres with a denser wind, even a reduction to 51 depth points
(with 61 impact parameters) is possible, without loss of significant
information. 

\begin{figure}
\resizebox{\hsize}{!}
  {\includegraphics[angle=90]{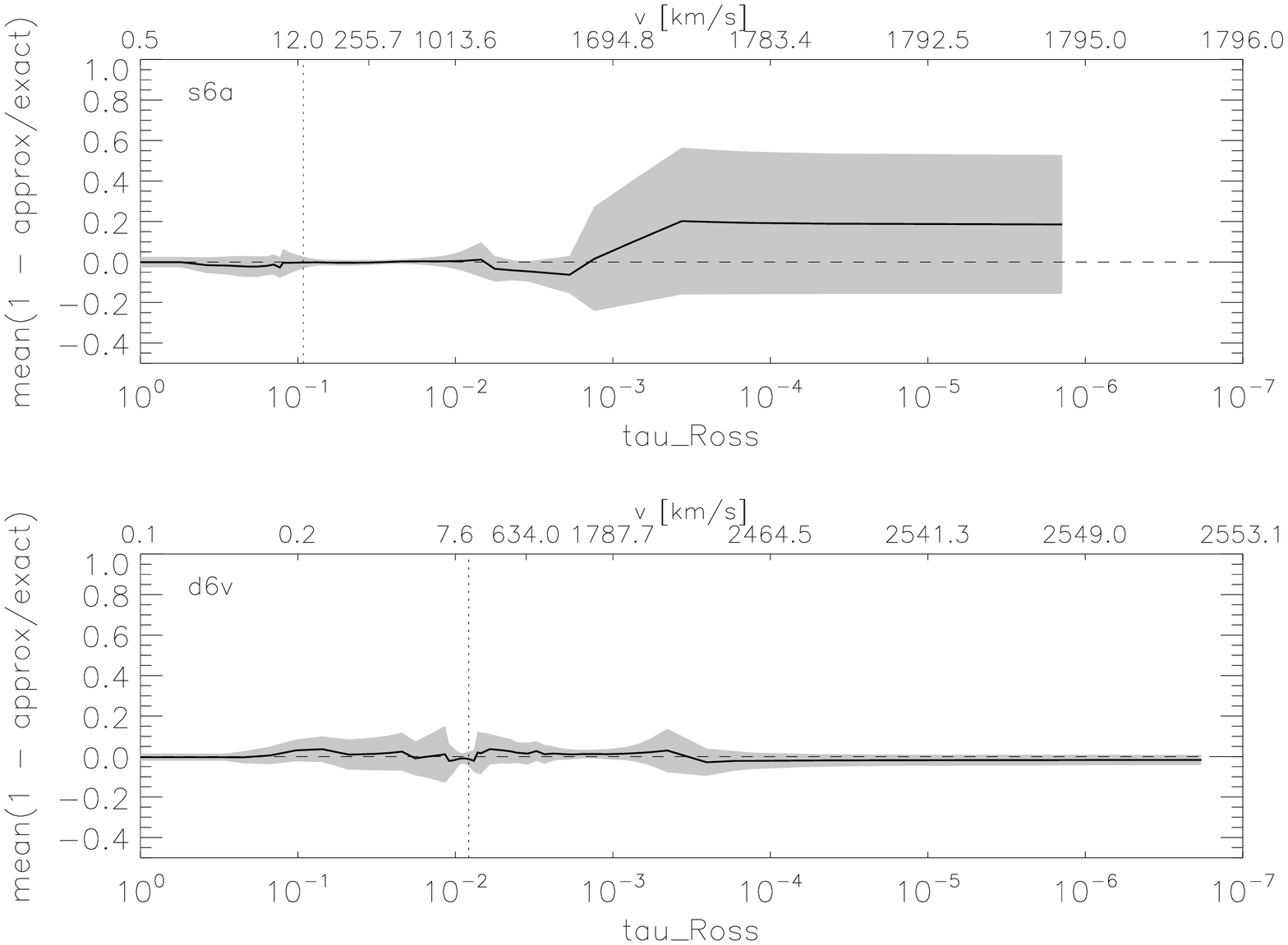}} 
\caption{Mean (relative) differences between the main ionization
fractions as predicted by our approximate NLTE description following
\citet{Puls05}, and our current ``exact solution'' (complete NLTE,
detailed CMF transport), as a function of $\taur \le 1$ and $v$
(in \kms). Displayed are the
models with the largest differences (s6a) and the smallest ones (d6v)
within our grid. The mean (see Eqs.~\ref{mean_dev_eq} and \ref{fkmax}) 
refers to 11 selected elements from
C to Ni, and its 1-$\sigma$ deviation is displayed in gray. The
vertical lines indicate the location of the sonic point.} 
\label{mean_dev}
\end{figure}

\begin{figure*}
\resizebox{\hsize}{!}
  {\includegraphics[angle=90]{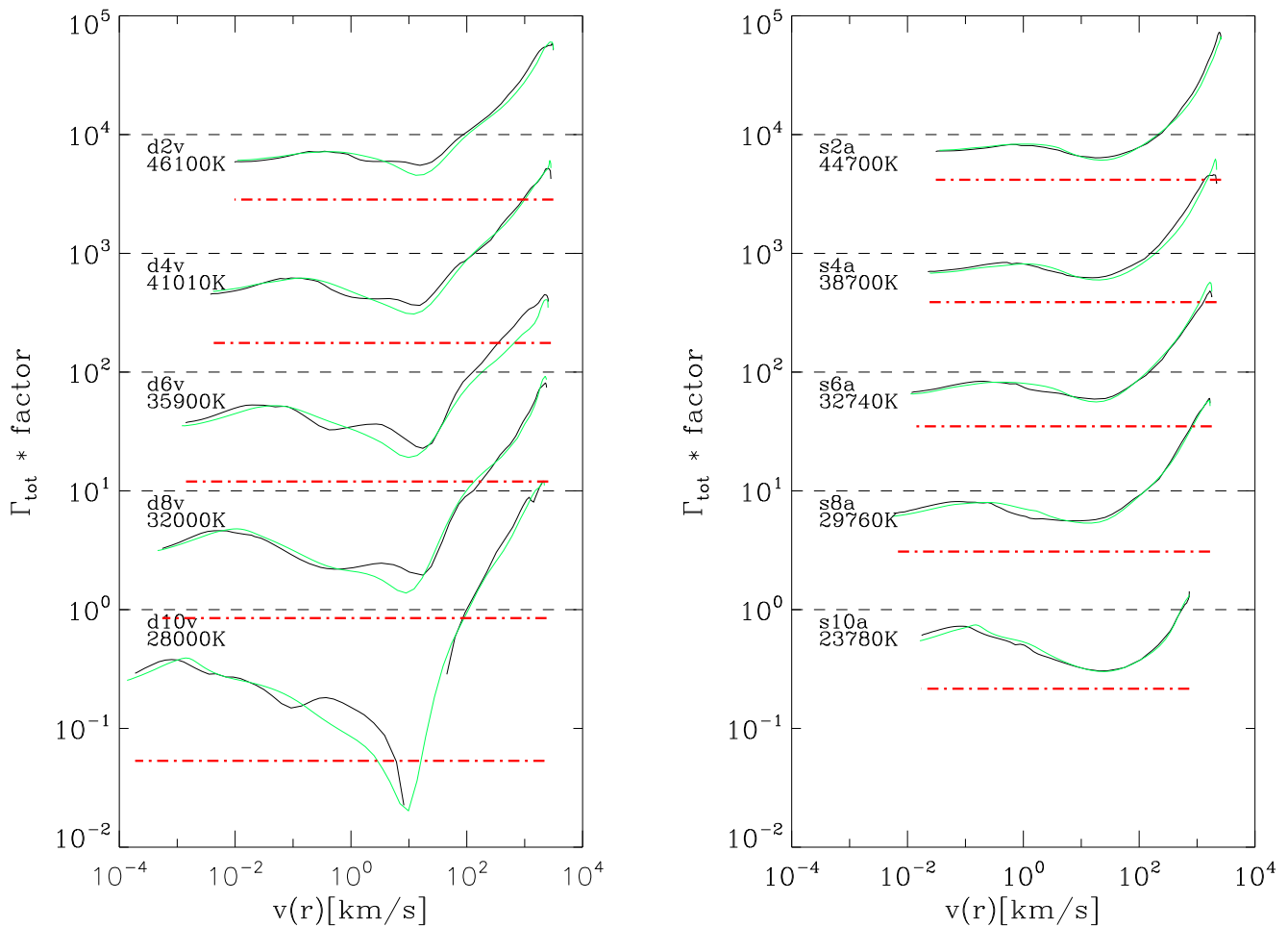}} 
\caption{$\Gamma_{\rm tot} = g_{\rm rad}^{\rm tot}/g_{\rm grav}$ for
all models from our grid, as a function of velocity. Left: dwarf
models; right: supergiant models. Black: results from {\sc fastwind} v11,
green: results from {\sc cmfgen}. To allow for a clear representation, 
all $\Gamma$-values have been multiplied with factors $10^i, i \in
[0,4]$, from bottom to top. The red dashed-dotted lines correspond to
$\Gamma_{\rm e} \propto \Teff^4/g$ for pure electron scattering, 
and the dashed lines indicate, for each model, 
the relation $\Gamma_{\rm tot} = 1$. For self-consistent models, $\Gamma_{\rm
tot} = 1$ should be located very close to the sonic point. See text.} 
\label{comp_grad}
\end{figure*}
\begin{figure*}
\resizebox{\hsize}{!}
  {\includegraphics[angle=90]{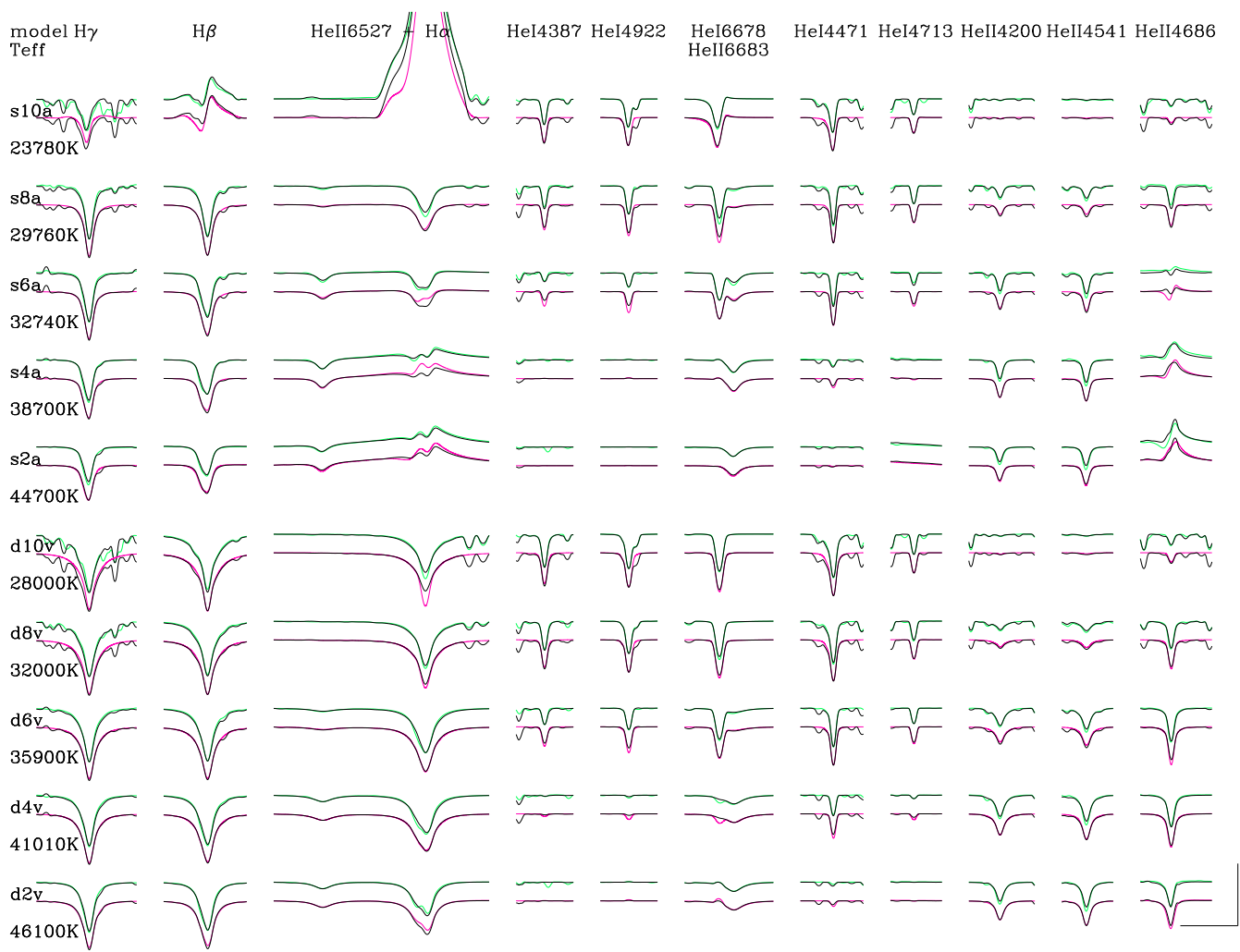}} 
\caption{Comparison of strategic H and He lines in the optical, for all
models from our grid. Black: {\sc fastwind} v11, green: {\sc cmfgen},
red: {\sc fastwind} v10, with no blends from other elements. The
marker in the lower right indicates a wavelength range of
20~\AA, and a vertical extent of 0.5 of the continuum. To enable a
better comparison, the spectra have been convolved with a rotational
broadening of $\vsini = 80 \kms$, and degraded to a resolving power of
10,000.}
\label{comp_hhe}
\end{figure*}

\subsection{Approximate versus detailed treatment} 

Our new method allows us to repeat a test already performed with an
earlier {\sc fastwind} version (cf. \citealt{Puls05}, their Figs.~5
and 6), namely to evaluate the reliability of our approximate NLTE
description. As explained earlier, this description is still in use
for the non-selected background elements, which play a certain role,
for example, for defining the (line-blocked) radiation field, and the
temperature structure. For this test, in Fig.~\ref{mean_dev} we
compare the mean relative differences between the main ionization
fractions predicted by the approximate method, and our current,
exact NLTE approach using a detailed CMF transfer,
\beq
\label{mean_dev_eq}
\langle \bigl(1-f_{k, {\rm approx}}^{\rm max}/
f_{k, {\rm detailed}}^{\rm max}\bigr)\rangle=
\frac{1}{N_k}\sum_{k \in [3,30]} \bigl(1-f_{k, {\rm approx}}^{\rm max}/
f_{k, {\rm detailed}}^{\rm max}\bigr),
\eeq
with $f_{k}^{\rm max}$ the maximum ionization fraction for element $k$,
\beq
\label{fkmax}
f_{k}^{\rm max}={\rm Max} \Bigl(\frac{n_{jk}}{n_k}\Bigl), \quad j \in [1,j_{\rm
high}(k)].
\eeq
Here, $n_k$ is the total population of element $k$, $n_{jk}$ the population of
ion $j$, and $j_{\rm high}(k)$ the highest ionization stage considered
for element $k$.

The mean itself is evaluated for our default set of $N_k = 11$
selected background elements (see Sect.~\ref{philosophy}), and is
displayed as a function of $\taur \le 1$ (since the differences for
larger \taur\, vanish anyhow, due to thermalization), and of velocity.
Due to the negligible computational effort, we can easily perform this
test, since at each iteration we anyhow calculate the approximate NLTE
occupation numbers for all background elements, before replacing those
for the selected elements by their exact counterparts\footnote{For
this specific comparison, we used only H and He as explicit elements,
such that nitrogen is treated within the selected background, and
could be included into the mean.}. After inspection of all our grid
models, it turned out that the largest differences occur for model s6a
(in the outer wind), and the smallest ones for model d6v. Both cases
have been displayed, including the 1-$\sigma$ scatter of the mean. For
all our models, the approximate treatment provides satisfactory
results in the photosphere, whereas larger deviations (with a mean
difference up to 20\%, and a large scatter) are possible in the outer
wind. Nevertheless, even these deviations are still tolerable, and we
conclude that our approximate NLTE approach is acceptable if one is
interested in gross effects such as line-blocking opacities, and maybe
even radiative line-accelerations, at least if derived via detailed
radiative transfer calculations. All this of course only if the
(pseudo-) continua are calculated in a reasonable way, comprising
line-blocking effects.

Without appropriate pseudo-continua, the ionization fractions would
certainly become erroneous (because of erroneous ionization integrals,
be them calculated in an approximate or exact manner). Thus we
have to check the differences between our previous (continuum-like
background opacities, observer's frame transport) and current
(detailed CMF-transport) approach. Again, for all our models, a fair
agreement is found, where a prototypical example is displayed in
Fig.~\ref{app_xj_iminus1}, left panel.

\subsection{The \HeI\, singlet problem} 
\label{singlets}

Already in our very first runs of the new program version, we
encountered the same problem as first described by
\citet{Najarro06b}, the so-called \HeI\, singlet problem, resulting
from a specific line-overlap effect between the \HeI\ resonance line
at roughly 584~\AA\ and a few close lying \FeIV\ lines (and,
potentially, lines from other elements). If the \FeIV\ lines have
oscillator strengths as found in current data-bases (on the order of
$\ga 10^{-3}$, for details, see \citealt{Najarro06b}), the line
overlap leads to a lower population (compared to the case without
overlap) of the \HeI\ $1s 2p ^1 P^{\rm o}$ level. This, in turn, is
the lower level of important diagnostic lines of the \HeI\ singlet
series in the optical, such as \HeI~4387, 4922, and 6678~\AA, and the
upper level of \HeI~2.058~$\mu$m in the K-band. Due to the lower
population, the optical lines become weaker (again: compared to the
case of no overlap), or even appear in emission, whereas the K-band
line becomes stronger. This behavior was particularly found in {\sc
cmfgen} models, whereas previous {\sc fastwind} versions produced
comparatively stronger optical lines (in agreement with observations),
because of the neglect of detailed line-interactions. Now, with the
new {\sc fastwind} version, very similar effects and \HeI\ singlet
profiles as in {\sc cmfgen} are predicted, telling us that both the
results from {\sc fastwind} v11 and {\sc cmfgen} and the involved
atomic data are consistent, and enforcing our confidence in the new
approach. To achieve consistency with observations, we ``cured'' the
problem in a similar way as suggested already earlier by
co-author F.N. (priv. comm), namely by reducing the (still quite
uncertain) oscillator strengths of the involved \FeIV\ transitions to
a lower value ($gf = 10^{-5}$, with $gf$ the product of oscillator
strength, $f$, and statistical weight of the lower level, $g$). With
such a reduction, the impact of the \FeIV\ lines decreases
significantly, and the resulting optical \HeI\ singlet lines
become stronger again, in agreement with the predictions by our
previous {\sc fastwind} version. Certainly, this problem needs to be
rechecked, when new atomic data calculations become available.
Here, we again warn about concentrating on these singlet lines
within quantitative spectroscopy: abundances, ionization conditions,
and micro-turbulent velocities, in combination with somewhat
insecure \FeIV\ transition frequencies, can affect the strength of
the overlap, and lead to additional uncertainties. Thus, we recommend
to prefer the results from the much more stable \HeI\ triplet lines.

\begin{figure*}
\begin{center}
  {\includegraphics[width=10cm, angle=90]{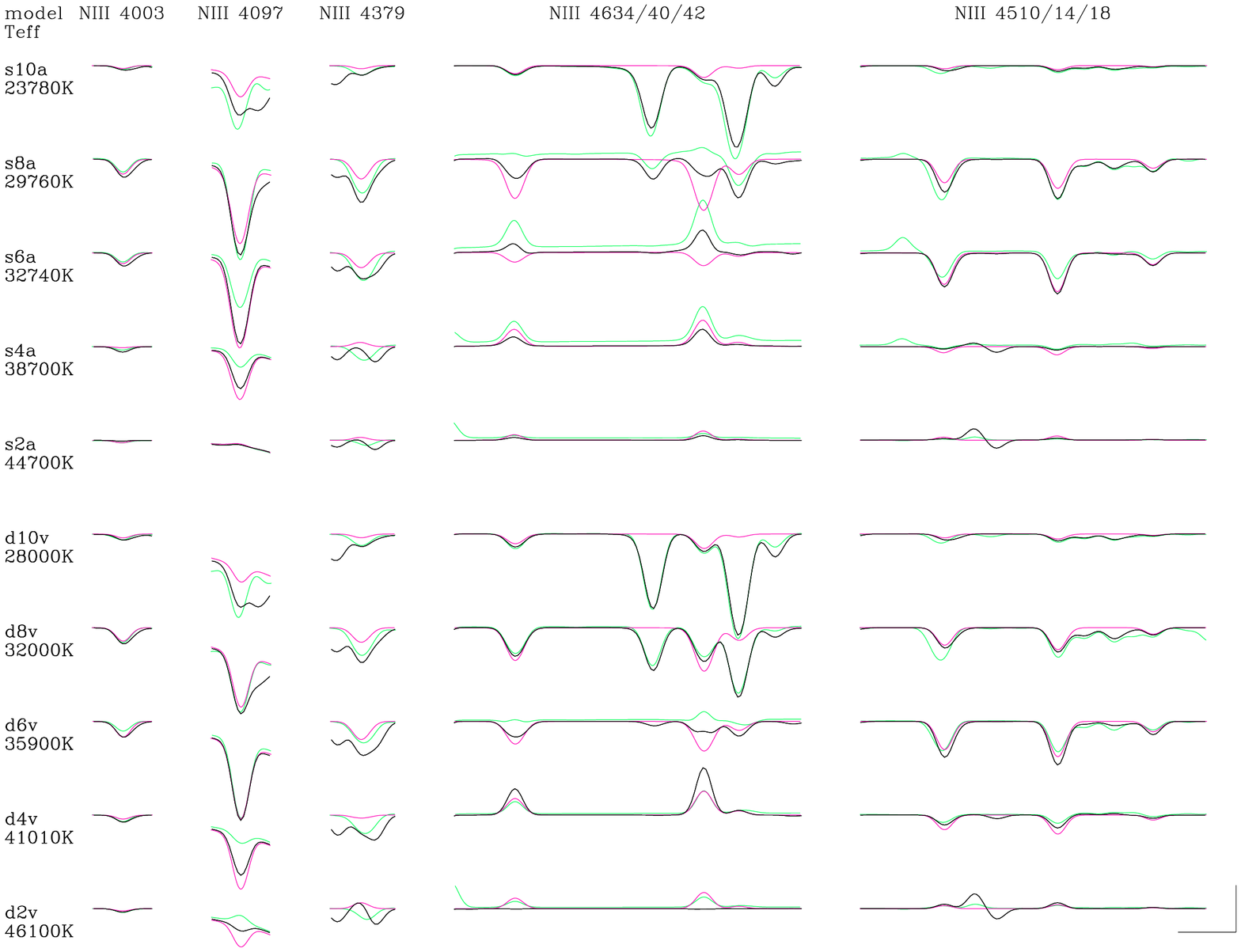}} 
\end{center}
\caption{As Fig.~\ref{comp_hhe}, but for strategic \NIII\ lines. Here,
the marker indicates a wavelength range of 2~\AA, and 0.2 of the
continuum. The positions of the individual components of \NIII\,\trip\
of the cooler objects can be clearly seen in the red {\sc fastwind}
v10 profiles. Since most lines are weak, no rotational broadening has
been applied.}
\label{comp_niii}
\end{figure*}

\subsection{Radiative acceleration}

Fig.~\ref{comp_grad} compares the total radiative acceleration as
calculated by {\sc fastwind} v11 and {\sc cmfgen}, measured in units
of gravitational acceleration, and as a function of velocity. Thus, it
displays Eddington's $\Gamma_{\rm tot}$. The red dashed-dotted
lines indicate the conventional Gamma-factor for electron scattering,
$\Gamma_{\rm e}$. For the supergiants, both codes agree almost
perfectly, while for the dwarf models, a certain deviation in the
transonic region is present. The major reason for this discrepancy is
most likely related to the different treatment of the transition
between photosphere and wind (in particular, corresponding velocities
and gradients), and to the different line lists. Even for these
models, however, the accelerations in the wind agree very well. 

For the coolest dwarf model (d10v), {\sc fastwind} predicts a
(slightly) negative total acceleration above the transition point, but
also in {\sc cmfgen} the total (still positive) acceleration is lower
than $\Gamma_{\rm e}$, indicating a larger inward- than
outward-directed line acceleration, due to negative fluxes from above
(resulting from strong wind lines that are not present in the
transition regime). 

Given that both codes use a largely different philosophy, and
different atomic data bases, the overall agreement is remarkable
though. We note that for all models, $\Gamma_{\rm tot} = 1$ is reached
only at substantial velocities (100~\kms or more), whereas, in
realistic wind-models, this should happen very close to the sonic
point (20 {\ldots} 25 \kms). Thus, the displayed models are far away
from being hydrodynamically self-consistent.  By iterating the
radiative acceleration and the resulting wind-structure to relax at
constant mass-loss rate, velocity structure, and $\Gamma_{\rm tot}
\approx 1$ at the sonic point\footnote{A small difference to the
unit-value results from pressure effects.}, \citet{Sundqvist19} used
the current {\sc fastwind} version to obtain such self-consistent
models. As it turned out, a quite steep velocity field in the
transonic region and the lower wind is required to fulfill the latter
condition (see also \citealt{KK17, Sander17} for similar approaches,
the latter performed with PoWR). Moreover, the negative
$\Gamma_{\rm tot}$-values found here for the coolest dwarf model
(invoking a $\beta$ velocity law) vanish when iterating for the
radiative acceleration. The characteristic ``force-dip'' 
(the decrease in acceleration before its increase in the wind) remains
always present though.

\subsection{The optical hydrogen and helium spectrum}
\label{HHe_optical}

In Fig.~\ref{comp_hhe}, we compare important diagnostic hydrogen and
\HeI, \HeII\ lines in the optical, in the upper sets with results
from {\sc cmfgen} (in green), and in the lower ones, with results from
the previous {\sc fastwind} version, v10 (in red). The HHe spectra
from our new {\sc fastwind} version have been derived by including
all overlapping lines that are present, whereas those from the old
version account only for H, He, and N components. Both {\sc cmfgen} and
{\sc fastwind} v11 models have been calculated with a diminished
influence of the \FeIV\ line(s) overlapping with \HeI~584 (see
Sect.~\ref{singlets}).

Again, the agreement (also for blends from the other elements) is
almost perfect, and even the shape of the \Ha\ and \HeII~4686 wind
emission coincides impressively. The only problem which was and still
is present refers to the cores of \HeII~$\lambda\lambda$4200-4541,
which, in the temperature range between 30 to 36 kK, are stronger in
{\sc cmfgen}, and lead to somewhat different effective temperatures
when analyzing hot star spectra by means of one code or the other.
Though this discrepancy has become slightly milder when comparing with
the new {\sc fastwind} version, the overall difference remains. The
agreement between the previous and current {\sc fastwind} version, on the
other hand, is excellent, and results from previous diagnostics should
remain (almost) unaltered. A tiny reduction in \Teff, on the order of
few hundred Kelvin (toward the lower values implied by {\sc cmfgen})
might be possible though, in the temperature range outlined above.

\subsection{The optical \NIII\ spectrum, and the 
formation of \NIII~\trip\ revisited}

In analogy to the previous section, in the following we compare
our current results for nitrogen, the third explicit element
considered, with results from alternative simulations, as an example
for elements where diagnostic lines in the optical are affected by
line overlaps in the EUV. In particular, we revisit the formation
of \NIII~\trip, which was already discussed by \citet{rivero11},
though with respect to our previous {\sc fastwind} version. 

Fig.~\ref{comp_niii} displays the comparison with {\sc cmfgen} (in
green) and {\sc fastwind} v10 (in red), and it is immediately clear
that the almost perfect agreement found for the H and He lines is no
longer present, though in most cases there is still a nice qualitative
concordance. Particularly \NIII~4003 and \NIII~\qua\ (from the quartet
system) are also in quantitative agreement, whereas \NIII~4379 is
strongly contaminated by \NII, \OII, and \CIII, which makes a clean
comparison with {\sc fastwind} v10 (and the spectroscopic analysis)
difficult. The often used diagnostic line \NIII~4097 (in the blue wing
of \Hd) is predicted to be much stronger at hotter temperatures 
(compared to {\sc cmfgen}) by both {\sc fastwind} versions.
Particular differences are present for the \mbox{(in-)famous} \NIII\ triplet
around 4640~\AA, where often the emission strengths predicted by either
of the three codes differ quite significantly. Only for the coolest
dwarf (d10v, d8v) and supergiant (s10a) models, where the triplet is
still in absorption, there is satisfactory agreement (even for the
neighboring, strong \OII\ lines at 4638.9 and 4641.8~\AA). The largest
discrepancies are found for models s8a and d6v, where {\sc cmfgen} predicts
weak emission, while {\sc fastwind} predicts weak (v11) or stronger
absorption (v10), and particularly for model s6a, where {\sc fastwind}
v10 predicts absorption, while v11 predicts moderate, and {\sc cmfgen}
considerable emission.

\begin{figure*}
\begin{minipage}{8.7cm}
\resizebox{\hsize}{!}
  {\includegraphics[angle=90]{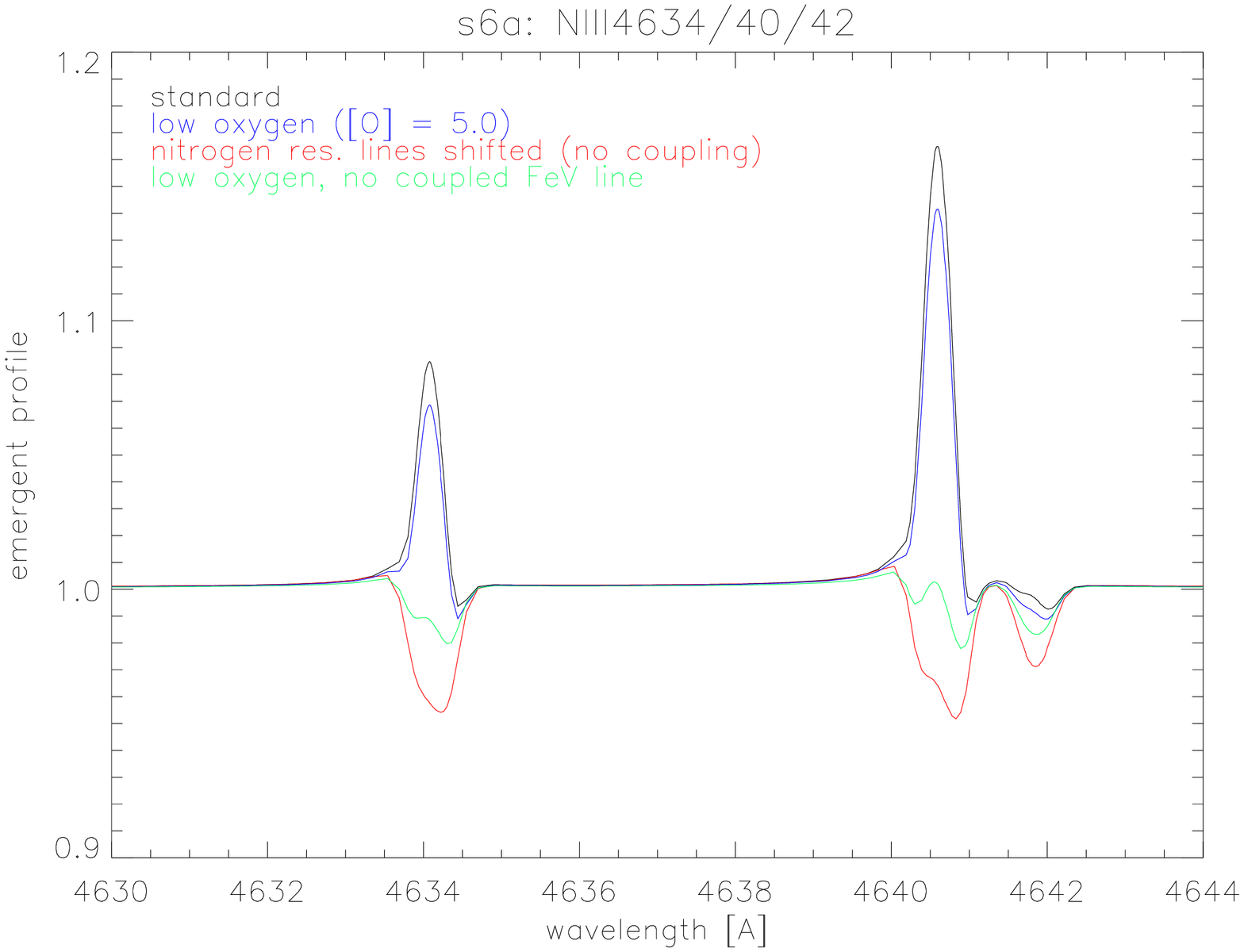}} 
\end{minipage}
\begin{minipage}{8cm}
\resizebox{\hsize}{!}
  {\includegraphics[angle=-90]{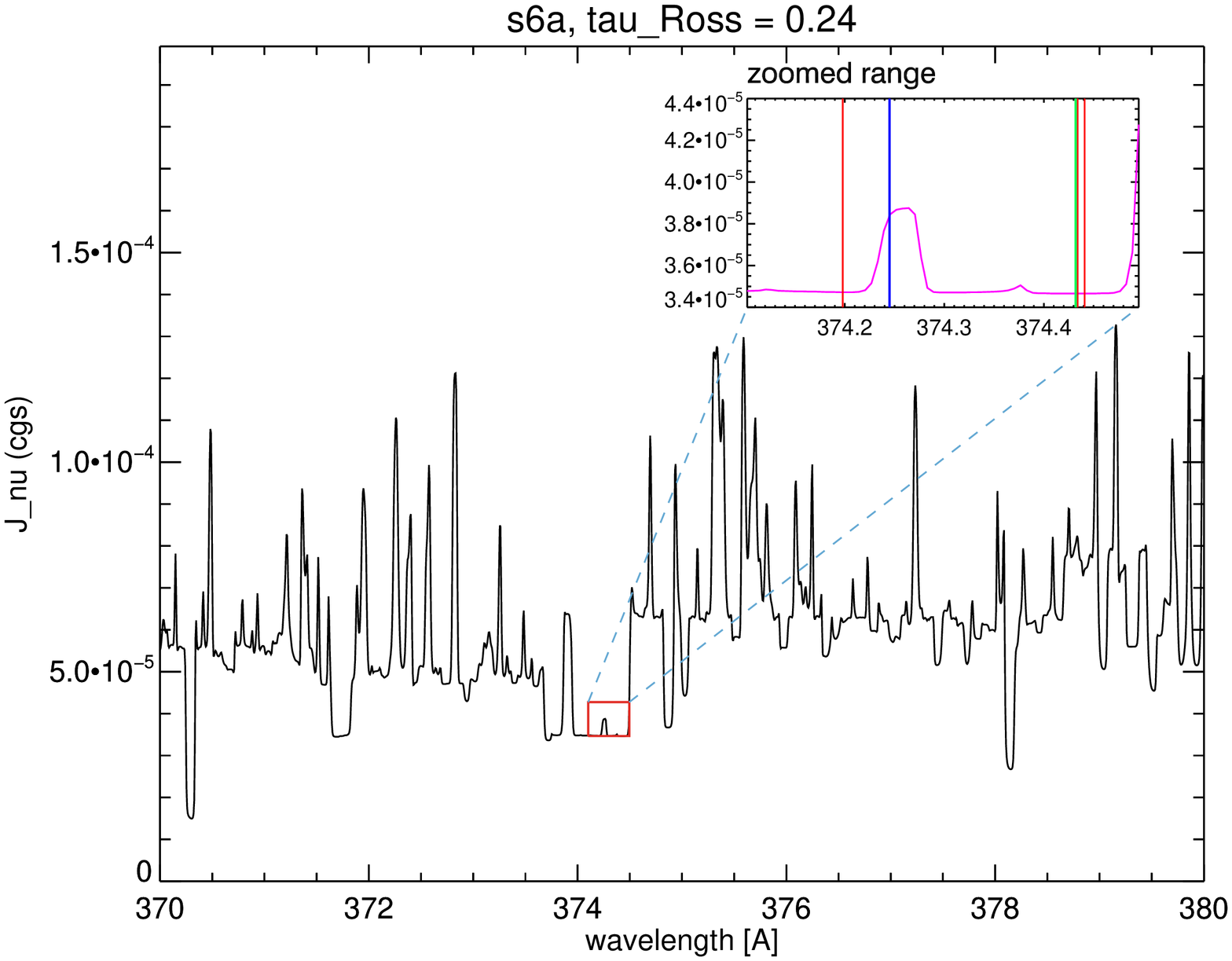}} 
\end{minipage}
\caption{On the formation of the \NIII\ \trip\ emission lines in hot
massive stars (here: model s6a, see Table~\ref{tabgrid}).  {\bf Left:}
Optical emission lines (no broadening, no degrading) for various
configurations of the EUV \NIII\ resonance lines (around 374~\AA)
coupled to the upper level ($3d$) of the optical transitions.  See
legend, and text. ''[X]'' means $\epsilon_{\rm X}$. {\bf Right:}
Corresponding EUV radiation field, $J_\nu(\taur=0.24)$. The insert
displays the decisive region, and participating lines. Red: \NIII\
resonance lines (three components); green: overlapping \OIII\
resonance line (out of a multiplet of 6 components); Blue: \FeV\ line
at 374.245~\AA. We note that 0.01~\AA\ correspond to 8~\kms.}
\label{s6a_niii_1}
\end{figure*}
\begin{figure}
\resizebox{\hsize}{!}
  {\includegraphics[angle=90]{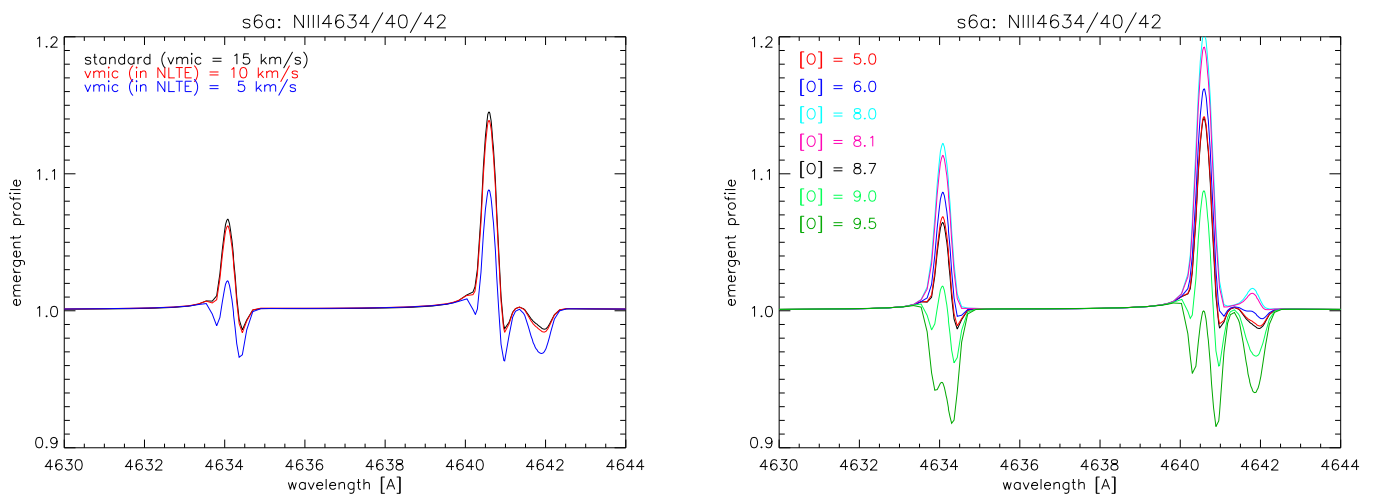}} 
\caption{As Fig.~\ref{s6a_niii_1}, left, but for different micro-turbulent
velocities in the NLTE model calculations. To allow for a meaningful
comparison of the impact on the occupations numbers alone, 
the micro-turbulent velocities in the formal integrals have been fixed at
15~\kms\ for all cases.}
\label{s6a_niii_vturb}
\end{figure}
\begin{figure}
\resizebox{\hsize}{!}
  {\includegraphics[angle=90]{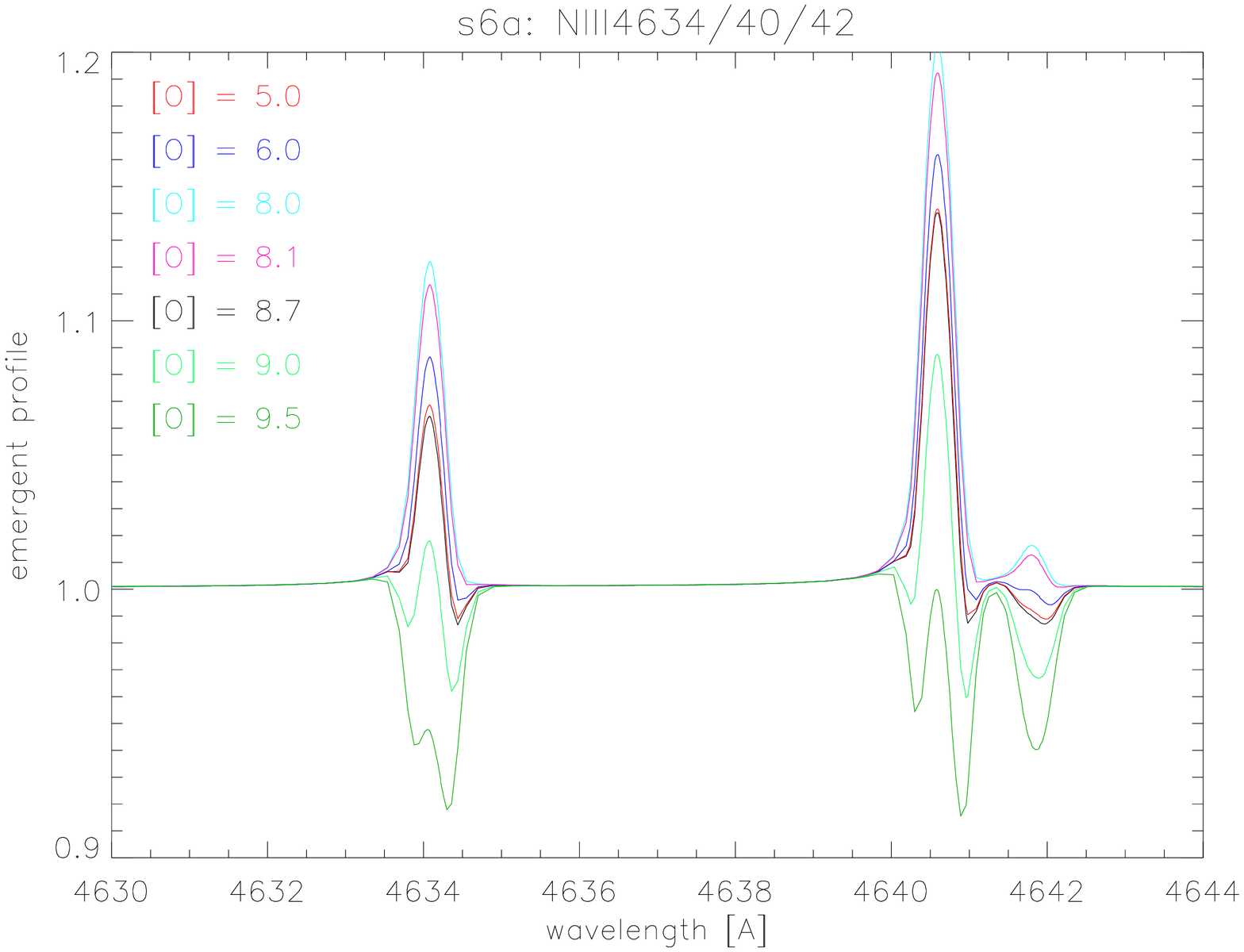}} 
\caption{As Fig.~\ref{s6a_niii_1}, left, but for different oxygen
abundances (see legend), and ``normal'' Fe content. We stress the
non-monotonic behavior. For increasing $\epsilon_{\rm O}$, the
emission strength at first increases (until $\epsilon_{\rm O} = 8.1$),
and then decreases, until strong absorption is produced (for
$\epsilon_{\rm O} = 9.5$).}
\label{s6a_niii_2}
\end{figure}

\begin{figure*}
\begin{center}
  {\includegraphics[width=10cm, angle=90]{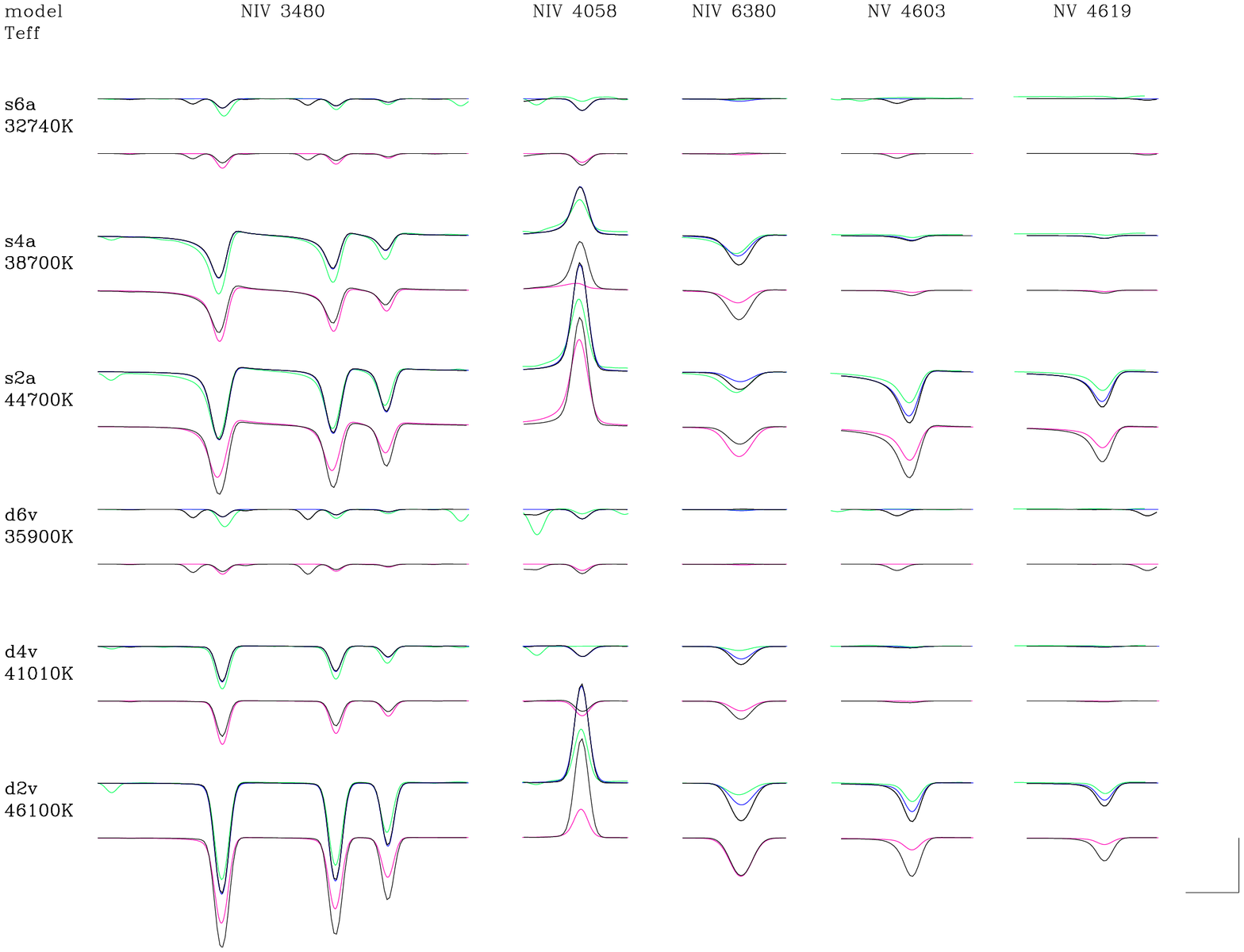}} 
\end{center}
\caption{As Fig.~\ref{comp_niii} (same scale), but for strategic \NIV\
and \NV\ lines of our hotter models. The blue profiles correspond
to models, where specific EUV transitions feeding the lower and upper
levels of \NIV~6380 and of \NV~$\lambda\lambda$4603-4619 
have been treated as isolated lines, i.e, ignoring blends. See text.}
\label{comp_nv}
\end{figure*}

Since model s6a was already scrutinized by
\citet{rivero11}, and displays the largest differences between v10 and
v11, we have re-investigated the line formation process of the triplet
lines using the new capabilities of v11.
In \citet{rivero11}, it was argued that the discrepancy between {\sc
cmfgen} (\NIII\ triplet in emission) and {\sc fastwind} v10 (\NIII\
triplet in absorption) is due to the line overlap between one
component (at 374.434 \AA) of the \NIII\ EUV resonance lines pumping
the upper level of the optical triplet transitions ($3d$), and one
component (at 374.432 \AA) of the \OIII\ resonance lines in the same
wavelength region (cf. insert of Fig.~\ref{s6a_niii_1}, right panel).
Since, at typical formation depths of the triplet lines, and for
atmospheric conditions similar to model s6a, the \OIII\ resonance
line source function is stronger than the \NIII\ one,
the pumping of \NIII\ $3d$ becomes more efficient than when the \OIII\ line
is absent (for details, see \citealt{rivero11}, their Sect.~7), and
the optical \NIII\ triplet appears in emission. If, on the other hand,
detailed line overlap effects are not included (as in
{\sc fastwind} v10), the triplet remains in absorption. We note
that the discussed effect is particularly strong for \Teff\ between 30
to 33 kK, whereas for hotter temperatures, the \OIII\ source function
becomes weaker than the \NIII\ one. At such temperatures, the pumping
by the \NIII\ resonance lines alone is sufficient to overpopulate the
$3d$ level and drive the \NIII\ triplet into emission.

Now, with the new version, all these effects are ``automatically''
accounted for, and, indeed, the corresponding optical lines appear in
emission (Fig.~\ref{s6a_niii_1}, left panel, black
profiles)\footnote{Though, compared to {\sc cmfgen}, to a lesser
degree.}. To check the arguments from \citet{rivero11}, we have
artificially shifted \NIII~374.434 by a small amount to avoid the
overlap, and, as predicted, the triplet components then appear in
absorption (red profiles). Our argumentation would also imply that, if
we alternatively decrease the oxygen abundance significantly (say,
down to $\epsilon_{\rm O} = 5$), the emission should vanish as
well. To our astonishment, however, the corresponding test (blue
profiles) did not confirm our expectation, and the lines remained in
emission. After carefully checking all processes, it turned out that,
in the decisive EUV wavelength range, there is one more strong \FeV\
line, which can couple with one of the other components of the \NIII\
resonance multiplet (at 374.198~\AA, see again insert), and, due to
the large mean intensity in this line, now plays the role of oxygen
when the latter is no longer present. Only if the \FeV\ line is excluded
from radiative transfer, the nitrogen triplet turns, for low
$\epsilon_{\rm O}$, into absorption (green). 

The lesson to learn from this exercise is that for certain transitions
which indirectly depend on EUV lines, there is always the chance that
line-overlap effects can have a large impact, because of the high line
density in the EUV regime. And since (as already argued with respect
to the \HeI\ singlet problem) the strength of the overlap strongly
depends on abundances, ionization conditions, micro-turbulent
velocities, and velocity field gradients, which control the line
optical depths and source functions, a variety of combinations can
lead to a variety of (non-monotonic) results. 

To examine the impact of \vmic, we display a comparison for three models
calculated with $\vmic = 15$~\kms\ (standard), 10~\kms (red), and
5~\kms (blue) in Fig.~\ref{s6a_niii_vturb}. To allow for a fair comparison of the impact on the
occupation numbers alone, in all three cases the formal integrals were
solved with the same (standard) value for \vmic. Obviously, the
differences between profiles from models with $\vmic = 15$ and
10~\kms\ are small, but for the lowest value, $\vmic = 5$~\kms\
(empirically not supported for a supergiant at 33~kK), the emission decreases
significantly. Here, because of the low micro-turbulent velocity, the
profile functions become considerably narrower, and the coupling between
the N and the Fe EUV resonance lines vanishes completely, whereas also
the coupling between the much closer N and O resonance lines
becomes weaker, inducing a reduction of emission strength.

As a final example for the (potential) non-monotonicity mentioned
above, we display, again for models s6a, the
reaction of the emission strength of the optical \NIII\ triplet on the
oxygen abundance in Fig.~\ref{s6a_niii_2}.
For abundances increasing until
$\epsilon_{\rm O} = 8.1$ (and ``normal'' $\epsilon_{\rm Fe}$), also the
emission strengths increase, though only moderately (\FeV!). For larger values, when
oxygen starts to dominate the line overlap even in the presence of \FeV, 
the emission begins to decrease, until, at $\epsilon_{\rm O} = 9.5$ (strongly
super-solar), the lines appear in strong absorption. In this case, the
source function of the oxygen resonance line, because of larger
optical depths also in connected transitions, has changed its
stratification in such a way as to prohibit effective pumping of the
$3d$ level.

Consequently, we advise to put, if at all, only low weight onto 
the triplet lines when analyzing the nitrogen spectrum; if it fits,
then fine, but if it does not fit, one might not conclude that there
is something ``rotten.'' After all, and as shown above, the emission
strength might depend, for example, on the oxygen and iron abundance,
and not on the nitrogen abundance (the quantity which needs to be
derived) alone, even if the impact of micro-turbulence is left
aside.

\subsection{The optical \NIV\ and \NV\ spectrum}

As a last comparison of optical spectra, Fig.~\ref{comp_nv} displays
important diagnostic lines from \NIV\ and \NV, for those models where
these lines are visible. For \NIV, the agreement between {\sc
fastwind} v11 and {\sc cmfgen} is satisfactory, though for the hottest
models the emission strength of \NIV~4058 and the absorption strength
of \NIV~6380 are larger when calculated via {\sc fastwind}. Compared
to the older version v10, the agreement is of similar quality; 
only for specific models (s4a and d2v) there are larger discrepancies.
Interestingly, there are no other (dominating) lines interacting
with the EUV resonance lines feeding the upper and lower
levels of \NIV~6380. This is visible when comparing with the profiles in
blue, calculated by treating these lines as isolated,
that is, discarding any direct overlap effect\footnote{
We note that to obtain a fair representation of \NIV~6380 in our
current standard approach (accounting for line overlaps), we 
use a specific treatment for some of the background lines
that are considered in a more approximate way, see Appendix~ \ref{no_upper}.}.

For close-to-solar nitrogen abundances, only the hottest models (d2v
and s2a) display well-developed optical 
\NV~$\lambda\lambda$4603-4619 profiles, sometimes with additional,
blue-wing wind absorption. These lines can serve as important
diagnostics for the effective temperatures of the earliest O-types
(for example, \citealt{rivero122}), and were shown, within {\sc fastwind}
v10, to agree with results from {\sc cmfgen} (also visible in
Fig.~\ref{comp_nv}).

On the other hand, the \NV\ lines (at least when comparatively strong)
do no longer agree with {\sc cmfgen} when calculated with the new
version v11. Indeed, they now are stronger (black versus green profiles
in Fig.~\ref{comp_nv}), and thus also stronger than predicted by v10
(black versus red). As for \NIV~6480, also the \NV\ profiles are barely
affected from EUV resonance line overlaps (black versus blue), and such
effects cannot be ``blamed'' for being responsible for the apparent
disagreement. As it turned out, the discrepancy is, if (i) compared to
{\sc cmfgen}, due to higher EUV fluxes around 266~\AA\ (the wavelength
of the 
\NV\ $2 \rightarrow 3$ transition pumping the lower level of the
optical \NV\ lines). (ii) If compared to v10, on the other hand, 
the discrepancy originates
from  higher temperatures (in v11) in the transition regime between
photosphere and wind. 
Consequently, v10 predicts lower \NV\ ionization fractions, and thus weaker
profiles. At our current knowledge, the agreement between v10 and {\sc
cmfgen} is just coincidental, and it is difficult to estimate which
prediction is closer to reality. More comparisons between theory and
observations for the hottest stars will certainly help to clarify this
issue. Until then, we warn about taking synthetic optical \NV\
line-profiles and strengths (from whatever code) at face value.
%
\begin{figure}
\resizebox{\hsize}{!}
  {\includegraphics[angle=90]{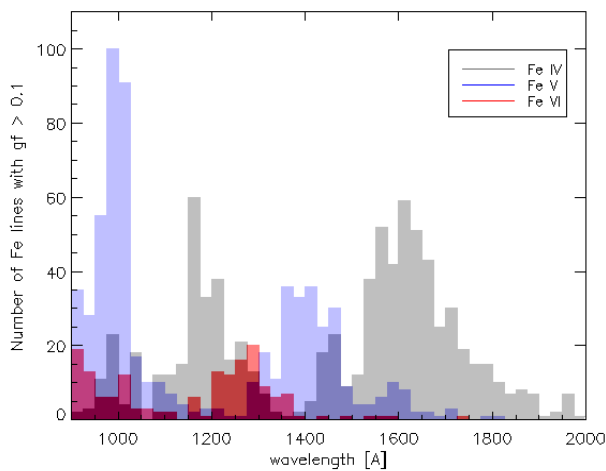}} 
\caption{Distribution of Fe\ {\sc iv,v,vi} lines with $gf \ge 0.1$
according to our line list, in the range between 900 to 2000~\AA. See
legend and text.} 
\label{linestat_fe}
\end{figure}

\begin{figure*}
\begin{center}
\begin{minipage}{9cm}
\resizebox{\hsize}{!}
  {\includegraphics[angle=90]{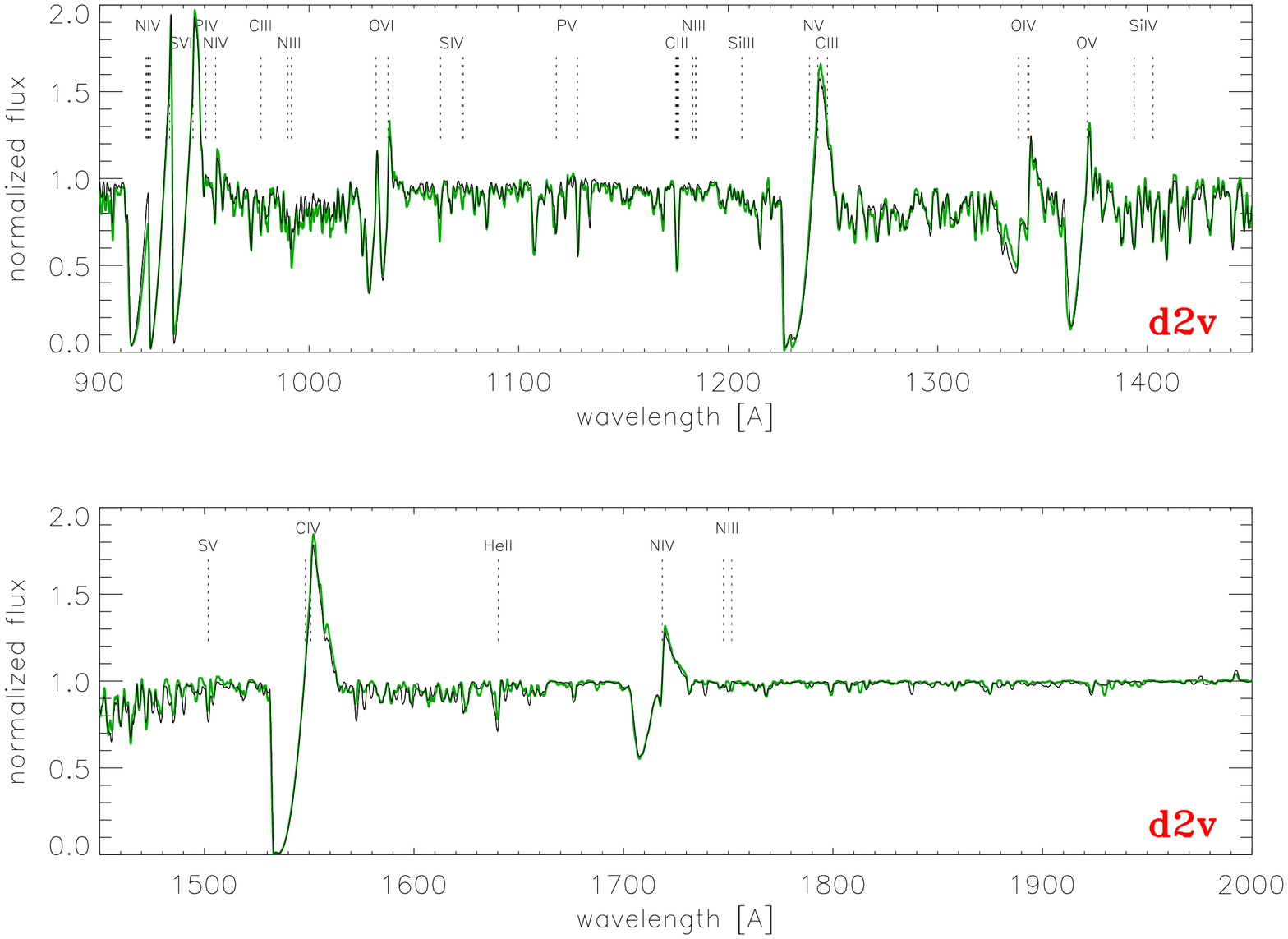}} 
\end{minipage}
\begin{minipage}{9cm}
\resizebox{\hsize}{!}
  {\includegraphics[angle=90]{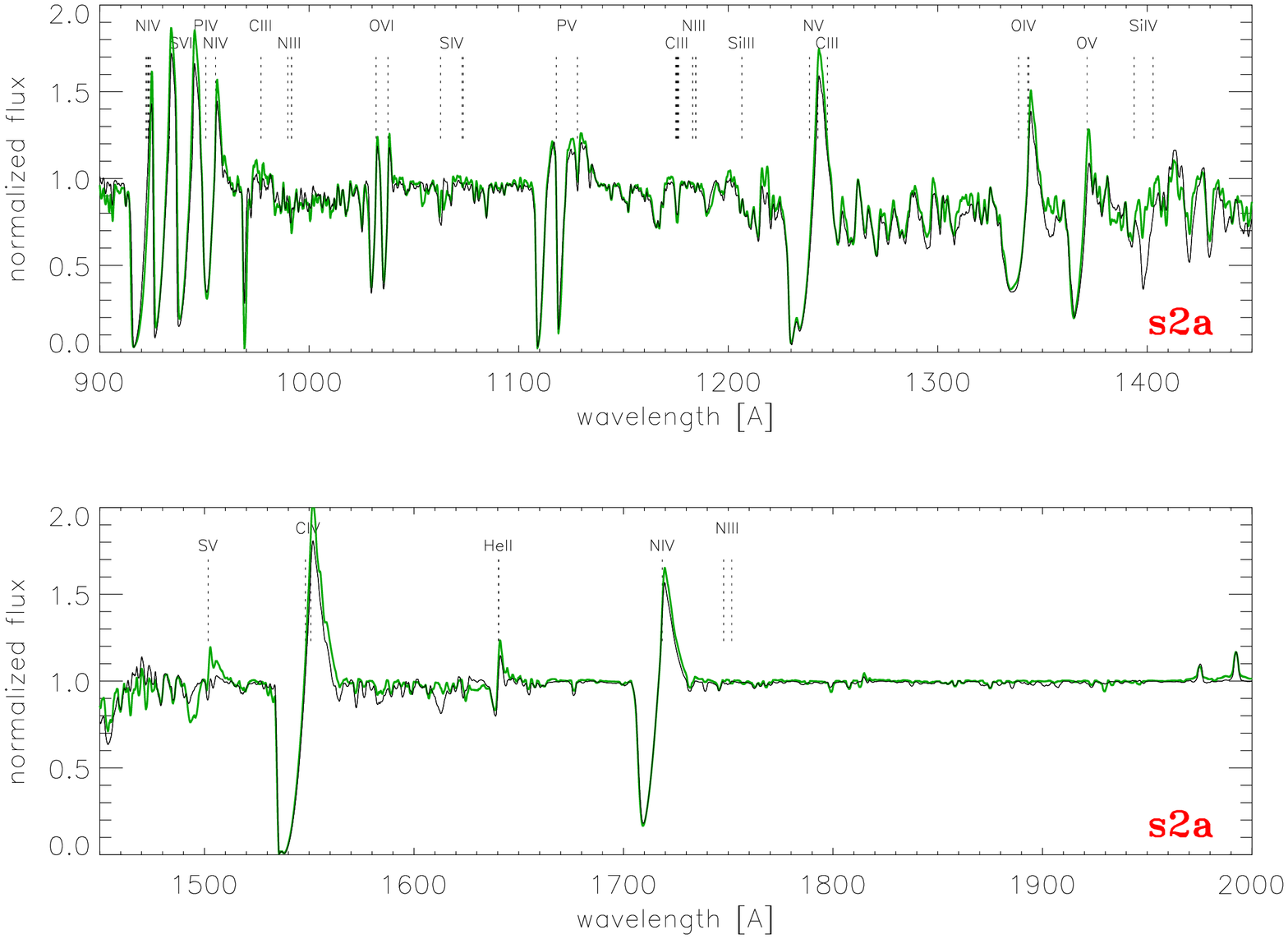}} 
\end{minipage}
\begin{minipage}{9cm}
\resizebox{\hsize}{!}
  {\includegraphics[angle=90]{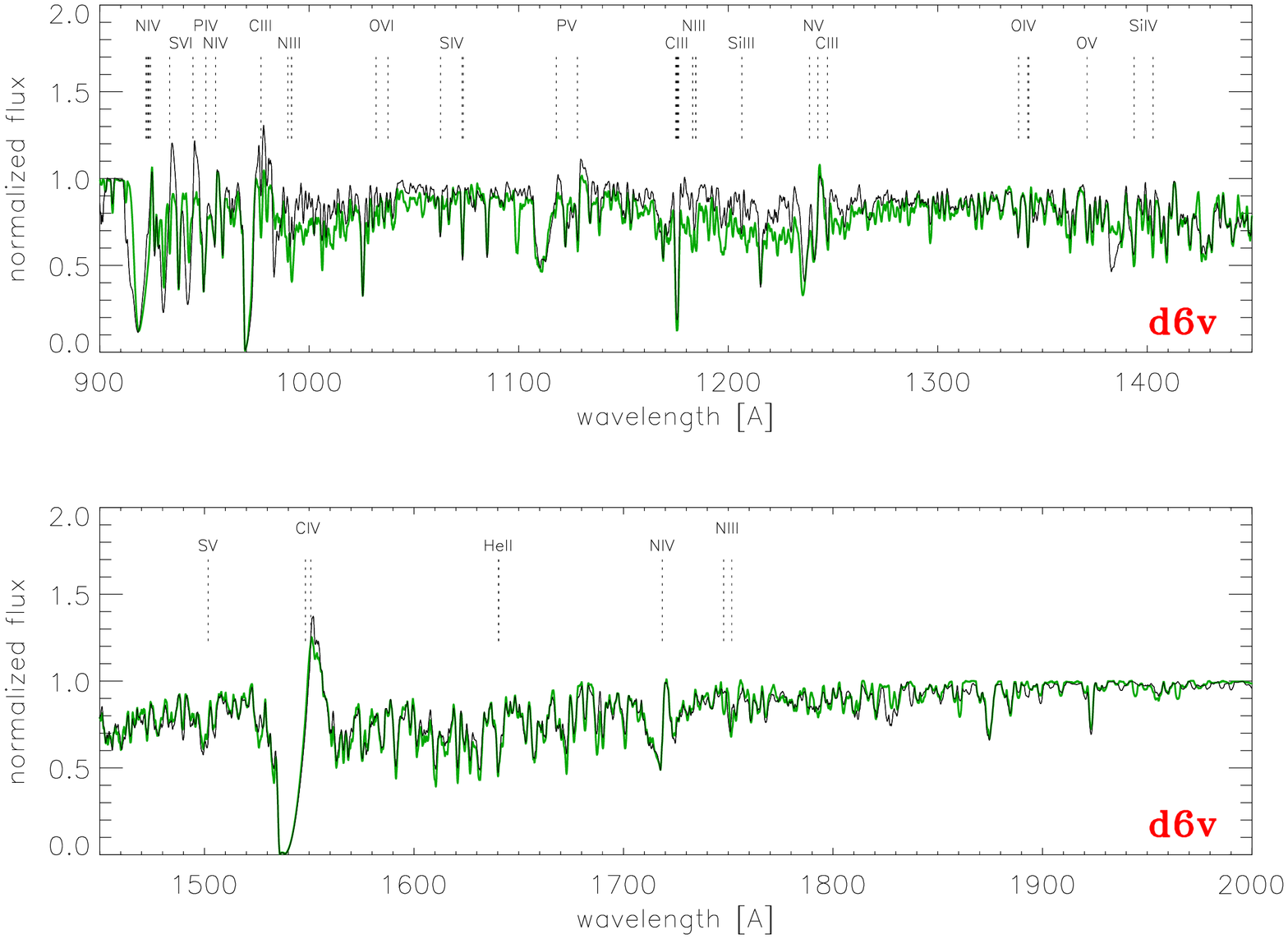}} 
\end{minipage}
\begin{minipage}{9cm}
\resizebox{\hsize}{!}
  {\includegraphics[angle=90]{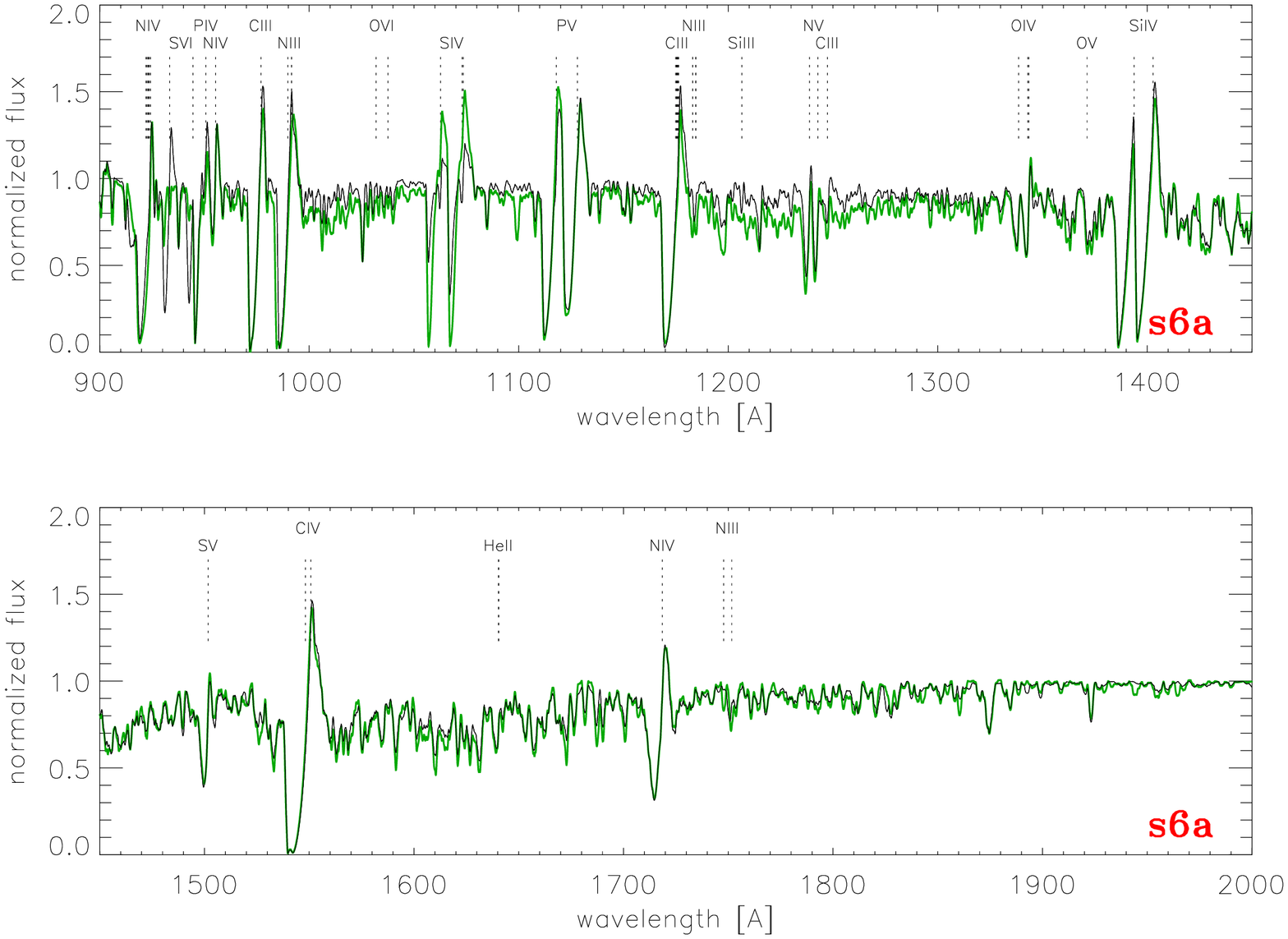}}
\end{minipage}
\begin{minipage}{9cm}
\resizebox{\hsize}{!}
  {\includegraphics[angle=90]{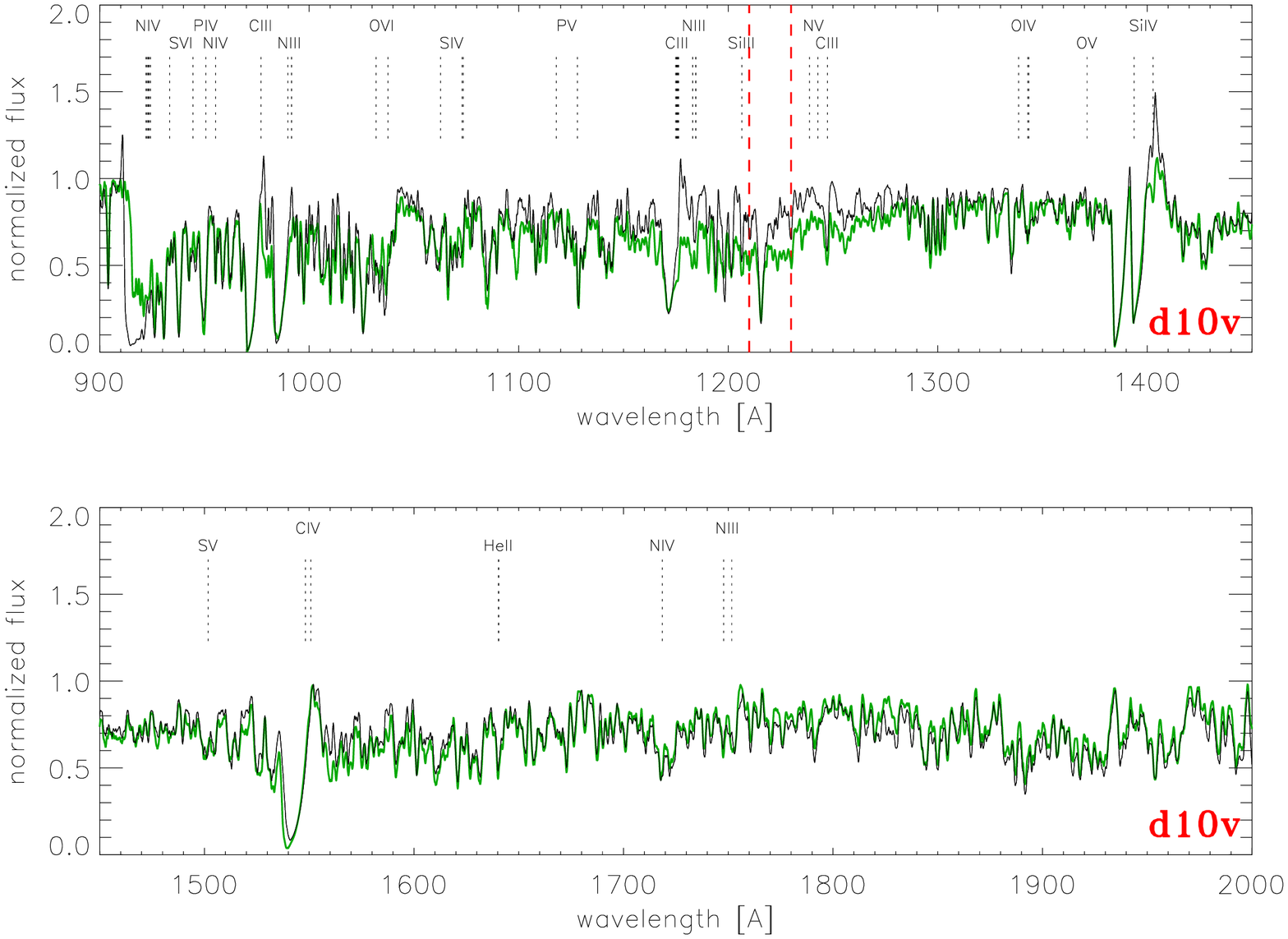}}
\end{minipage}
\begin{minipage}{9cm}
\resizebox{\hsize}{!}
  {\includegraphics[angle=90]{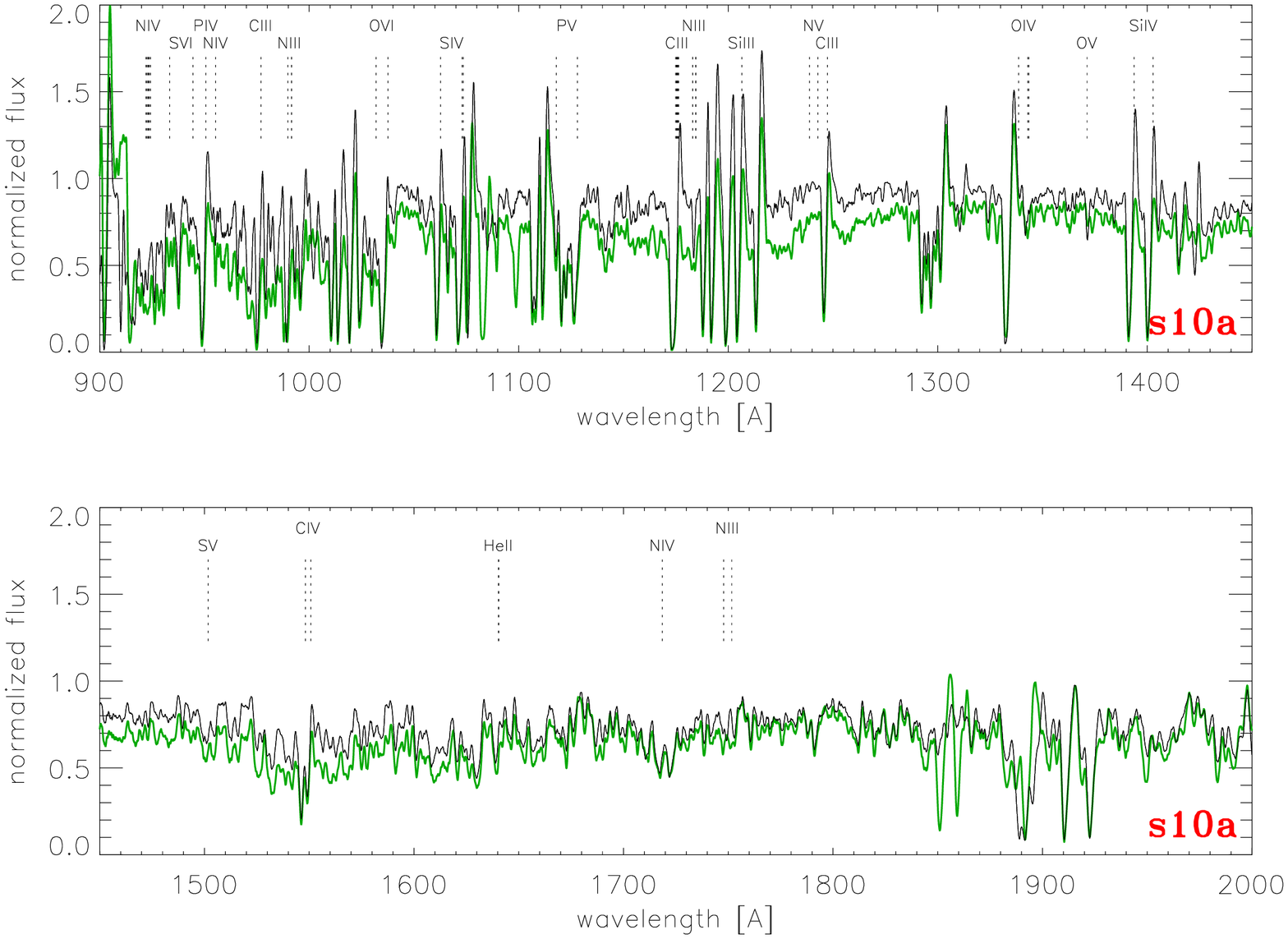}}
\end{minipage}
\end{center}
\caption{Sythetic UV spectra from {\sc fastwind} (black) and {\sc
cmfgen} (green), for hot, intermediate, and cool O- and early B-star models. To
allow for an easy comparion, all spectra have been convolved with
\vsini = 200~\kms. For model d10v, the range enclosed by red lines
is detailed in Fig.~\ref{d10v_1200}.}
\label{uv_spec}
\end{figure*}

\begin{figure*}
\resizebox{\hsize}{!}
  {\includegraphics[angle=90]{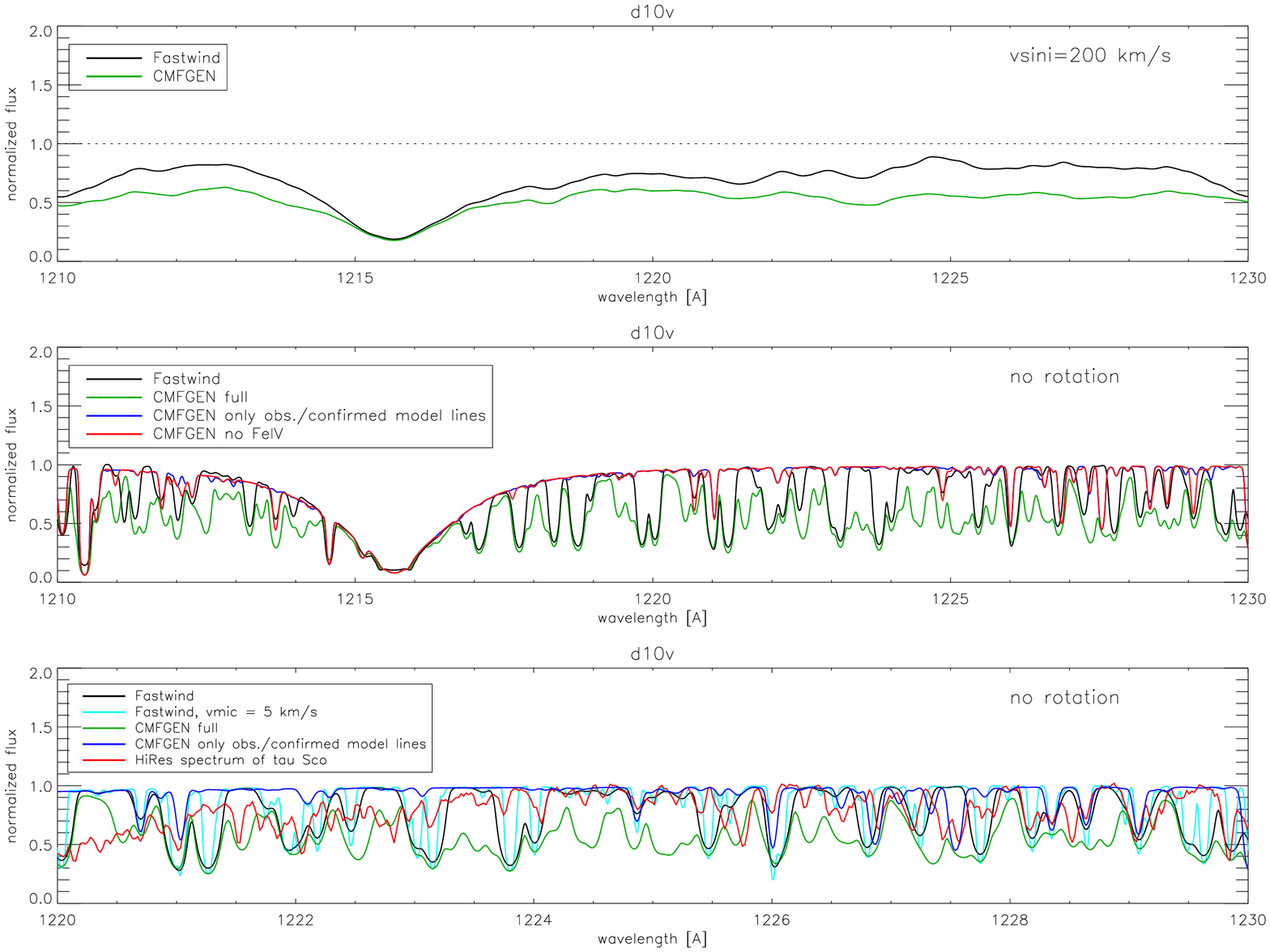}} 
\caption{Close-up of the region between 1210 or 1220 and 1230~\AA, for
model d10v. In addition to the ``standard'' spectra from {\sc
fastwind} and {\sc cmfgen} (with \vmic\ = 15~\kms, as in
Fig.~\ref{uv_spec}), we display results from a {\sc fastwind} formal
solution with \vmic\ = 5~\kms, and two further sets where
specific lines have been discarded from the formal solution in
{\sc cmfgen}, namely either all \FeIV\ lines, or all transitions 
(i.e., from all elements) between energy levels that are
predicted, but not observationally identified. For color coding, see
legend. All spectra have been calculated with an intrinsic resolution
of 0.01~\AA\ or better.  In the upper panel, they have been convolved
with \vsini = 200~\kms, as in Fig.~\ref{uv_spec}, whereas in the lower
ones no broadening has been applied.}
\label{d10v_1200}
\end{figure*}

\subsection{The UV range}
\label{uv_range}

In Fig.~\ref{uv_spec} below, we finally compare the region between 900 to
2000~\AA, for the hot (d2v, s2a), intermediate (d6v, s6a) and cool
(d10v, s10a) O- and early B-star regime. In the individual panels (two per
model), we have indicated the rest wavelengths of important
transitions from carbon to sulfur ions. 
The distribution of the ubiquitious
Fe lines is displayed in Fig.~\ref{linestat_fe}, for the stronger
lines ($gf \ge 0.1$) of ionization stages {\sc iv, v}, and {\sc vi}
relevant for our model grid. As is well-known, the line number
decreases with increasing ionization stage, and, except for a common
peak around 1000~\AA, the different ions culminate in different
ranges: \FeIV\ around 1200~\AA, close to \Lya, and between 1500 to
1700~\AA; \FeV\ between 1350 and 1500~\AA; and \FeVI\ between 1200 and
1375 \AA. Though Ni has similar line-densities, the impact on the
(photospheric) background opacity is weaker, due to a lower abundance 
(factor $\sim$18 lower than Fe for solar conditions).

To allow for a meaningful comparison, all spectra in
Fig.~\ref{uv_spec} have been convolved with \vsini\ = 200~\kms, since
otherwise the multitude of narrow Fe lines would hamper such an
effort. This is shown in Fig.~\ref{d10v_1200}, where a close-up of
the region between 1210 and 1230~\AA\ is displayed for convolved and
non-convolved spectra.

For the hottest models, the agreement between {\sc fastwind} v11 and
{\sc cmfgen} is almost perfect (also regarding \FeV\ and \FeVI) , and
only for the supergiant model (s2a), some discrepancies for \SV,
the region close to \SiIV, and the pseudo-continuum around 1600~\AA\ are visible. We note
that, in these models, \OVI\ is clearly present in the wind (though far
away from being saturated), even without any X-ray radiation field
(cf. Sect.~\ref{xrays}, and \citealt{Carneiro16}). 

Proceeding toward the intermediate models, the only ``light'' element
which disagrees is sulfur, where {\sc fastwind} indicates more \SVI\
and less \SIV, whereas \SV\ coincides. In both codes, \OVI\ has become
purely photospheric, and the only additional differences refer to the
iron forest around 1000 and 1200~\AA, where {\sc cmfgen} predicts
considerably more absorption. On the other hand, the region longward
of \CIV\ (dominated by \FeIV) agrees perfectly.

Considerable differences are present for our coolest models.
Though the wind lines display a satisfactory agreement (with larger
wind emission in {\sc fastwind}, due to a weaker contamination by the
background), the iron forest around 1000~\AA, and particularly around
1200~\AA\ (denoted by red dashes) is significantly stronger in the
d10v model from {\sc cmfgen}. Indeed, the pseudo-continuum around
\Lya\ is shifted to a level of $\sim$60\% of the true continuum,
compared to a level of $\sim$80\% in {\sc fastwind}
(Fig.~\ref{d10v_1200}, upper panel). For the corresponding supergiant
model, s10a, such a discrepancy is even present throughout the almost
complete range shortward of 1700~\AA.

Before providing further details, we stress that in concert with most
strategic wind lines, also the \PV\ line profiles perfectly agree for
all our grid models, including those that are not displayed here. This
implies that the predicted ionization fractions of phosphorus coincide
perfectly, strengthening our confidence in using this line as a
diagnostic tool to constrain the amount of velocity-space
porosity induced by optically thick clumps (\citealt{fullerton06, oskinova07, Sundqvist11,
Sundqvist14}).

In Fig.~\ref{d10v_1200}, we now investigate the origin of the
discordance between the pseudo-continua within the cooler (but, for
specific ranges, also the intermediate) models, by means of a close-up
into the wavelength range between 1210 and 1230~\AA, and model d10v. 
Whereas, in {\sc cmfgen}, more or less the complete range around \Lya\
is affected by photospheric line absorption, the spectrum predicted by
{\sc fastwind} coincides only for the stronger lines. In between these
lines, however, many line-free regions are visible, comprising roughly
50\% of the total range. These differences are responsible for the
different pseudo-continuum flux-levels mentioned earlier.

Since the hotter models (including d4v and s4a) do not display such a
discrepancy (if at all, {\sc fastwind} predicts more \FeV\ absorption
around 1600~\AA), it is most likely that the origin relates to \FeIV.
This expectation is confirmed when calculating (within {\sc
cmfgen}) the formal integral excluding all \FeIV\ lines (red
spectrum in the middle panel). In this case, only very few lines
are still present in the considered spectral range, and the pseudo-continuum
does not become depressed. 

We have checked our Fe model atom, and realized that (at least) the
$4d$ levels and corresponding transitions are missing. This certainly
needs to be further investigated (work in progress). However, we
also realized that just in the considered range around 1200~\AA, the
majority of lines synthesized by {\sc cmfgen} (including a multitude
of very weak, overlapping ones) are due to transitions between energy
levels that have been theoretically predicted, but, until to-date, not
been observationally identified (in the following denoted as
``non-observed levels''). In the range above 1700~\AA, where {\sc
cmfgen} and {\sc fastwind} spectra agree, such transitions play only a
minor role. When excluding all transitions between non-observed energy
levels in Fig.~\ref{d10v_1200}, the result is quite similar to
the case when excluding \FeIV\ completely (blue versus red spectra
in the middle panel). We thus conclude that it is the multitude of
such transitions that is responsible for the strong continuum
depression. 

Nevertheless, and irrespective of this discrepancy between synthetic
spectra, there is another issue: For the displayed comparison, we have
deliberately chosen model d10v (and not s10a, where the discrepancy is
even larger), since (i), with \Teff = 28~kK, it is not as contaminated
by \FeIII\ as model s10a (\Teff $\approx$ 24~kK), and (ii), we can
compare our results also to the high resolution {\sc copernicus}
spectrum of $\tau$~Sco (B0.2 V), a well known object with very low
rotation rate (close to zero, for example, \citealt{nieva12}). The latter
condition enables to record most individual photospheric lines that
are actually present. The spectra, with a nominal resolution of
0.05~\AA, have been taken from \citet{RogersonUpson77}. We note that
$\tau$~Sco is only slightly hotter than our grid model d10v 
(\citealt{Repo05, Marcolino09, martins12, nieva12}), which, in
conjunction with its higher gravity, should result in similar
ionization conditions.

From the lower panel of Fig.~\ref{d10v_1200}, red spectrum, it is
evident that many of the actually observed lines are much weaker than
predicted by both {\sc cmfgen} and {\sc fastwind}. To demonstrate
that this is not an effect of too large a micro-turbulent velocity
(adopted as \vmic = 15~\kms\ in our ``standard'' models), the turquois
spectrum in the lower panel\footnote{Synthesized from a {\sc fastwind}
formal solution.} has been calculated with \vmic = 5~\kms, a value
inferred from an optical analysis of $\tau$~Sco \citep{nieva12}.
Evidently, the discordance with the observations is still present.
Moreover, various predicted lines are even absent in the
observations. Wrong normalization is not likely an issue, at least
around 1228-1230~\AA, since the strengths of those lines that are
simultaneously present in theory and observations are quite
similar\footnote{The lower continuum in the observations at roughly
1220~\AA\ is due to the red \Lya\ wing, being dominated by
interstellar hydrogen.}.

One might argue that $\tau$ Sco is not well-suited for the above
comparison, because of its strong and complex magnetic field
\citep{Donati06}. Indeed, the UV wind-lines seem to be significantly
affected by the magnetic field \citep{Petit11}, but the photospheric
lines in the optical (from H, He, and various metals) are prototypical
for non-magnetic stars of the given spectral type (for example, \citealt{nieva12}).
We thus expect that also the photospheric UV Fe-lines should be
representative for normal conditions. Moreover, we checked $\tau$
Sco's UV spectrum against the one from 10 Lac \citep{Brandt98}, a
non-magnetic star of similar spectral type (O9V).  Both spectra turned
out to be very similar, displaying only few lines in the range
around \Lya. (We note that the few, actually present lines are
somewhat shallower and broader in 10~Lac, because of its non-vanishing
projected rotational speed).

Thus, it is quite probable that both codes might overestimate
the \FeIV\ line blocking (for such models where \FeIV\ is
significantly populated) in certain wavelength regions, since many 
of the predicted lines from non-observed energy levels might not be
present in reality. Of course, detailed comparisons with observations
for a variety of objects are required to substantiate such a
hypothesis, and, in case, to allow us to improve the atomic
models and line-lists.

\section{X-rays from wind-embedded shocks}
\label{xrays}

Obviously, reliable UV diagnostics require a sufficient 
decription of the ionization balance of relevant atomic species. This
balance can be significantly affected by X-ray emission from
wind-embedded shocks. Already the first X-ray satellite observatory,
{\sc einstein}, has revealed that O-stars are soft X-ray sources
\citep{ Harnden79, Seward79}. Many subsequent and recent studies
(particularly using {\sc chandra} and XMM-{\sc newton}) have improved
our knowledge on corresponding features and processes (for a brief
summary, see \citealt{Carneiro16}). This X-ray (and EUV) emission is
widely believed to originate in wind-embedded shocks, which in turn
should be related to the so-called line-deshadowing instability,
``LDI'' (\citealt{LS70, ORI, OCR88, Owocki94b, Feldmeier95}).
Due to both direct and Auger ionization (where, under prototypical
conditions, the latter process only affects \NVI\ and \OVI, see
\citealt{Carneiro16} and references therein), the ionization
equilibrium for all ions with edges $\la$~350~\AA\ can be modified,
particularly in the intermediate and outer wind. This has not only
consequences for (E)UV metal lines, but also for optical lines with
levels pumped by such transitions (see previous sections, and
\citealt{MartinsHillier12} for the specific case of optical carbon
lines), and even for HeII~1640 and HeII~4686 (again,
\citealt{Carneiro16}).

Corresponding processes have been implemented into various
unified atmosphere codes designed for the analysis of hot stars (see
Sect.~\ref{intro}), both to allow for a meaningful analysis of
affected lines, and also to calculate the background opacities (from
the cool wind material) in the X-ray regime, required for the 
diagnostics of X-ray emission lines.  The distribution of the shocks
and their emission is usally estimated by means of parameterized
models (with various degrees of complexity regarding their cooling
zones, \citealt{Hillier93, Feldmeier97b, OwockiSundqvist13}),
described by input quantities such as filling-factors, $\fv(r)$, and
shock (front) temperatures, $T_{\rm s}(r)$.

\subsection{``Unified'' volume filling factors} 
Also in {\sc fastwind} v10, such X-ray emission from
wind-embedded shocks has been implemented, and we keep this treatment
in v11. The implementation itself has been detailed by
\citet{Carneiro16}, together with a study on the consequences of this
emission for the X-ray, UV and optical spectrum.

These authors also discuss specific scaling relations
for the X-ray luminosity as a function of $\mdot/\vinf$, but note a
certain discrepancy between their description (based on
\citealt{Feldmeier97b}) and the study by \citet[hereafter
Ow13]{OwockiSundqvist13}. In particular, the latter predicts different
scaling relations than presented by \citet{Carneiro16} and earlier
work by \citet{OwockiCohen99}. In this section, we try to unify both
investigations, and provide corresponding results.

The major difference between the approaches by \citet{Feldmeier97b}, 
related earlier work \citep{Hillier93}, or follow-up studies
(including \citealt{Carneiro16}) and Ow13 roots in the expression for
the effective X-ray emissivity. Whereas in the former approach, this
quantity scales with $\rho^2$ for both radiative and adiabatic cooling
zones (at least for standard assumptions), it scales with $\rho$ for
radiative and with $\rho^2$ for adiabatic cooling in Ow13,
giving rise to different dependencies between radiation field and wind
density. This might be important for several applications, since for
stars with not too thin winds radiative cooling prevails in the major
part of the wind.  

The basic reason for the difference is related to the (radiative)
cooling length, $l_{\rm s}$ (with subscript ``s'' for shock\footnote{Here
and in the following, we mostly follow the notation by Ow13, but use
$\fv$ instead of $f_{\rm V}$, to discriminate the volume filling factor of
X-ray emitting material from the volume filling factor associated with
``cold,'' overdense material (= clumps).}), which
in \citet{Feldmeier97b} does not explicitly enter the emissivity,
since at each point in the wind the emissivity is averaged over the
shock cooling zone, such that the length cancels out (their Eq.~2).
Ow13, on the other hand, accounts for the actual size of the cooling
zone(s) when calculating the X-ray luminosity (their Eqs.~6 and 18),
where $l_{\rm s}$ varies as $l_{\rm s} \propto \rho^{-1}$. In this approach then,
the effective volume filling factor from an ensemble of shocks results
in
\beq
\label{fv_owo}
\fv(r) = \frac{1}{1+r_{\rm s}/l_{\rm s}} \frac{\dd N_{\rm s}(r)}{\dd \ln r} := 
\frac{n_{\rm s}(r)}{1+r_{\rm s}/l_{\rm s}},
\eeq
where $r_{\rm s}$ is the position of the shock front, and $N_{\rm s}(r)$ the
cumulative number of shocks until radius $r$. In other words, for a
constant number of new, emerging shocks, $n_{\rm s}(r) = const$, the volume
filling factor remains constant only if $l_{\rm s} \gg r_{\rm s}$,
that is, for adiabatic
cooling\footnote{Under the conditions considered by Ow13, radiative
and adiabatic cooling rates are equal if $l_{\rm s}=r_{\rm s}$ , whereas,
following \citet[Eq.~9]{Feldmeier97b}, 
$t_{\rm c}/t_{\rm f} \approx  5 (v(r)/u_{\rm j}(r)) \cdot (l_{\rm s}/r_{\rm s})$,
with radiative cooling time $t_{\rm c}$, dynamical wind flow time
$t_{\rm f} = r/v(r)$, and jump velocity, $u_{\rm j}(r)$.}. 
Consequently, the X-ray
emissivity, $\eta_{\rm x}(r) \propto \fv\, \rho^2$, depends on $\rho^2$, due
to the cooling function alone. On the other hand, for $l_{\rm s} <
r_{\rm s}$, in the radiative regime, $\fv \propto n_{\rm s}(r) l_{\rm s}/r_{\rm s} \propto
n_{\rm s}(r)/(\rho r_{\rm s})$, such that
the effective emissivity depends linearly on $\rho$.

Although derived in a physically strict way by Ow13, 
the radiative limit of Eq.~\ref{fv_owo} can be simply explained as
follows:
The total volume $V_{\rm s}$ of an X-ray emitting ensemble of radiative shocks,
each with cooling length $l_{\rm s} < \Delta r$, and located inside a shell
of volume $V = 4\pi r^2 \Delta r$, can be estimated as 
\beq 
V_{\rm s} \approx \Delta N l_{\rm s} r^2 \Delta \Omega. 
\eeq
Here, $\Delta N = N_{\rm s}(r+\Delta r)-N_{\rm s}(r)$ 
is the number of shocks in between $r$ and $r+\Delta
r$, and $\Delta \Omega$ is the
solid angle subtended by one shock, for simplicity assumed to be
equal for all shocks inside $\Delta r$. The effective volume filling
factor is then
\beq
\fv(r) = \frac{V_{\rm s}}{V}=
\frac{\Delta N l_{\rm s} r^2 \Delta \Omega}{4\pi r^2 \Delta r} = 
\frac{\Delta \Omega}{4\pi} \frac{\Delta N}{\Delta r} l_{\rm s} \approx
 {const} \cdot\frac{\dd N_{\rm s}}{\dd \ln r} \frac{l_{\rm s}}{r}.
\eeq
This is the radiative limit of Eq.~\ref{fv_owo}, corresponding to
shocks with a comparatively small cooling length, due to a large density,
and proportional to both the shock distribution and $\rho^{-1}$.
Now, if (in the outer wind, or generally for winds of low
density) the density becomes as low as to result in $l_{\rm s} \ge r_{\rm s}$, adiabatic
cooling takes over, and the radiative cooling length does no longer
play any role. Then (for details, see Ow13), the volume filling
factor becomes solely controlled by the shock distribution, and the
emissivity depends on $\rho^2$. 
\renewcommand{\arraystretch}{1.2}
\begin{table}
\caption{Stellar, wind, and X-ray emission parameters common to all
our $\zeta$~Pup-like models. The first row presents the adopted
photospheric parameters, similar to those derived by \citet{Najarro11}
and \citet{bouret12}. $\YHe = N_{\rm He}/N_{\rm H}$ is the Helium
abundance by number, whereas $(Z/Z_\odot)_{\rm A}$ is the mass
fraction of element A, normalized to the corresponding abundance from
\citet{asplund05}, and taken from \citet{Cohen10} (but see text). The
second row (left part) displays the wind parameters (for a wind
adopted as smooth), with $\beta$ the exponent of the typical wind
velocity law. The right part of the second row displays the X-ray
emission parameters (cf. \citealt{Carneiro16} for details) common to
all models, with maximum shock temperature $T_{\rm s}^\infty = T_{\rm s}(r
\gg \Rstar)$ reached far out in the wind. $u_{\rm j}^\infty$ is the
jump velocity, $u_{\rm j}(r)$, at large distances from the star, where the
latter has been parameterized following $u_{\rm j}(r) =
u_{\rm j}^\infty(v(r)/\vinf)^{\gamma_{\rm x}}$, with exponent $\gamma_{\rm
x}$ (see also \citealt{pauldrach94c}). We note that $T_{\rm s}(r) \propto
u_{\rm j}^2(r)$.}
\label{tab_zeta_pup}
\tabcolsep1.8mm
\begin{tabular}{ccccccc}
\hline 
\hline
\Teff[K]  & \logg & \Rstar/\rsun & \YHe & $(Z/Z_\odot)_{\rm C}$ & $(Z/Z_\odot)_{\rm N}$ &
    $(Z/Z_\odot)_{\rm O} $  \\
40,000 & 3.6 & 19.5 & 0.14 & 0.08  &  5.0  &  0.20\\
\hline
\end{tabular}
\tabcolsep0.9mm
\begin{tabular}{ccc||cccc}
\mdot[\msunyr] & \vinf[\kms] & $\beta$ & $u_{\rm j}^\infty$[\kms] &
   $T^\infty_{\rm s}$[K] & $\gamma_{\rm x}$ & \Rmin/\Rstar \\
8.0 $\cdot 10^{-6}$ & 2250. & 0.9  & 625. & 5.6 $\cdot 10^6$ & 1.  & 1.5\\
\hline
\hline
\end{tabular}
\end{table}
\renewcommand{\arraystretch}{1.0}
\renewcommand{\arraystretch}{1.2}
\begin{table}
\caption{Volume filling factors for four different model series. The
first row quantifies the constant filling factor adopted in the
first series, whereas rows 2 to 4 refer to the power-law shock
distribution, Eq.~\ref{n_s}.}
\label{tab_fvol}
\begin{tabular}{cclc}
\hline 
\hline
series &  approach  &   &  $p$ \\  
1  & $\fv = const$ & \fv = 3.6 $\cdot 10^{-3}$ & -- \\
2  & $\fv$ from Eqs.~\ref{fv_owo} and \ref{n_s}  &  $n_{\rm so}$ = 0.1   & 0 \\
3  &         ``      &  $n_{\rm so}$ = 2.7   & 1 \\
4  &         ``      &  $n_{\rm so}$ = 38.   & 2 \\
\hline
\end{tabular}
\end{table}
\renewcommand{\arraystretch}{1.0}

Thus, it is basically the assumption of a spatially constant volume
filling factor (as adopted in \citealt{Carneiro16}) that differs from
the \textbf{result} by Ow13 on $\fv$, and it is easy to unify both approaches.
At first, we checked that the (radiative) cooling lengths as quantified by
\citet{Feldmeier97b} (and used by {\sc fastwind}) and by Ow13 are
consistent: indeed, the basic dependencies 
\beq
\frac{l_{\rm s}}{r_{\rm s}} \propto \frac{T_{\rm s}^2}{\rho_{\rm ws} r_{\rm s}} \propto
\frac{T_{\rm s}^2}{\mdot}r_{\rm s} v_{\rm ws},
\label{l_over_r}
\eeq
with $\rho_{\rm ws}$ and $v_{\rm ws}$ the wind density and wind velocity at
the shock front, respectively, and adopting a cooling function
$\propto T^{-1/2}$ in the relevant temperature range (for example,
\citealt{Raymond76, Schure09}) -- are identical, and only the numerical
fore-factors are different, by a factor of three (with $l_{\rm s}$ from Ow13 being
the larger one).

In the following simulations, we neglected the potential effects of
thin shell mixing (which can be easily included if required), and
introduced a new input option, allowing us to replace the previous
(spatially constant) volume filling factor by Eq.~\ref{fv_owo}. Again following Ow13, we
adopt
\beq
n_{\rm s}(r) = n_{\rm so}\left(\frac{\Rmin}{r}\right)^p,
\label{n_s}
\eeq
where $n_{\rm so}$ and $p$ are additional input parameters, and $n_{\rm so}$
is adapted until a specified $L_{\rm x}$ value is reached. The input
quantity \Rmin\ (denoted as $R_{\rm o}$ by Ow13) is the onset of shock
formation and X-ray emission,
typically on the order of 1.5 \Rstar\ (cf. \citealt{Carneiro16} and
references therein), and other input parameters describing the shock
temperature just after the discontinuity are used in agreement with
the description in \citet[Sect. 2.1]{Carneiro16}. Finally, and
somewhat inconsistent, we keep our previous method to switch from the
radiative to an adiabatic post-shock temperature and density
stratification when the cooling and the dynamical wind flow times are
identical.

All following simulations have been performed with our new {\sc
fastwind} version including complete CMF transfer, and photospheric
and wind parameters in most cases similar to those derived for the
O4I(f) supergiant $\zeta$~Pup, see Table~\ref{tab_zeta_pup}. For
simplicity, and to allow for an easy comparison with analytical
predictions and earlier work, the wind is adopted as smooth, with a
mass-loss rate (if not specified differently) that matches the \Ha\
wind emission. This mass-loss rate is a factor of four higher
than derived from analyses accounting for optically thin clumping
\citep{Najarro11, bouret12}. CNO abundances have been been taken from
\citet{Cohen10}\footnote{We note that these numbers, particularly the
nitrogen abundance, are different from the values provided in the 
analysis by \citet{bouret12}.}, again for consistency with earlier
work.

\begin{figure}
\resizebox{\hsize}{!}
  {\includegraphics[angle=90]{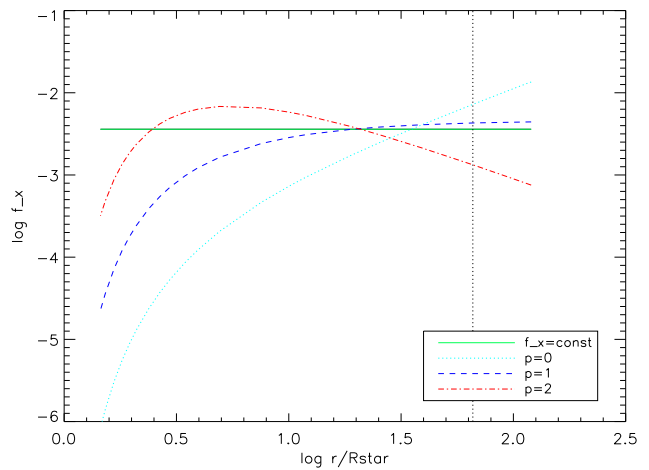}} 
\caption{X-ray volume filling factor, $\fv$, as a function of $\log
r/\Rstar$, for different shock distributions parameterized according
to Eqs.~\ref{fv_owo} and \ref{n_s} (see legend, and
Table~\ref{tab_fvol}), and stellar and wind parameters for our
$\zeta$~Pup-like model (Table~\ref{tab_zeta_pup}). The vertical
dotted line denotes the region where cooling and wind flow times are
equal, $t_{\rm c} = t_{\rm f}$. Due to the large mass-loss rate, the ratio
$l_{\rm s}/r_{\rm s}$ remains smaller than unity throughout the wind.}
\label{plot_fvol}
\end{figure}

Fig.~\ref{plot_fvol} displays the run of the volume filling factor for
four different X-ray emission models (see Table~\ref{tab_fvol}), using
either a constant volume filling factor, $\fv = const$, or a volume filling
factor consistent with Ow13, and different shock distributions, with
$p \in [0,1,2]$ according to Eqs.~\ref{fv_owo} and \ref{n_s}.  All values
for $\fv$ and $n_{\rm so}$, respectively, have been chosen such that the
resulting X-ray luminosity (in the range 0.1 to 2.5 keV) becomes 
$L_{\rm x}/L_{\rm bol} = 10^{-7}$, a prototypical value for massive O-stars
(\citealt{chlebo91, Sana06}).

Since, due to the large mass-loss rate, the ratio $l_{\rm s}/r_{\rm s}$ remains
below unity troughout the wind for this specific parameter set,
the effective volume filling factor behaves as 
\beq
\fv(r) \rightarrow n_{\rm s}(r) \frac{l_{\rm s}(r)}{r} \propto n_{\rm s}(r) \frac{T_{\rm s}^2(r)}{\mdot}r
v(r) \propto r^{1-p} v(r),
\eeq
which leads, for roughly constant shock temperatures, to a steep
increase $\propto r v(r)$ for $p=0$, a moderate increase $\propto
v(r)$ for $p=1$, and to increasing and later on decreasing values
$\propto v(r)/r$ for $p=2$, in accordance with Fig.~\ref{plot_fvol}.
\begin{figure}
\resizebox{\hsize}{!}
  {\includegraphics[angle=90]{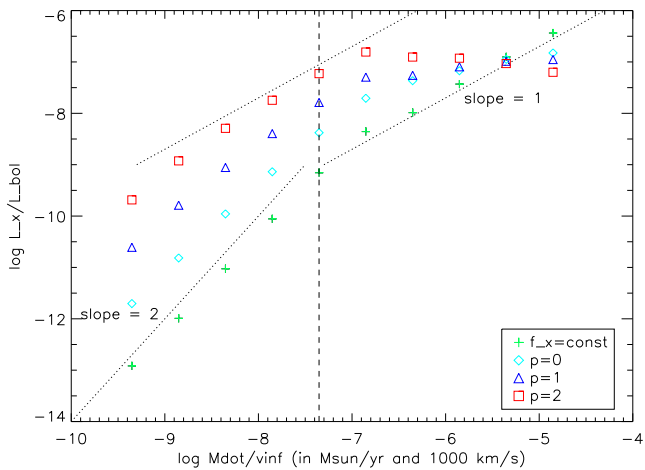}}
\caption{X-ray luminosity (in the range 0.1 to 2.5 keV, and in units
of $L_{\rm bol}$) as a function of $\mdot/\vinf$, for stellar models
with photospheric and wind parameters as in Table~\ref{tab_zeta_pup},
but mass-loss rates ranging in between $10^{-9} {\ldots}
10^{-4.5} \msunyr$, and different assumptions on $\fv(r)$ (see
legend, and Table~\ref{tab_fvol}). 
The dotted lines display linear relations (with respect to the log-log
scaling), with slopes $m$ = 1 and 2, to guide the eye. Models
to the left of and at the vertical dashed line are
optically thin in the X-ray emitting region at all contributing
frequencies, i.e., $R_1 < \Rmin$ for energies between 0.1 and 2.5 keV.
See text.}
\label{plot_lx}
\end{figure}

\subsection{Testing predicted scaling relations for the X-ray luminosity} 
Using our new ``unified approach,'' we can now re-check the scaling of the
X-ray luminosity with wind-density parameter $\mdot/\vinf$, and
compare it with the predictions by Ow13. Fig.~\ref{plot_lx} shows the
result of such an analysis, where we display our findings when applying
the different parameterizations for the effective volume filling
factor as introduced above. For all four models series ($\fv=const$, and
$p \in [0,1,2]$, see Table~\ref{tab_fvol}), we used the same photospheric and wind parameters as
displayed in Table~\ref{tab_zeta_pup}, but varied, within each series,
the mass-loss rate in the range $10^{-9} {\ldots} 10^{-4.5} \msunyr$.
To allow for a clear comparison, we did not vary the value of
\vinf, thus in the following we actually check only the reaction of
$L_{\rm x}$ versus \mdot\ for the case $\vinf=const$.

If we denote the typical radius where the wind becomes optically thick
in X-rays (due to ``cold'' material) by $R_1$, and $R_{\rm a}$ as 
the border between radiative and adiabatic cooling 
(in the current framework defined by $l_{\rm s}(R_{\rm a}) = R_{\rm a}$), we can
summarize the predictions by \citet{OwockiCohen99} and Ow13 as follows
(again discarding potential thin-shell mixing):

\smallskip
\noindent
\textit{I. Optically thin conditions} ($R_1 < \Rmin$).\\ 
a) For a constant volume filling factor, $\fv = const$, or purely
adiabatic cooling ($R_{\rm a} < \Rmin)$, the X-ray luminosity should
scale as $L_{\rm x} \propto (\mdot/\vinf)^2$.\\
b) For a volume filling factor according to Eq.~\ref{fv_owo}, and for
radiative and adiabatic cooling with $R_{\rm a} > \Rmin$, a scaling
via $L_{\rm x} \propto (\mdot/\vinf)$ is predicted.

\smallskip
\noindent
\textit{II. Optically thick conditions} ($R_1 > \Rmin$).\\
a) For a constant volume filling factor,  $L_{\rm x} \propto (\mdot/\vinf)$.\\
b) When \fv\, is defined according to Eq.~\ref{fv_owo}, and for radiative and
adiabatic cooling with $R_1 < R_{\rm a}$, the X-ray luminosity
should follow $L_{\rm x} \propto (\mdot/\vinf)^{1-p}$.\\

\smallskip
\noindent
Comparing these predictions with Fig.~\ref{plot_lx},
we see a fair agreement. For constant volume filling factors (case Ia:
``old'' approach, green plus signs), there is only the division
between optically thin (``slope=2'') and optically thick (``slope =
1'') conditions, independent of radiative or adiabatic cooling, and
in agreement with the former predictions by \citet{OwockiCohen99} and
simulations by \citet{Carneiro16}. The transition takes place at the
dashed line, which displays the maximum wind density until which the
X-ray radiation field remains optically thin in the region above $\Rmin$
for all considered frequencies. A ``slope=2'' is also present for the
other series with \fv\, according to Eq.~\ref{fv_owo}, as long as the
wind-density is so low that $R_{\rm a} < \Rmin$. However, for these models
the slope changes toward unity when the wind density increases, even
if the wind is still optically thin for X-rays (case Ib). This is
particularly visible for the model with $p=2$ (red squares). 

\begin{figure*}
\begin{minipage}{9cm}
\resizebox{\hsize}{!}
  {\includegraphics[angle=90]{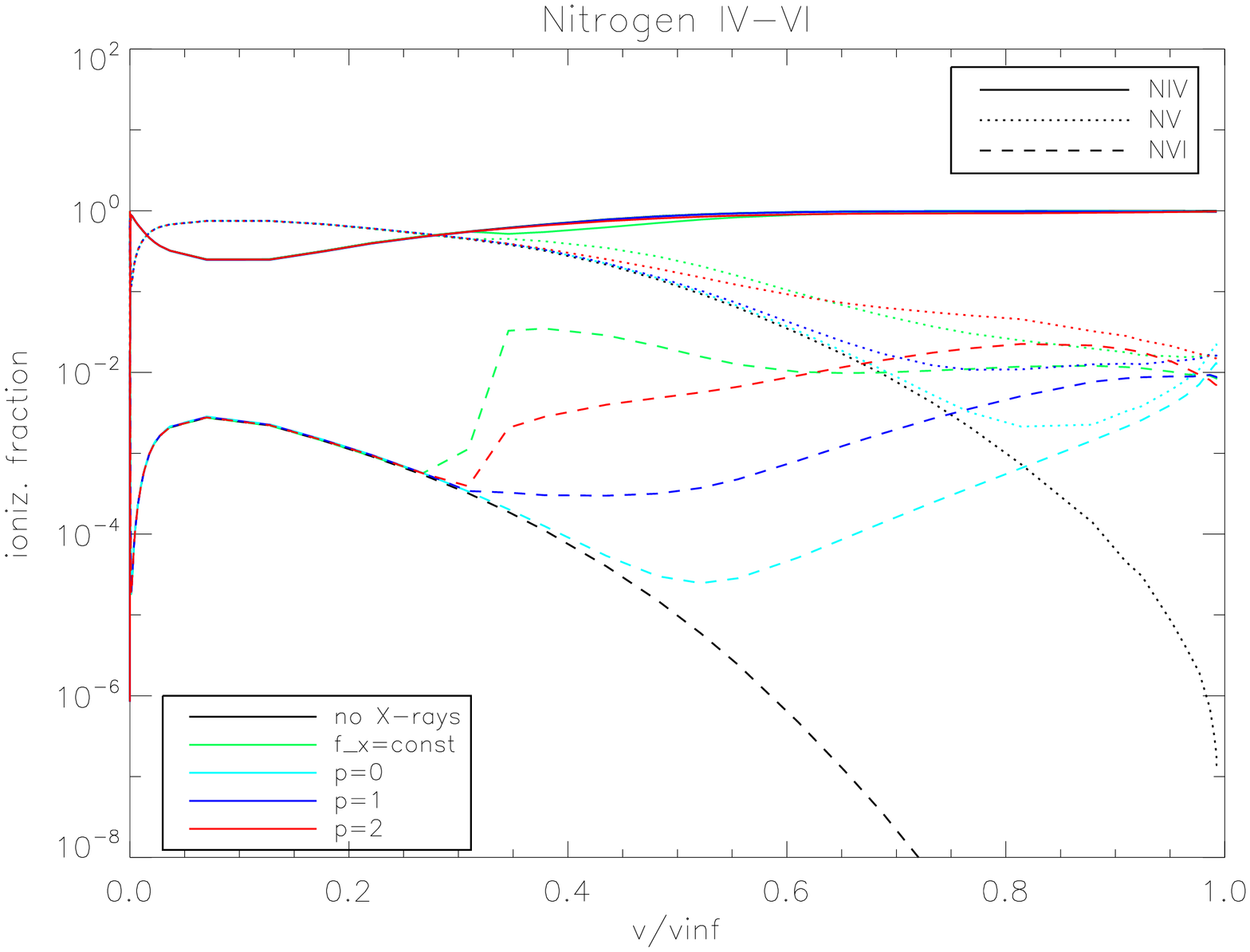}} 
\end{minipage}
\begin{minipage}{9cm}
\resizebox{\hsize}{!}
  {\includegraphics[angle=90]{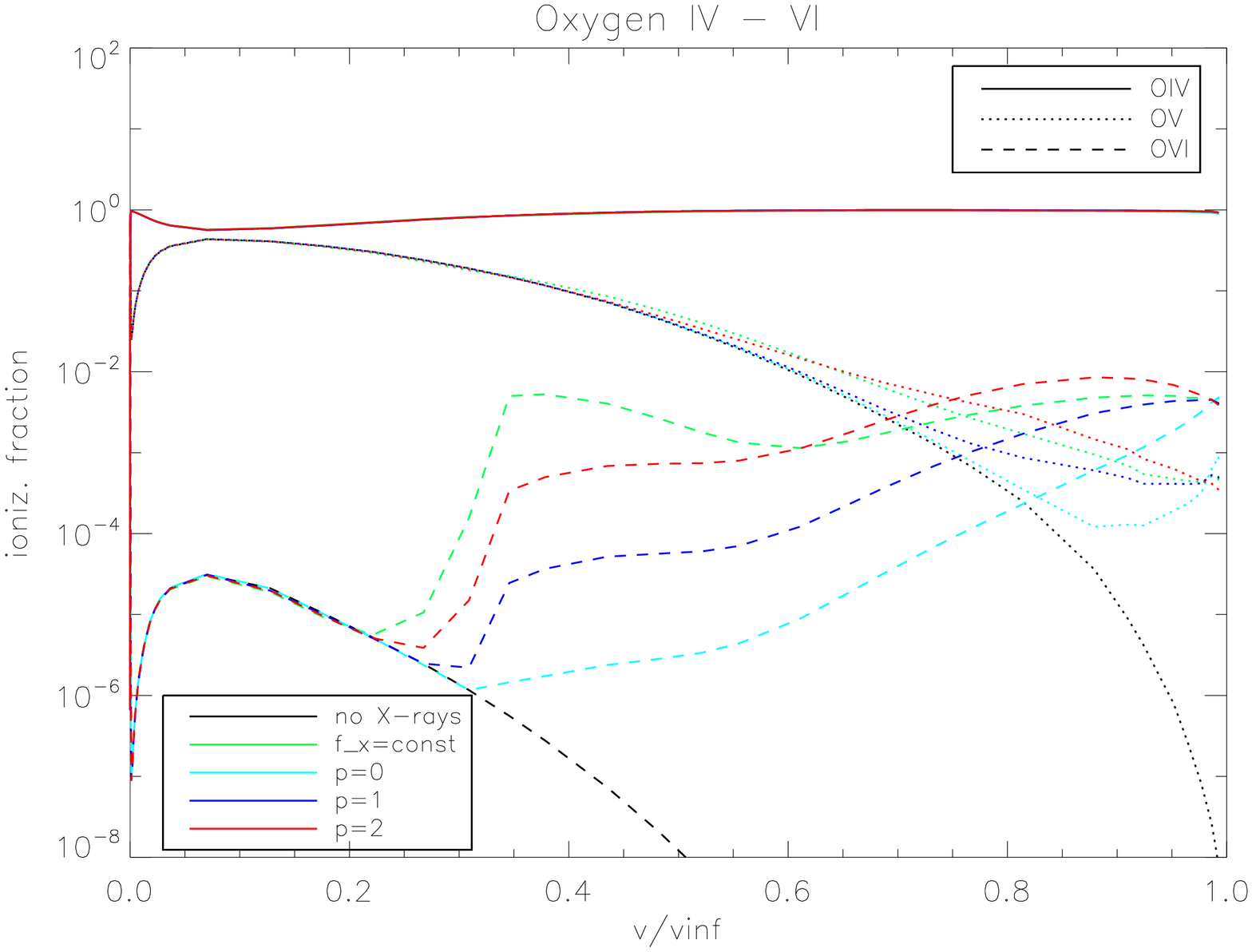}} 
\end{minipage}
\caption{Ionization fractions of 
nitrogen and oxygen, as a function of $v(r)/\vinf$, for the same stellar model 
and shock distributions as in Fig.~\ref{plot_fvol}.
For line-coding, see legends.}
\label{no_ifrac}
\end{figure*}

\begin{figure*}
\resizebox{\hsize}{!}
  {\includegraphics[angle=90]{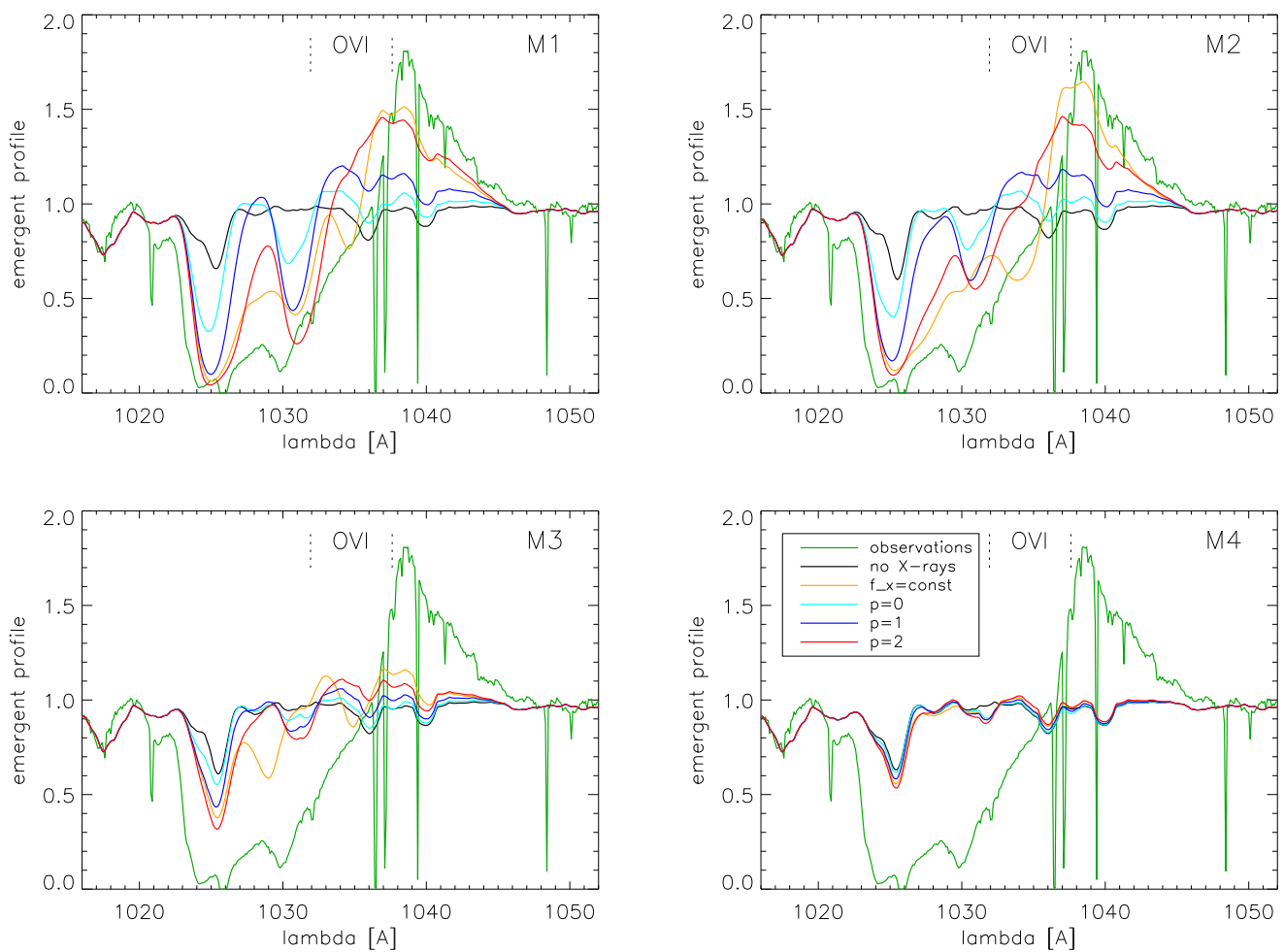}}
\caption{Synthetic line profiles for the UV resonance doublet of \OVI,
for the same stellar model and different shock distributions as in
Fig.~\ref{plot_fvol} (M1, upper left), and for three other
models (M2 to M4) differing in mass-loss rate and clumping
properties (see Table.~\ref{tab_zp_models}). Same color coding as in
Fig.~\ref{no_ifrac}. Profiles from the smooth-wind models M1 and M2 are
close to being identical, though mass-loss rates differ by a factor of four.
All spectra have been calculated with a depth dependent
``micro-turbulence'' (proportional to $v(r)$, with a minimum of 15
\kms, and a maximum of 0.1\vinf), and convolved with \vsini=220~\kms.
For comparison, the observed COPERNICUS spectrum (at a resolution of
0.2\AA, \citealt{Snow77}) is displayed in dark green. The dotted lines
at the top display the rest-frame frequencies of the doublet
components. The absorption feature around 1215\,\AA, visible in
all synthetic spectra without significant \OVI\ contribution, is
stellar \Lyb. Model M1 and M4 and corresponding spectra (for
$\fv = const$) are very similar to those displayed (and discussed) by
\citet{Zsargo08b} in their Fig.~1.}
\label{OVI_zp}
\end{figure*}

When the wind becomes optically thick (beginning to the right of the dashed
line), the slope changes again. For
constant volume filling factors (case IIa), toward a slope of unity,
and for the models with varying filling factor, at least in the spirit
predicted by case IIb, toward $1-p$. In particular,
for $p=2$, the slope becomes
negative, that is, for increasing mass-loss the X-ray luminosity
decreases; for $p=1$, the slope becomes (almost) constant; and 
for $p=0$  it remains positive, though flatter than analytically predicted. 

The reason for the latter deviation between predicted and simulated
slope in the regime to the right of the vertical dashed line is at
least two-fold. 
First, the calculated luminosity is an integral over a large range of
frequencies. Whereas to the left of the dividing line, the radiation
field is optically thin at all contributing frequencies, it becomes
optically thick at all frequencies only for the highest wind-density
models considered. In between, the lower energies are optically thick,
whereas the higher ones are still optically thin. For example, for
$\log \mdot/\vinf = -6.35$ in the units of Fig.~\ref{plot_lx}, the
radiation field is still optically thin for wavelengths below 15\AA\
(0.83 keV). Thus, in many cases the luminosity consists of a
combination of optically thin and thick radiation, contrasted to the
analytic limits.
Second, we note that part of the quantitative discrepancies 
might be caused by numerical issues, related
to the less reliable flux determinations in models which are still
optically thick at the outer boundary.

\subsection{Impact on ionization fractions, and the \OVI\ resonance
doublet} 

To date, it is still unclear which kind of parameterization of the
shock distribution is more consistent with real wind conditions. To
this end, a careful analysis of UV P Cygni resonance lines (in
particular, \OVI) might be valuable. Though this is beyond the scope
of this paper, we have checked the influence of different
\fv-stratifications onto the ionization balance, and onto the \OVI\ line
profile (see also \citealt{Carneiro16} for the case of $\fv = const$).
In Fig.~\ref{no_ifrac}, we display the changes introduced to the
ionization fractions of nitrogen and oxygen by X-ray emission, for the
same model (similar to $\zeta$ Pup) as discussed before, and
various \fv-stratifications. The results from a corresponding model
without any X-ray emission is displayed in black. We concentrate here
on nitrogen and oxygen, since, for example, carbon and phosphorus are barely
affected in the considered parameter range (see also \citealt{Carneiro16}).

\begin{table}
\caption{Specific properties of the four models underlying the
\OVI\ resonance line profiles displayed in Fig.~\ref{OVI_zp}. Model 
M1 is identical with our previous test model characterized in
Table~\ref{tab_zeta_pup}, and the other models differ only in
mass-loss rate and clumping properties (\fcl\ is the clumping factor).
Model M4 has photospheric and wind parameters close to those
inferred for $\zeta$~Pup. For models M3 and M4, the clumping factor
below $v(r)=0.1 \vinf$ increases linearly from unity to the displayed
value.}
\label{tab_zp_models}
\begin{tabular}{ccl}
\hline 
\hline
model & mass-loss rate [\msunyr]   & \fcl  \\  
M1  &  $8.0 \cdot 10^{-6}$ & 1 (smooth) \\
M2  &  $2.0 \cdot 10^{-6}$ & 1 (smooth) \\
M3  &  $2.0 \cdot 10^{-6}$ & 4 for $v \ge 0.1 \vinf$ \\
M4  &  $2.0 \cdot 10^{-6}$ & 16 for $v \ge 0.1 \vinf$ \\
\hline
\end{tabular}
\end{table}

Within nitrogen and oxygen, the largest changes occur for \NVI\ and
\OVI, and are present from the onset of X-ray emission on. On the
other hand, \NV\ and \OV\ change only in the outer wind, and the lower
ionization stages ({\sc iii} -- not displayed, and {\sc iv}) 
are not affected at all. Though
particularly \OVI\ is increased by many decades for all
\fv-stratifications considered (explaining the origin of the
well-known ``super-ionization,'' \citealt{Snow76,
LamersMorton76, Hamann80}), different shock distributions result in
quite different ionization fractions, where, particularly in the lower
and intermediate wind, these fractions follow the strength of \fv\
(cf.~Fig.~\ref{plot_fvol}). Though this behavior should allow for
tight constraints when comparing with observations, we note that
for such an objective all other parameters need to be known quite
precisely, in particular regarding abundances and
wind inhomogeneities. 

Indeed, when comparing the resulting synthetic \OVI-doublets
with the corresponding {\sc copernicus} spectrum for $\zeta$ Pup in
Fig.~\ref{OVI_zp}\footnote{We stress that no fit has been aimed
at in any of the displayed panels.}, a strong dependence on the
adopted wind structure becomes visible. Before commenting on the
impact of different shock distributions, we first concentrate on this
dependence on wind structure.

As already outlined before, our standard model
(Table~\ref{tab_zeta_pup}) adopts a mass-loss rate that is a factor of
four larger than the observationally derived one when accounting for
optically thin clumping. This should certainly affect the profiles:
for example, and at least for optically thin clumping, lines with an
opacity scaling linearly with $\rho$, such as
\CIV~$\lambda\lambda$1548-1550 (but not \OVI), should scale with the
``actual'' mass-loss rate, (almost) independent of the adopted clumping factor,
\fcl\ (for example, \citealt{Puls08} and references therein).
On the other hand, and based on analytic
considerations, \citet{Zsargo08b} highlighted a completely
different relation for those lines where Auger-ionization (due to X-ray
emission) dominates, as often (but not always, see
\citealt{Carneiro16}) the case for the superionized
\OVI\ resonance line discussed here.
Then, the profiles should become (almost) independent on mass-loss
rate for smooth winds, and should depend on the adopted clumping
stratification alone if inhomogeneous winds were considered.

To check and demonstrate whether these predictions are followed by
{\sc fastwind}, we synthesized the \OVI-doublets for our standard
model (M1), and three additional ones, with wind-parameters
provided in Table~\ref{tab_zp_models}. For these additional models, we
calibrated the \fv\ and $n_{\rm so}$ parameters as to obtain the same
X-ray luminosity as present in model M1. Model M2 is still a
smooth model, but with a mass-loss rate corresponding to the
observationally derived value (a factor of four lower compared to
M1). Indeed, and as predicted by \citet{Zsargo08b}, the profiles
for both smooth-wind models M1 and M2 are very similar,
despite the large difference in \mdot.  When subsequently increasing
the clumping factor in models M3 and M4 (that is, compressing the
wind mass into smaller and smaller clump-volumes), the
profile-strengths decrease (again as predicted, because of increased
recombination inside the clumps). We note that the M4 profiles
differ only weakly from a corresponding model without any X-ray
emission at all. When finally comparing our profiles for models M1
and M4 to those resulting from the ({\sc cmfgen} incl. X-ray)
simulations\footnote{With stellar, wind, and X-ray parameters very
similar to ours, and assuming spatially constant X-ray volume filling
factors.} by \citet[their
Fig.~1]{Zsargo08b}, we find an extremely close agreement, also here
proving the consistency between {\sc fastwind} and {\sc cmfgen}.

Among the displayed models, M4 is the one closest to results
from analyses of $\zeta$~Pup, performed in the UV (though excluding
\OVI), optical, NIR, and radio domain, and assuming optically thin
clumping (\citealt{Puls06, Najarro11, bouret12}). From the severe
discrepancy between observations and simulations, one might conclude
that there is also a severe problem with the inferred parameters.
However, \citet{Zsargo08b} have flagged the importance of the
inter-clump medium for a correct description of the \OVI\ resonance
line formation. They showed that when relaxing the assumption of
a void inter-clump medium, and accounting for reasonably low, but
non-zero densities in such regions, its contribution is sufficient to
explain the observed profile strength. Since corresponding simulations
require a two-component, NLTE description, this is beyond the scope of
{\sc fastwind}'s current capabilities. Instead, we consider the 
mass-loss independent, smooth-wind case. For our purposes, this
should be close enough to the conditions in an inhomogeneous wind,
when a low-density inter-clump medium covers a large volume, and the
contribution from the overdense region is negligible. In particular,
and to investigate the dependence on different shock-distributions, in
the following we concentrate on the profiles for models M1 or M2.

Without any X-rays (black), there is no signal at all, whereas for all
simulations including X-rays, a P Cygni profile is clearly visible.
Since the absorption troughs are not saturated, the absorption
strength is proportional to the ionization fraction, and hence to the
run of \fv\ (see above). We note that corresponding normalization
factors are well-constrained, since the adopted X-ray luminosity is
close to the observed one for $\zeta$ Pup, $\log (L_{\rm x}/L_{\rm bol})
\approx -7.15 {\ldots} -6.85$ \citep{Berghoefer96a,
pauldrach01, Zsargo08b}. For the considered models, a constant volume
filling factor, or a filling factor that is rather similar (here:
$p=2$), result in profiles that are closest to observations. In all
cases, however, the line emission from the lower wind is too
large, and/or the absorption is too low, which might be cured by a
combination of a lower \Rmin\ value and modified X-ray filling factors
in those regions, and, ultimately, by an appropriate clumping law
including a consistent description of the interclump medium. From
the absorption strength at high velocities (where all \fv\ values are
similar), it might be concluded that both our description and the used
abundances are reasonable. And indeed, the oxygen abundance adopted in
our X-ray emitting models, $(Z/Z_{\odot})_{\rm O}$ = 0.2, is close to
the value derived by \citealt{bouret12}, $(Z/Z_{\odot})_{\rm O}$ =
0.25.
      
\section{Summary and conclusions}
\label{summary}

The work presented in the previous sections can be summarized as
follows: We have updated our unified, NLTE atmosphere and spectrum
synthesis code, {\sc fastwind} v10, by including -- via a new program
module -- a detailed comoving frame transfer covering the most
important wavelength range (current default: from 200~\AA\ to
10,000~\AA, where these limits can be modified by the user). The new
version is called {\sc fastwind} v11. Both explicit and the most
important background elements (called selected) are handled with
similarly high precision, and only the non-selected elements are
treated within our previous, approximate NLTE approach, though also
here the required radiation field is taken from the detailed (though
re-mapped) CMF solution. In the default version, both the ray-by-ray
solution for the specific intensity, and the moments equations are
solved subsequently.

The major part of computation time is spent by the ray-by-ray
solution, which mostly depends on the adopted micro-turbulent
velocity, \vmic. Total turnaround times including the final
observer's frame formal integral to calculate emergent fluxes and
normalized spectra), are on the order of 1.3 to 1.9 hours for $\vmic =
15 \kms$, and 3 to 4.5 hours for $\vmic = 5 \kms$, for an Intel Xeon
2.7 GHz processor, where the lower values refer to models with H and He
as explicit elements (fast convergence), and the higher values to a
combination of explicit elements that is slowly converging, due to
complex line-overlap processes (for example, H,He,N). The required RAM
amounts to a comparatively low value of 1.6 GB.

We have carefully compared our results with alternative ones from {\sc
cmfgen} and our previous version, {\sc fastwind} v10, by means of a
comprehensive model grid covering early B- to hot O-type dwarfs and
supergiants. We confirm previous results on the deviations between an
exact and the approximate NLTE treatment following \citet{Puls05},
which amount, on average, in between few to 20\% for the main ionization
fractions, though with a significant scatter. 

The total radiative acceleration as calculated by {\sc cmfgen} and
{\sc fastwind} v11 agrees almost perfectly for the supergiant models,
whereas for dwarf models there is a certain discrepancy in the
transonic region, presumably related to different velocity fields in
this regime, and to different line-lists. The overall remarkable
agreement (already mentioned in a first announcement of v11,
\citealt{Puls17}) justifies its usage within the calculation of
self-consistent wind models, as performed by \citet{Sundqvist19}. In
particular, the strong decrease in the acceleration in the transonic
region, predicted by both {\sc cmfgen} and {\sc fastwind} for late
O-dwarf models (here: d10v), might be responsible (at least in part)
for the weak-wind effect in cooler dwarfs (lower observed mass-loss
rates than resulting from ``simplified'' theories, for example
\citealt{Puls08} and references therein). This conjecture is
corroborated from corresponding results from the self-consistent wind
models quoted just above.

Also the optical H and He spectra agree in most cases very well, and only
for \HeII~$\lambda\lambda$4200-4541 there are, similar as for {\sc fastwind} v10,
certain differences in the line-cores. Morever, and most promising, we
find the same \HeI\ problem as identified by \citet{Najarro06}, and
have patched it in a similar way. The agreement between the old and
the new {\sc fastwind} is even better, and an analysis with both codes
should result in parameters which are close to being identical.
This finding allows, for example, for still using the older v10 approach
whenever a pure HHe analysis in the optical is sufficient, by simply
switching back to this version via the corresponding option inside the
code (see Sect.~\ref{philosophy}).

With respect to diagnostic nitrogen line profiles in the optical, the
agreement is still acceptable, though for specific transitions the
differences are no longer insignificant. The largest difference
occurs for \NV~$\lambda\lambda$4603-4619 (when present), where our
new version predicts stronger absorption.

We have re-investigated the formation of \NIII\ \trip, and confirmed
the basic findings by \citet{rivero11}. In particular, the agreement
with {\sc cmfgen} could be improved, because the shortcomings of the
previous v10 (neglect of line-overlap effects) are no longer
present in v11. However, we have also identified one additional, \FeV\
line in the EUV, which participates in the formation of the \NIII\
emission lines. In combination with the adopted size of \vmic, and in
dependence of Fe, O, and N abundances, the emission strengths 
of these lines can become quite non-monotonic. 

Overall, we warned the reader against uncritically relying on the \HeI\ singlets, the
\NIII\ emission lines, and \NV~$\lambda\lambda$4603-4619, when
performing quantitative H-He-N spectroscopy in the optical, and to put
a lower weight onto these lines, compared to others. The impact of
\vmic\ (allowing or prohibiting specific line-overlaps) needs to be
always kept in mind.

With the new {\sc fastwind} version, we are now able to synthesize
large continuous portions of the spectrum, in particular the UV range.
Also here, the agreement with {\sc cmfgen} is mostly excellent, and
the only relevant differences concern the sulfur (resonance) lines
(future checks required!), and the pseudo-continuum of the coolest
models within our grid, which in the {\sc cmfgen} models is
considerably more depressed. This depression (particularly around
\Lya) is almost exclusively due to \FeIV\ lines connecting energy
levels that are not observationally confirmed. Indeed, when comparing
with observations, many of these lines seem to be absent, and one of
the future tasks is to carefully check this issue, which certainly is
not only an ``academic'' one. (i) As extensively discussed in the
previous sections, line-overlap in the EUV (particularly with Fe
lines) can significantly affect the population of various levels, and
thus the absorption and emissions strengths of diagnostic lines.
Hence, it is of prime importance to check whether such Fe-lines are
actually present, or not. (ii) When investigating the UV spectrum of
fast rotators, the definition of a reliable continuum level becomes
problematic, if the line density was indeed as high as predicted by
current models. In this case, normalization procedures might result,
in specific wavelength regimes, in an erroneous continuum level
(Fig.~\ref{d10v_1200}, upper panel, and particularly the predicted
flux distribution for model s10a, Fig.~\ref{uv_spec}), which in turn
might lead to erroneous conclusions when comparing with theoretical
models that do not include a consistent metallic background. (iii) If,
on the other hand, \FeIV\ lines play a lesser role in specific models
compared to others (as here for the case {\sc fastwind} versus {\sc
cmfgen}), the pseudo-continuum is closer to unity, and the
emission peaks of P~Cygni lines appear higher in a normalized
spectrum. Thus, the reliability of the \FeIV\ background can directly
affect the wind diagnostics. Obviously, not only \FeIV, but also lower
ionization stages (\FeIII\ and \FeII\ in B- and A-type supergiants)
need to be investigated as well.  

In the last section of this work, we unified the seemingly
discordant results on the X-ray emission from wind-embedded shocks
when using the models by either \citet{Feldmeier97b} or
\citet{OwockiSundqvist13}. As it turned out, the assumption of a
depth-independent X-ray filling factor, when combined with the former
approach, is the origin of this discordance. By alternatively applying
the depth-dependent expression provided by \citet{OwockiSundqvist13},
both approaches could be unified. Scaling relations for the X-ray
luminosity as a function of wind density, as calculated by our new 
{\sc fastwind} version, agree well with the analytical expression
provided by \citet{OwockiSundqvist13}. A first comparison of the
super-ionized \OVI\ resonance line in the UV with observations
indicated that this line is well suited to obtain clues on the shock
distribution, but only if a meaningfull description of the inter-clump
medium has become available.

Aside from a solution of the problems discussed throughout this paper,
the next steps in our further program development are obvious. On the
programing side, we might think about parallelization. This should
be possible for the most time-consuming part, the ray-by-ray solution,
when the outermost loop extends over p-rays instead of over 
wavelengths. Such a modification would increase the required memory,
and it might be better to keep our present strategy of calculating
different models in parallel, which is suitable at least for the
construction of model grids. When applying on-the-fly optimization
strategies such as genetic algorithms, a shorter wall-clock time due
to parallelization might be advantageous though.

Regarding future updates of the involved physics, a rather simple
improvement concerns the implementation of optically thick wind
clumping, which was already done in {\sc fastwind} v10 \citep{SP18},
and needs only to be included into the new routines of v11 (and tested!).
Most importantly, however, is an improvement of our Fe (and Ni) model
atoms, to include higher lying levels into the detailed NLTE treatment.
To date, it is not clear whether we can continue with our current
approach of packing suitable levels, or whether we will completely
switch to a more general super-level approach for these elements.
Finally, it will be of highest relevance to extend our presently
considered wavelength range into the NIR, to be prepared for the
upcoming observations of next-generation telescopes.

\begin{acknowledgements} 
We gratefully thank our anonymous referee for very constructive
comments and suggestions. Particular thanks to the late Adi
Pauldrach for providing us with the WM-{\sc basic} data base, and to John
Hillier for making {\sc cmfgen} available. 

F.N. acknowledges financial support through Spanish grant
ESP2017-86582-C4-1-R (MINECO/FEDER) and from the Spanish State
Research Agency (AEI) through the Unidad de Excelencia ``Mar\'ia de
Maeztu''-Centro de Astrobiolog\'ia (CSIC-INTA) project No.~MDM-2017-0737.

J.O.S. acknowledges support by the Belgian Research Foundation Flanders
(FWO) Odysseus program under grant number G0H9218N, as well as from
the KU Leuven C1 grant MAESTRO C16/17/007.

K.S. would like to thank the DAAD (WISE) Scholarship and 
Inspire Fellowship (DST) programs for financial support, and colleagues at 
the USM Munich for hosting during an internship.\\
\newline
This paper is dedicated to Adi Pauldrach, who left us much too early.
We will always remember his personality, his laughter, his dedication
to soccer, and his invaluable contributions with respect to line-driven
winds, NLTE-modeling of massive star atmospheres, and manifold
applications. May he rest in peace.
\end{acknowledgements}

\bibliographystyle{aa}
\bibliography{puls_fwII.bib}

\appendix
\section{Approximation of the incident intensity required for the CMF
transport}
\label{app_iminus}

\begin{figure*}[h]
\begin{minipage}{9cm}
\resizebox{\hsize}{!}
  {\includegraphics[angle=90]{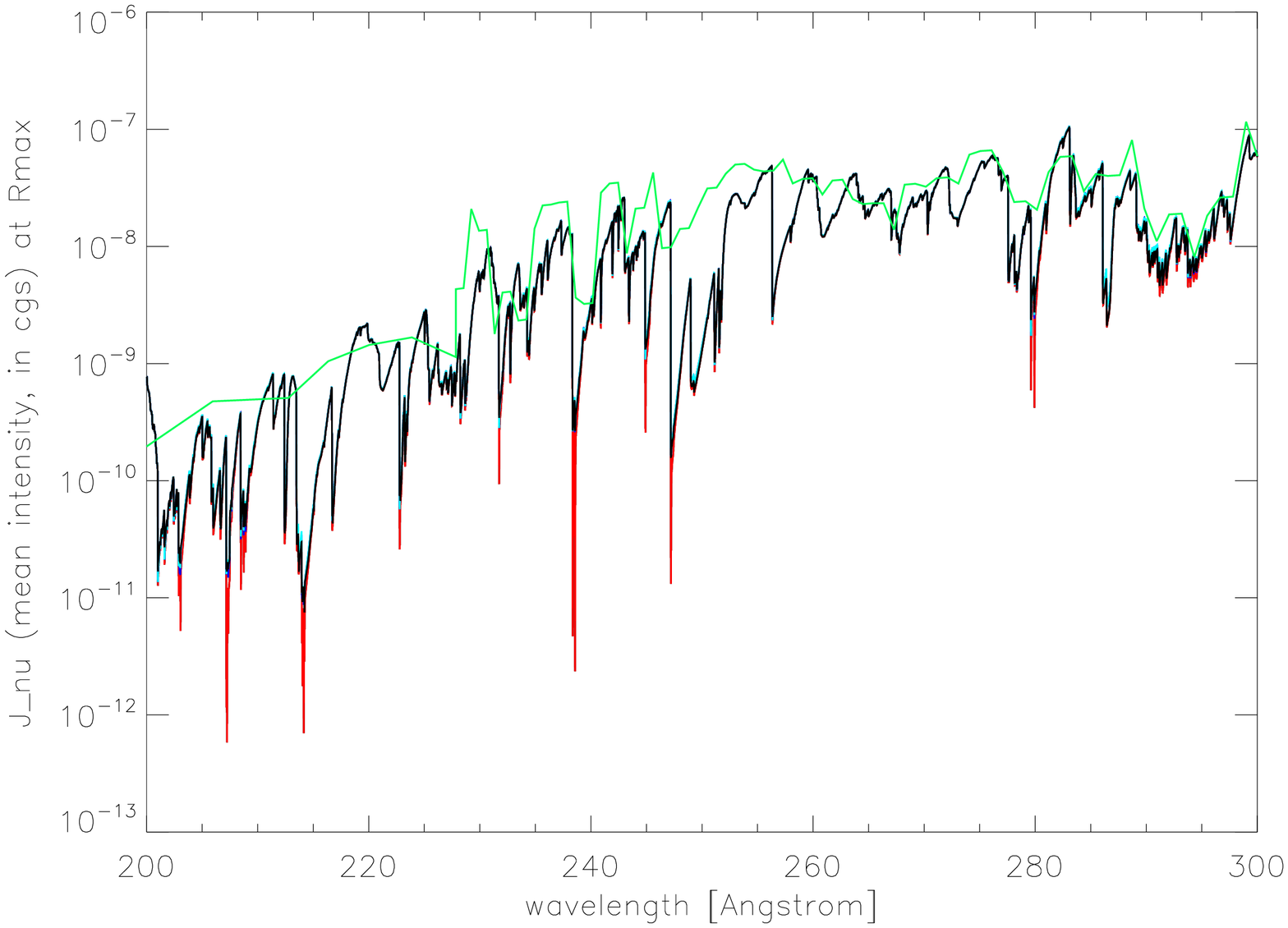}} 
\end{minipage}
\begin{minipage}{9cm}
\resizebox{\hsize}{!}
  {\includegraphics[angle=90]{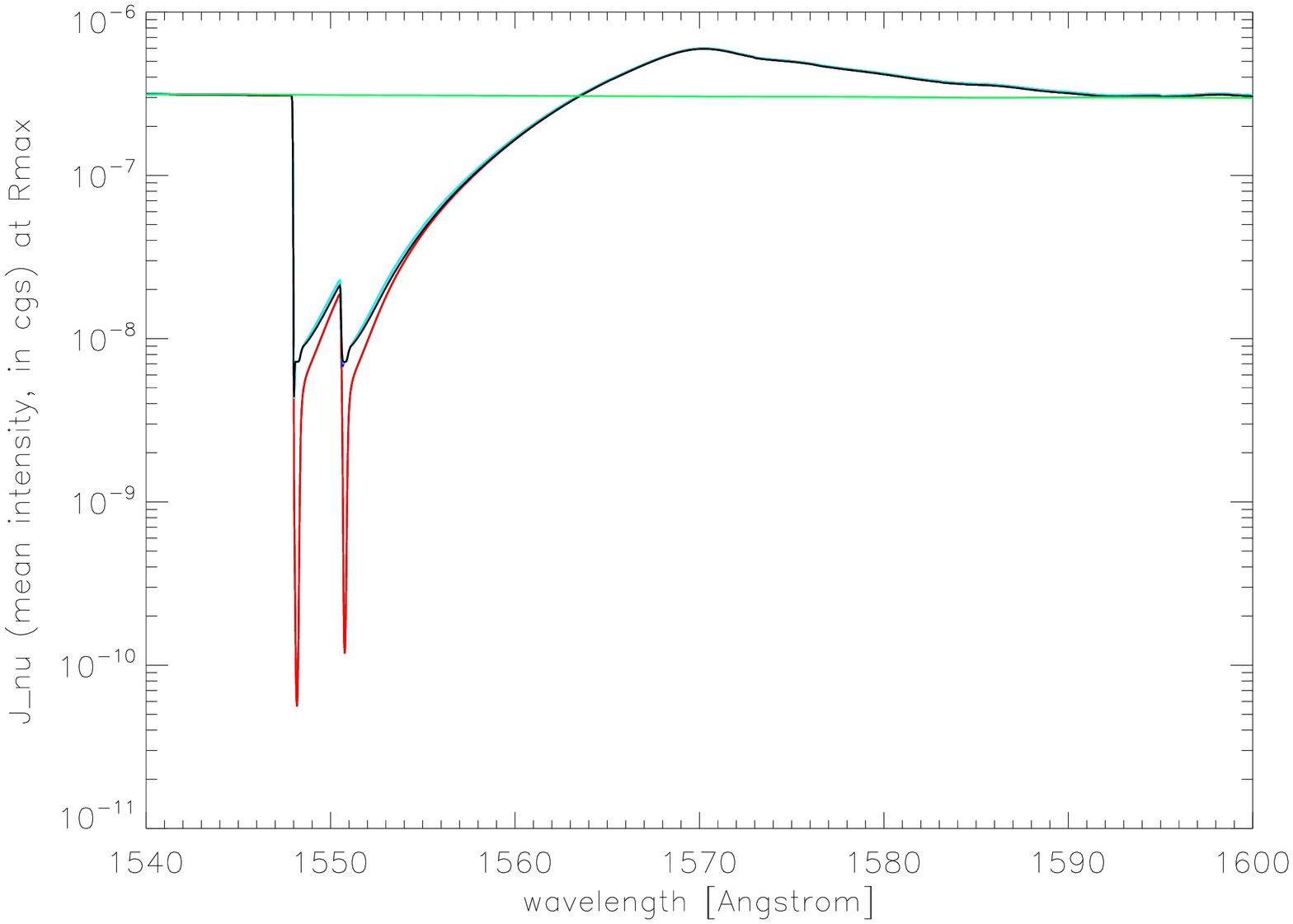}} 
\end{minipage}
%
%
\caption{CMF mean intensity at \Rmax ($\approx 120~\Rstar$), for a hot
dwarf model with \Teff = 55,000 kK and a large mass-loss rate, 
calculated with different approximations for the incident intensity,
$I^-$, and different wavelength regimes. All intensities have been
taken from the same iteration step, namely the first step after the
detailed CMF-treatment has been initiated. Color-coding -- black:
complete incident intensity including frequency derivative, as
described in the text; red: $I^- = 0$; dark blue (mostly blended by
turquoise): $I^-$ specified only for lines and continua of significant
optical depth, gradient neglected; turquoise: $I^-$ specified for all
frequencies, but gradient neglected; green: approximate
pseudo-continuum, as used in the previous {\sc fastwind} version, and
still at highest and lowest wavelengths. Left panel: EUV range between
200 {\ldots} 300 \AA; right panel: frequency range around the \CIV\
resonance doublet in the UV.}
\label{app_xj_iminus1}
\end{figure*}
\begin{figure}
\resizebox{\hsize}{!}
  {\includegraphics[angle=90]{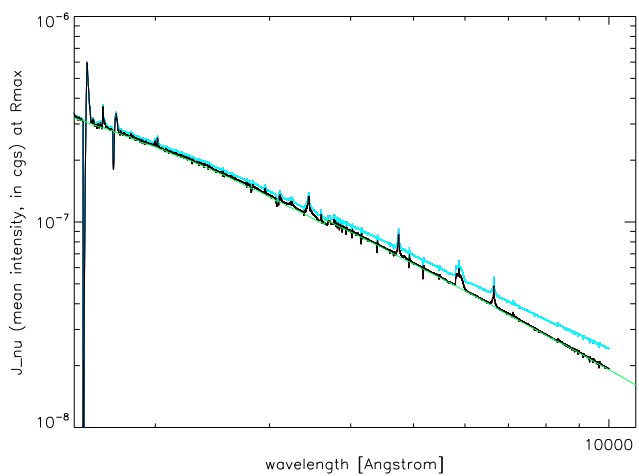}} 
\caption{As Fig.~\ref{app_xj_iminus1}, but for $\lambda = 1,500
{\ldots} 11,000~\AA$. The impact of the frequency derivative
(turquoise versus black) is clearly visible. Without inclusion, no smooth
transition between detailed and approximate treatment (at
$\lambda_{\rm max}$ = 10,000 \AA) would be possible. Setting
$I^-$ to zero at all frequencies (red curve in
Fig.~\ref{app_xj_iminus1}) results in a flux-distribution identical to
the black one in the optical range, whereas specifying $I^- \ne 0$
only for optically thick lines and continua, and neglecting the
frequency gradient (dark blue curve in Fig.~\ref{app_xj_iminus1}), results
in fluxes similar to the turquoise ones.} 
\label{app_xj_iminus2}
\end{figure}
In this appendix, we describe our approximation for the incident
intensity, $I^-$, at \Rmax, the outer boundary of our computational
domain ($\approx 120~\Rstar$). As already mentioned in
Sect.~\ref{basic_cmf}, and to minimize the computational effort, we do
not ``extend'' the atmosphere in the ray-by-ray solution toward
larger radii, as suggested by \citet{hilliermiller98}. The latter
procedure avoids the specification of a non-zero incident intensity,
$I^{-}(\Rmax)$, which otherwise would need to be accounted for, as
long as the outer atmosphere is not completely optically thin (and 
which is difficult to estimate). Indeed, a reasonable choice of $I^{-}
\ne 0$ is often essential for a reliable radiative transfer solution
(and related quantities such as occupation numbers) when the outer
boundary is not extended beyond the actual grid.

Instead of such an extension (to reach optically thin conditions and
then set $I^{-} =0$), we developed and tested an approximate
description of $I^-(\Rmax)$, and its frequency derivative. Also the
latter needs to be specified in the CMF equations, because of the
presence of the term $\partial v/\partial x = \partial u/\partial x
-\partial I^-/\partial x$ at the outer boundary, with Feautrier
variables $u$ and $v$, and $x$ the CMF-frequency in suitable
units. For example, and as also done in the following, one might
define $x$ (in the neighborhood of a line with transition frequency
$\nu_0$) as a CMF frequency displacement, measured in units of
maximum Dopplershift, 
\beq 
\label{Dopplerunits} 
x=\frac{\nu-\nu_0}{\Delta \nu_\infty},\qquad \Delta
\nu_\infty=\frac{\nu_0 \vinf}{c}  
\eeq
(for the general background and further details on CMF transfer, we
refer to \citealt{mihalasbook78} and \citealt{hubenybook14}).

If, on the other hand, an appropriate incident intensity would be
neglected at all, particularly the line cores would become too deep
(cf. Fig.~\ref{app_xj_iminus1}, red versus black), and optically thick
continua too weak. 
Our development of an appropriate boundary condition required
a number of subsequent steps, which are summarized in the following,
and which are (partly) displayed in Figs.~\ref{app_xj_iminus1} and
\ref{app_xj_iminus2}.

\paragraph{Optically thick continuum.} By using the (static) equation
of transfer (accounting for the fact that in most cases the frequency
shift between outermost grid-point and infinity is negligible), the 
incident intensity can be approximated (to zeroth order in source
function, and assuming $I^-(\nu,z \rightarrow \infty,p) = 0$) by
\beq
\label{iminus_cont}
I^-(\nu,\zmax,p) \approx S_c
(\nu,\Rmax)\bigl(1-\exp(-\tau_c(\nu,\zmax,p))\bigr),
\eeq
with CMF-frequency $\nu$, and continuum source function and optical depth, $S_c$ and $\tau_c$,
respectively, where $\tau_c$ can be approximated by (analytically)
extrapolating the opacities from \Rmax\ to infinity.

\paragraph{Optically thick line cores.} Here we use a Sobolev-like
approach (following \citealt{Lucy71, Puls91}), namely neglecting the
spatial derivatives in the CMF equation of transfer, and assuming that
one line dominates the opacity and source function (a generalization to
more than one line is possible, and implemented into our code). 
This results 
(see also \citealt{Puls20}) in
\beqa
\label{iminus_line}
I^-(x,\zmax,p) &\approx&  I^-(\xmax,\zmax,p)
                \exp\bigl(-\tau_L(x,\Rmax,\mu)\bigr) + \nonumber \\
               &+& S_L(x,\Rmax)(1 -
	       \exp\bigl(-\tau_L(x,\Rmax,\mu)\bigr),
\eeqa
with line source function $S_L$, and line optical depth
\beq
\tau_L(x,\Rmax,\mu) \approx \tau_S(\Rmax,\mu) \times \int_x^{\xmax} \phi(x) \dd x.
\eeq
$\tau_S$ is the well-known Sobolev optical depth (depending on line
opacity, transition frequency, radial and tangential velocity
gradient, and cosine of angle between $z$- and radial direction,
$\mu=z/r$), and $\phi(x)$ the appropriate, normalized profile
function. Finally, $\xmax = n_{\rm max} v_{\rm Dop}/\vinf$ denotes the
frequency shift at the blue wing of the transition, with $n_{\rm max}
\approx 3 {\ldots} 5$, in dependence of line-strength, and
$v_{\rm Dop}$ is the effective Doppler speed, comprising the thermal and the
micro-turbulent contribution. As long as there is no close (blueward)
line, the first term in Eq.~\ref{iminus_line} can be neglected
compared to the second, local one. 

\medskip
\noindent
Both approximations as outlined above (for optically thick continua
and/or lines) can be united, by basically summing up the optical
depths, and calculating a total source function (details are beyond
the scope of this appendix).

The impact of including this approximate incident intensity can be
seen by comparing the line-cores with and without this term (still
neglecting its frequency derivative) in Fig.~\ref{app_xj_iminus1},
where the red ($I^- = 0$) and dark blue (coinciding with the
turquoise) graphs display the resulting CMF mean intensities evaluated at
\Rmax. Without $I^-$, the line cores become too deep,
while including $I^-$ as specified above cures this problem
(independent of the specific atmospheric model considered).

\paragraph{Optically thin conditions.} If, however, only these terms
were accounted for (and even if accounting also for their frequency
derivatives), certain atmospheric models might display a problem. 
This is exemplarily shown in Fig.~\ref{app_xj_iminus2}, for the case
of a hot atmosphere with \Teff = 55,000~K, and a large mass-loss rate:
longward from the ``last'' strong UV resonance line (\CIV), when all
lines and the continuum have become optically thin at the outer
boundary, the CMF mean intensity begins to deviate from our previous,
pseudo-continuum approach (in green, see Sect.~\ref{philosophy}), and
this trend continues until the upper limit of our detailed CMF
treatment is reached, at $\lambda_{\rm max}$ = 10,000 \AA. One
might argue that our new solution is simply more accurate than the
previous one, because of the inherent approximations within the
latter. However, (i), our previous approach has been carefully tested,
by comparing the results for large model grids with corresponding {\sc
cmfgen} and WM-{\sc basic} solutions, and particularly the optical
flux distributions turned out to be very similar (see
\citealt{Puls05}, their Figs. 8 and 15). After all, those fluxes are
dominated by electron-scattering, and by bound-free and free-free
opacities and emissivities from H and He, which are not affected from our
opacity sampling approach in v10. (ii) Test-calculations using v11 with $I^-
= 0$ at all frequencies resulted in optical fluxes almost identical to
those from v10. Thus, there {\bf is} a problem, and one might
speculate whether this problem is related to processes neglected
thus far.

Indeed, and after many tests, it turned out that the discrepancy is
due to large inward directed intensities at high impact parameters ($p
\rightarrow \Rmax, \mu \rightarrow 0$) arising within the
solution for the last strong line. These intensities did not
``decay'' when proceeding toward longer wavelengths in the CMF
transport, resulting in overestimated mean intensities close to the
outer boundary, and contaminating the solution throughout the
complete optical range. The reason for these large intensities
relates to the large value of $I^-$ at the red end of the last
optically thick line core. Without appropriate means, this large value
cannot decrease within the frequency transport, as long as the
opacities are low (cf. Eq.~\ref{iminus_line}, first term), which is
often true for the considered outer region and directions.

In reality, however, there {\bf is} a decay, due to the fact that, for
wavelengths redward from the last line core, the illumination from
above is controlled by resonance zones that move outward as the
wavelengths increase, into the region outside the computational
domain, far beyond \zmax\ (the extended region in {\sc cmfgen}).
Consequently, $I^-$ decreases from its maximum value (at the red edge
of the last optically thick line core) to zero, and this decrease, in
combination with the corresponding frequency derivative, leads to an
effective negative source term at the outer boundary, enabling more
realistic solutions (black versus dark blue and turquoise in
Fig.~\ref{app_xj_iminus2}).

Thus, at first we have to define the frequency range over which $I^-$
declines to zero (when the resonance zone has reached ``infinity''), from its
starting value $I^-_{\rm red}$, where ``red'' should denote the red
edge of the previous line core, $x = x_{\rm red} = - \xmax$.  
Without any (continuum) absorption and
emission, the specific intensities remain constant along the
characteristics of the partial differential equations decribing
the CMF transfer, and for inward directed radiation, this leads to
(for example \citealt{Puls20})
\beq
I^-(x,\mu v(r)/\vinf)=I^-(x+\Delta x,\mu v(r)/\vinf+\Delta x).
\eeq
At the outer boundary, $v(r)/\vinf$ is close to unity. Thus, for
``core rays'' with $p \le \Rstar$ and $\mu(\zmax) \rarrow 1$, $\Delta
x$ is small, since the outermost resonance zone which can give rise to
a non-zero incident intensity must fulfill the condition $\mu
v(r)/\vinf+\Delta x = 1-\epsilon + \Delta x := 1$ (in the limit of
negligible thermal+micro-turbulent speeds, and with $\epsilon$ a small
number). However, for large impact parameters (which had been
identified as the ``problematic'' rays), $\mu$ is quite low, and we
find that 
\beq 
\label{deltax} 
\Delta x \approx 1-\mu(\zmax) = 1-\frac{\zmax}{\Rmax}, 
\eeq 
which is on the order of unity, and thus corresponds to dozens to
hundreds of thermal (including micro-turbulence) Doppler shifts. 

Summarizing, for $x$-values\footnote{As a reminder, the frequency
displacements considered here, $x_{\rm red} {\ldots} x$, are negative, with
$x_{\rm red} \ge x$, whereas $\Delta x \ge 0$.} 
close to the red edge of the previous line
core, the outer boundary is illuminated by an intensity close to
$I^-_{\rm red}$, whereas for frequencies lower than $x= x_{\rm red} -
\Delta x$, there is no longer any resonance zone which could
illuminate the outer boundary at $x$, and $I^-(x)$ must be zero (if
the continuum is optically thin).

To simplify our solution scheme, we assume that $I^-(x)$ decreases
linearly from $I^-_{\rm red}$ at $x= x_{\rm red}$ to zero at $x=
x_{\rm red}-\Delta x$, with $\Delta x$ from Eq.~\ref{deltax}, which
allows us to specify both $I^-(x)$,
\beq
\label{iminus_thin}
I^-(x) \approx I^-_{\rm red} \bigl(1 - \frac{x_{\rm red} - x}{\Delta
x}\bigr), \quad x \in [x_{\rm red}, x_{\rm red}-\Delta x],
\eeq
and its frequency derivative in an
easy and computationally fast way.  As argued above, the corresponding
terms are (almost) irrelevant for core-rays, but for rays with large impact
parameter, which moreover have a large weight for the mean intensity,
they are significant, and lead, overall, to a perfect representation
of the run of $J_\nu(\Rmax)$ (black versus green relations in
Fig.~\ref{app_xj_iminus2}). We note that it is indeed the gradient of
$I^-$ which is the decisive component: using the $I^-$ term as
estimated above alone (turquoise relation) does not mitigate the problem.

Under certain conditions, the continuum might be optically thick at 
$x= x_{\rm red}$. Then, the same strategy as above might be applied,
but $x_{\rm red}$ replaced by the frequency where the continuum
becomes optically thin, and $I^-_{\rm red}$ replaced by the
corresponding quantity using Eq.~\ref{iminus_cont} (or an appropriate
combination with overlapping line processes).

\paragraph{Line overlap effects.} A final complication might arise
because of strong lines seperated by less than $\Delta x = 1$, as
present, for example, for most UV resonance doublets. In this case, the
incident intensity for the red component needs to be modified, due to
the direct illumination by the blue one, which corresponds to a
non-negligible first term in Eq.~\ref{iminus_line}. There are
certainly better approximations, but in our current approach we simply
use the maximum of $I^-(x)$, either calculated from 
Eq.~\ref{iminus_thin}, with $I^-_{\rm red}$ derived at the red edge of
the blue component, or from the local term proportional to the source
function in Eq.~\ref{iminus_line}\footnote{We note that in this (and
other) cases care has to be taken, since $x$ refers
to the frequency displacement with respect to a specific line, here to
the blue and red component, respectively.} This
approach has not led to any obvious problems to date.

\medskip \noindent Collecting all our findings, there are two basic
possibilities to deal with $I^-(x,\zmax,p)$. Either, one adopts $I^- =
0$ at all frequency points, and accepts certain inconsistencies in the
line cores and optically thick continua; or one has to provide an
adequate $I^- \ne 0$ including its frequency derivative, even
for optically thin conditions, at least within the range $x_{\rm red}
{\ldots} x_{\rm red}-1$ to the red of all optically thick line cores.
Both formulations ensure that the detailed CMF solution and the
pseudo-continuum treatment are consistent within the optical
regime, and particularly at $\lambda_{\rm max}$. However, only a
physically correct prescription with $I^- \ne 0$ allows for a similar
consistency around $\lambda_{\rm min}$, since in the latter regime the
continuum is often optically thick beyond \Rmax, and many optically
thick lines are present as well. Moreover, if one aims at a solution
of the CMF transfer using the moments equations, the second method,
exclusively used in our current {\sc fastwind} version, is certainly
better suited, since otherwise the required ratios of intensity
moments often display an abrupt change at the outer boundary, giving
rise to numerical artifacts.

\section{Convergence behavior}
\label{convergence}

\begin{figure*}[h]
\begin{center}
\begin{minipage}{8cm}
\resizebox{\hsize}{!}
  {\includegraphics[angle=90]{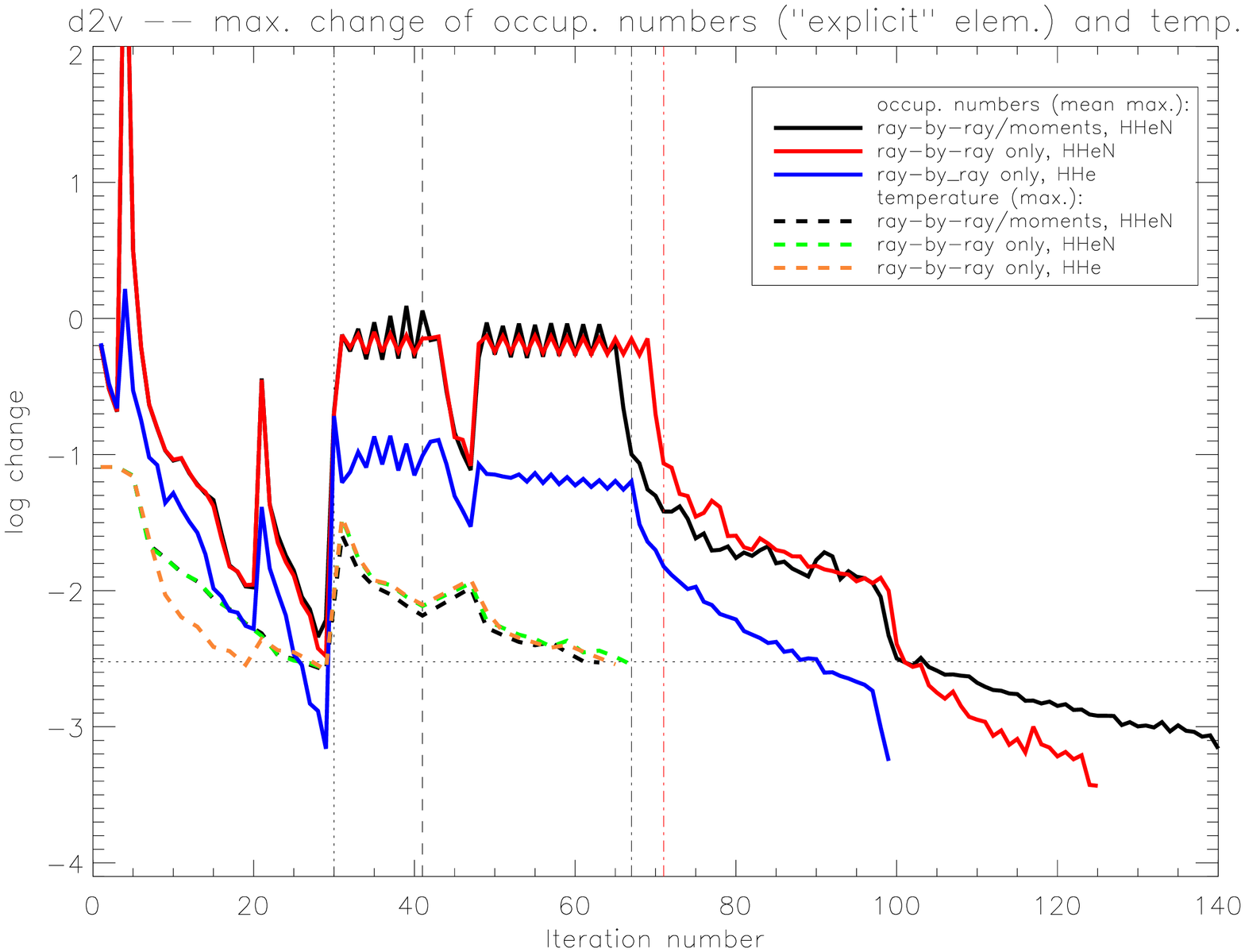}} 
\end{minipage}
\begin{minipage}{8cm}
\resizebox{\hsize}{!}
  {\includegraphics[angle=90]{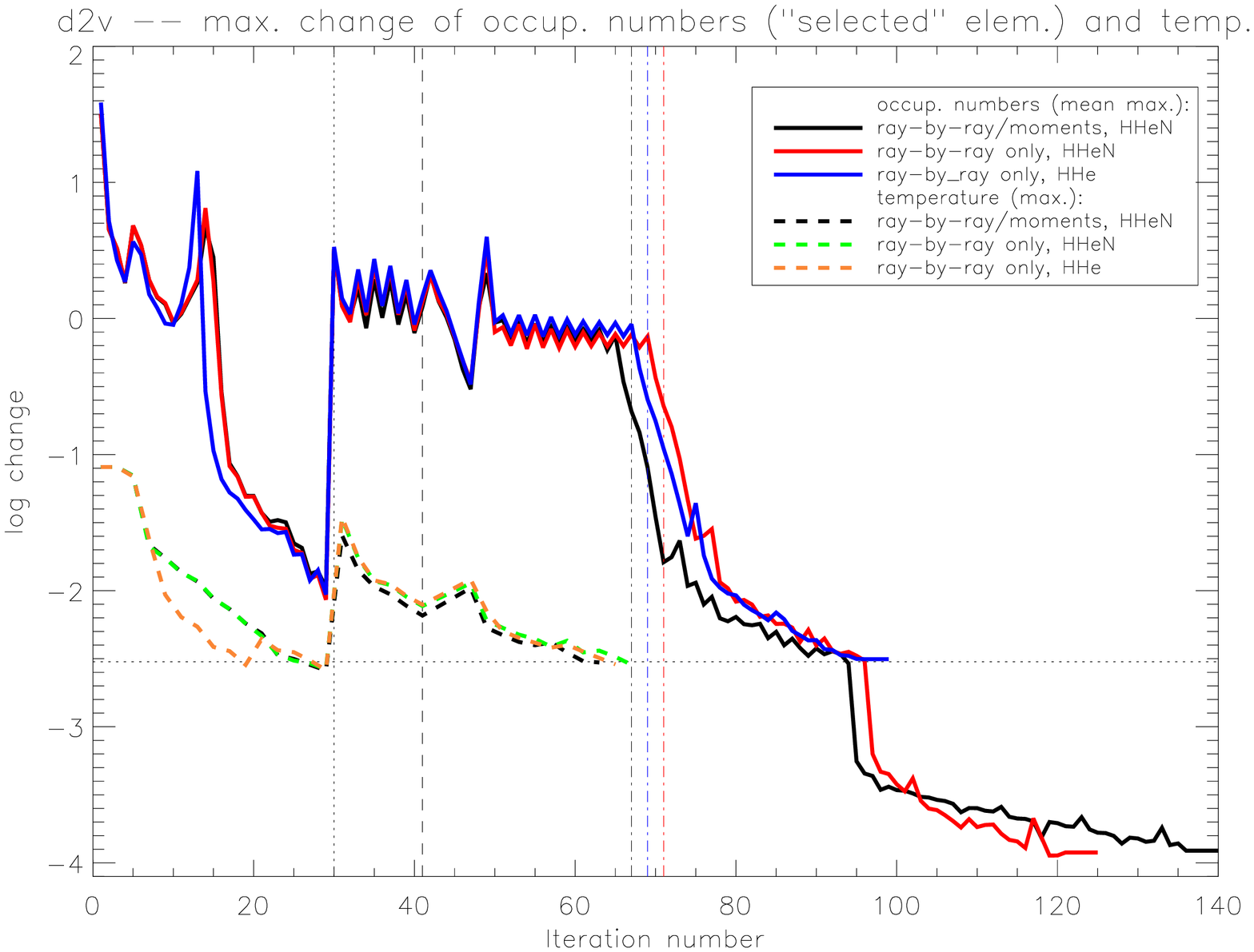}} 
\end{minipage}
\begin{minipage}{8cm}
\resizebox{\hsize}{!}
  {\includegraphics[angle=90]{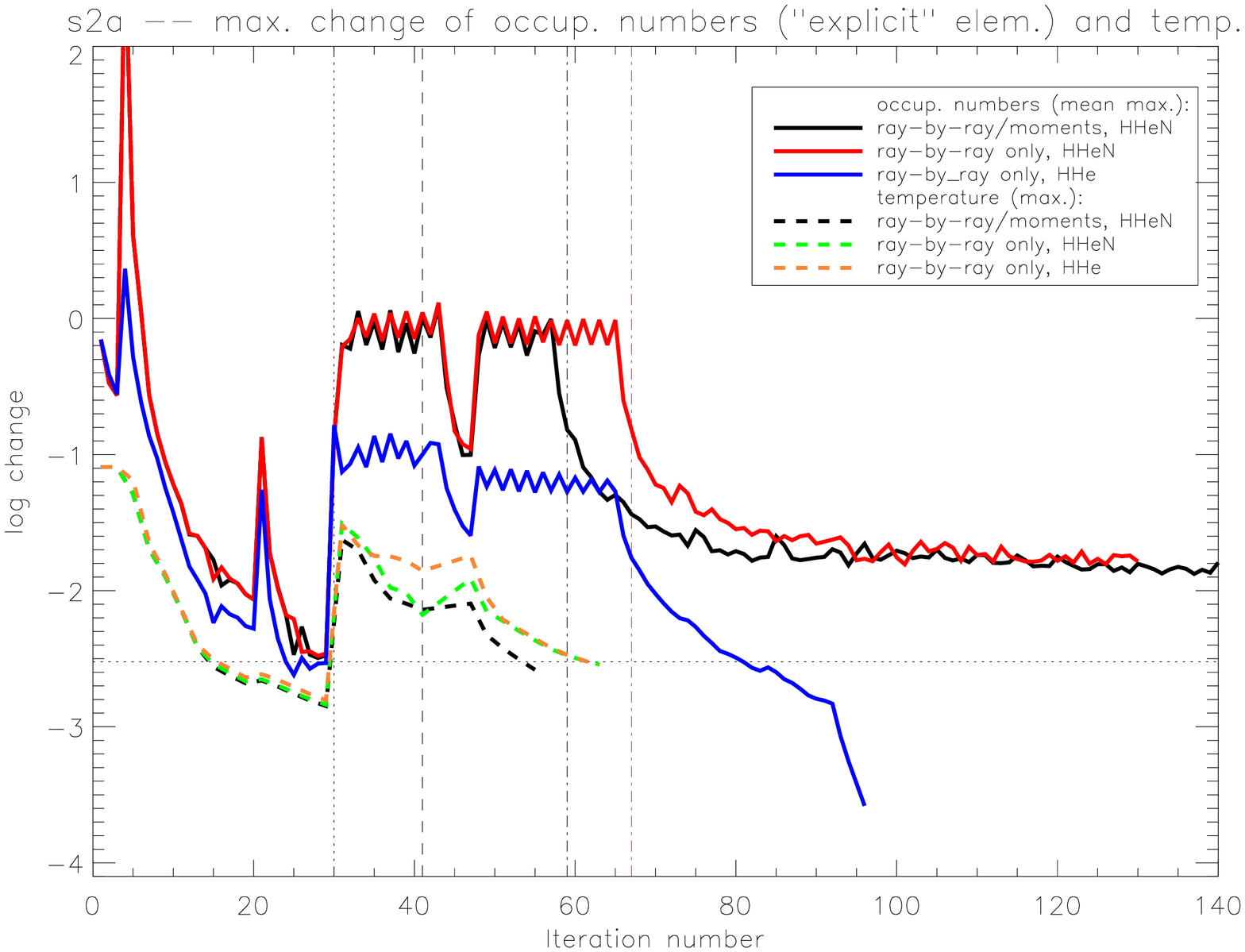}} 
\end{minipage}
\begin{minipage}{8cm}
\resizebox{\hsize}{!}
  {\includegraphics[angle=90]{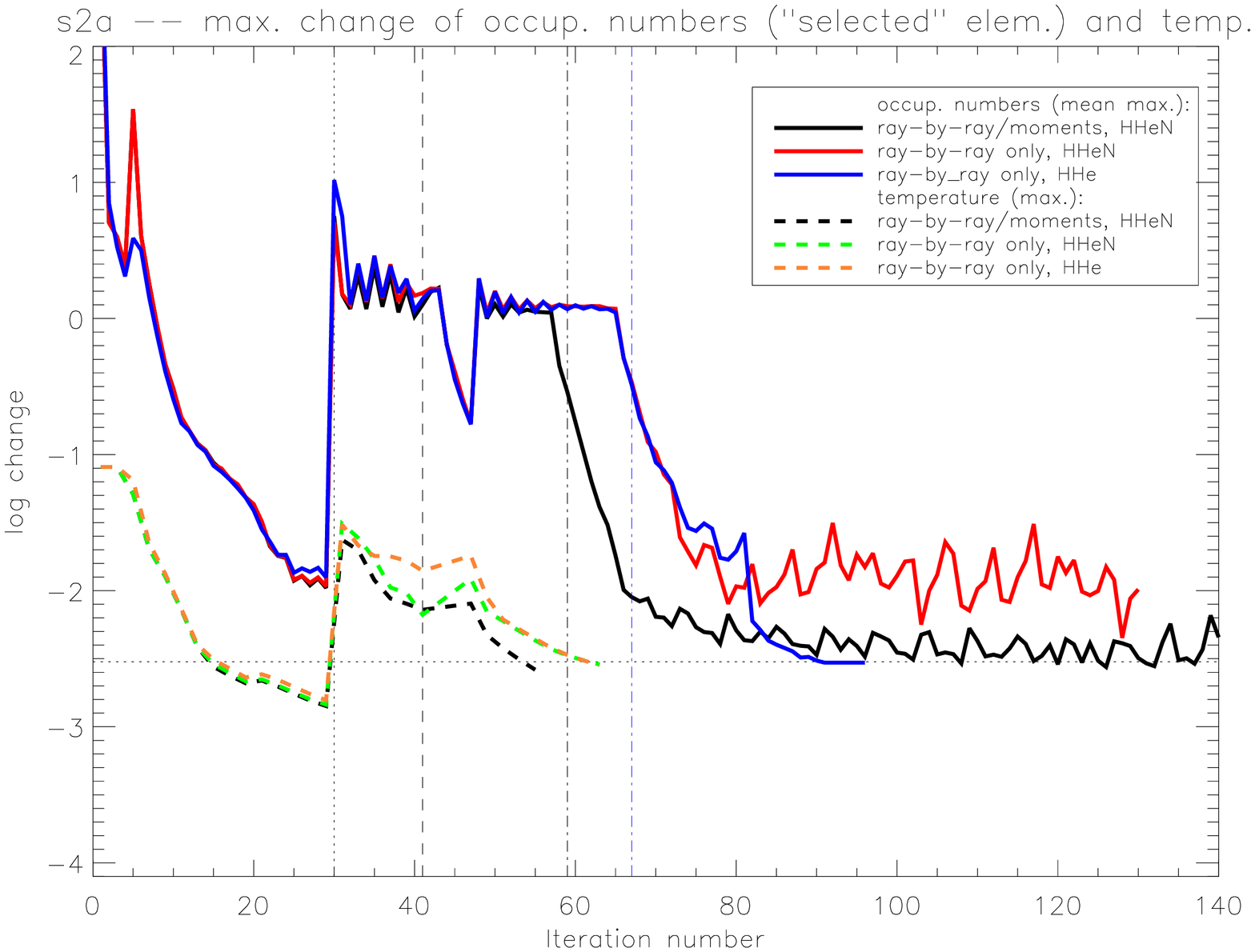}} 
\end{minipage}
\end{center}
\caption{Prototypical convergence behavior for three different kinds
of model calculations (see legend and text). Displayed is the maximum
relative change of temperature (regarding all grid-points, dashed)
within two consecutive iteration steps, and the corresponding mean of
the maximum change within all specified occupation numbers, where the
averaging has been performed over all depth-points with $\taur \le 1$.
Left panels: mean maximum change within occupation numbers from
explicit elements (HHeN in black and red, HHe in blue); right panels:
mean maximum change within occupation numbers from selected background
elements. Upper and lower panels: convergence behavior for our
hottest dwarf (d2v, see Table~\ref{tabgrid}) and supergiant (s2a)
model, respectively. Our convergence criterion for the temperature is
displayed by a horizontal dotted line. The vertical lines display the
onset of the detailed CMF calculations within $\lambda_{\rm min}$ and
$\lambda_{\rm max}$ (dotted), the update of the photospheric structure
(dashed), and the onset of the Ng-extrapolation scheme for the C,N,O
resonance lines (dashed-dotted). See text.} 
\label{conv}
\end{figure*}

The convergence behavior for various set-ups offered by our
implementation is displayed in Fig.~\ref{conv}. In particular, we
compare our default procedure (ray-by-ray solution plus moments
equations, black) with an approach using the ray-by-ray solution
alone. The impact of different sets of explicit elements is also
compared. In the former approach, we used H, He, and N, as explict
elements\footnote{Such a set-up, in connection with the previous {\sc
fastwind} version, has been employed, for example, by \citealt{rivero12,
grin17}, and \citet{markovapuls18}, for a quantitative nitrogen
spectroscopy of massive stars in the LMC and the Galaxy.}, whereas in
the latter, we either considered the same elements (red), or H and He
only (blue).

Fig.~\ref{conv} allows us both to explain our basic strategy, and to
discuss two prototypical situations with respect to convergence
behavior. In all panels, we display the maximum (regarding all
grid-points) relative change of temperature (dashed) within two
consecutive iteration steps, and the corresponding radial mean of the
maximum change within all specified (cf. Sect.~\ref{basic_cmf})
occupation numbers. In the latter case, we provide the (radial) mean,
to suppress local maxima which sometimes occur, but we average the
maxima per grid point only over the most important region with $\taur
\le 1$, since for larger optical depths the occupation numbers begin
to thermalize, and corresponding changes become very low (typically,
on the order of $10^{-4 {\ldots} -5}$). If included into the
averaging, this region would bias the result.

Before concentrating on the convergence properties, we explain our 
basic strategy, by means of the upper panels corresponding to our
hottest dwarf model (d2v), with parameters provided in
Table~\ref{tabgrid}. The left and right panels display the changes
within the explicit and the selected background elements. 

During the first 30 iterations, we follow our previous {\sc fastwind}
version, namely performing 10 to 20 iterations with Sobolev line
transfer for all elements, and then switching to the CMF approach for
the explicit and most important background transitions, using in
parallel the pseudo-continuum approach. As displayed, the convergence
is very fast, both for occupation numbers and temperature. If we would
continue with this approach, as done in {\sc fastwind} v10, the
complete model would have converged after a total of 40 to 60
iterations. In the new version, however, we switch to the detailed CMF
calculations (in between $\lambda_{\rm min}$ and $\lambda_{\rm max}$),
at iteration \#30, indicated by the dotted vertical line. Immediately,
the occupation numbers and the temperature structure change
significantly, not the least because the Rosseland optical depths
change at each grid point (detailed calculation versus sampling), though
typically by only few percent. After a while, the temperature begins
to stabilize again, and we perform an update of the photospheric
structure (particularly accounting for the updated radiative
acceleration), indicated by the dashed vertical line. As already
explained in \citet{rivero12}, one such update is usually sufficient,
as long as it is performed not too early. From this point on, the
temperature stabilizes again, and we consider it as converged, if the
maximum changes are below 0.3\% (indicated by the dotted horizontal
line). Ideally, one would iterate the temperature and occupation
numbers in parallel, until a final, common convergence has been
reached (as done in models employing a complete linearization). In our
method, however, even small changes in temperature directly couple to
changes in occupation numbers. Consequently, we would need a much
larger number of iterations until the convergence criterion has been
reached, though the final changes in temperature (compared to our
standard treatment) would be marginal. Indeed, after the temperature
has been considered as converged and remains fixed, also the
occupation numbers stabilize (for the inspected model, quite quickly),
and we consider the model as converged when the mean maximum changes
fall below $10^{-3}$ to $10^{-4}$. {Moreover, in each iteration we
check the flux-conservation. For all models computed thus far, this
condition remains fulfilled (typically at the one to two percent
level) also after temperature convergence, that is,
after the temperature has been fixed, and only the occupation numbers
are allowed to vary.

To increase the final convergence, we apply a
Ng-extrapolation \citep{Ng74} for the source functions of important
resonance lines (here: from C, N, O); extrapolating the source
functions from all line transitions is computationally prohibitive,
since, to perform such an extrapolation, the values from at least
three prior iteration steps would need to be stored.

At first we note that, for all three set-ups, the temperature
convergence is very similar, which is also true for the occupation
numbers from the selected elements. For the explicit elements, on the
other hand, the mean maximum deviations for the HHeN models are larger
than for the HHe model, which indicates that the convergence of
nitrogen is more problematic than that of H and He, at least when a 
detailed atomic model is used, and all levels are inspected for
convergence. Nevertheless, for this hot dwarf model,
the occupation numbers from all elements eventually converge. 

This is no longer true for the second model displayed in the lower
panels, a model for a hot supergiant (s2a). Here, a well-behaved
convergence is present only for the model with HHe as explicit
elements (in blue). Interestingly, the selected elements,
in particular, N, seem to show a better final convergence, compared to
the case when N is treated as an explicit element (left lower panel,
in red). However, this is mostly due to our recipe for evaluating the
maximum changes. As explained in Sect.~\ref{basic_cmf}, for explicit
elements we inspect all levels, because of our requirement for high precision. 
For selected elements, we only account for changes
in the ionization fractions though. Since the convergence problems (see
below) mostly concern the excited levels of N (connecting the triplet
lines around 4640~\AA), these problems do not show up in the displayed
maximum changes. Thus, N (selected), in concert with the other
selected elements, seems to converge without problems, contrasted to
the case when N has been treated as an explicit element.

On the other hand, both models with HHeN (in black and red)
display an oscillatory behavior (with respect to maximum changes) in
the last iterations, close to the one-percent level. Before providing
further details, we note that also here the temperature has converged
after roughly 60 iterations, which is also true for the other models
discussed in the current study (Sect.~\ref{comparison}), except for
model s8a. Independent from set-up, the latter requires roughly 110
iterations to achieve a converged temperature structure;
interestingly, the corresponding {\sc cmfgen} model has similar
difficulties, which indicates that it is located in a parameter regime
where specific processes can easily push the ionization equilibrium
and/or optical depths into different directions.

The origin of the apparently bad convergence of the occupation numbers
is mostly related to line-overlap effects (in rest-frame, or
wind-induced), particularly between strong lines of C, N, O, and Fe
(either between different elements, or between different ions of the
same element, \citealt{rivero11, MartinsHillier12}), discussed
in more detail in Sect.~\ref{comparison}. In those cases
where one of the overlapping lines strongly dominates, the overlap
does not lead to specific problems (in this case, the source function
of the weaker component will adapt to the source function of the
stronger one). If, however, both lines have a similar strength, an
oscillatory behavior becomes possible, as visible for model
s2a (both in the explicit and selected elements, which are coupled via
such line overlap effects). Until to-date, we found no real means to
improve the situation, that is, to ensure convergence for all levels at
all depth-points. Moreover, analytic considerations based on
simplified but similar conditions indicated that such oscillations are
possible indeed, independent of the specific iteration scheme. 
Thus, we investigated how the occupation numbers and
the synthetic spectrum are affected by different start models,
different extrapolation schemes, different damping procedures, and so
on.  As a major conclusion, we found that a large number of iterations
(on the order of 140, as displayed in Fig.~\ref{conv}) is usually
sufficient to obtain a more or less unique solution, with only few 
oscillating levels over a restricted spatial domain. In this case, the
emergent spectra become very similar for the various test models
(with maximum differences on the order of few percent in the peak
heights or depths of few lines), independent of convergence history: due
to the oscillatory behavior of the affected levels, there is only a
limited range over which they vary, and most other levels can and
indeed do converge. Additionally, the oscillation pattern, if present,
often occurs only inside the intermediate wind, above 0.1~\vinf,
such that photospheric profiles, even if in emission due to NLTE
effects, are not affected at all.

\section{More technical issues}
\label{more_details}

\subsection{Re-mapping of radiation field, and non-coherent electron
scattering} 
\label{remap}

To save computational time, we re-map the mean intensities resulting
from our detailed CMF calculations (with $N_f$ frequency points, see
Eq.~\ref{N_f}) onto a coarser grid, which in the very first iteration
steps serves as a frequency grid for the continuum and
pseudo-continuum transport. Depending on considered elements
(ionization edges to be resolved), this grid comprises between 1000 to
2000 frequency points. The re-mapping is done in such a way that the
frequency integral of the mean intensity between two coarse mesh
points, $\nu_i, \nu_{i+1}$, remains conserved, such that
$\int_{\nu_i}^{\nu_{i+1}} J_{\rm coarse}(\nu) \dd \nu =
\int_{\nu_i}^{\nu_{i+1}} J_{\rm fine}(\nu) \dd \nu$. The re-mapped
mean intensities are used to calculate corresponding integrals for the
ionization and recombination rates (see also \citealt{hilliermiller98,
pauldrach01, rivero11} for the specific case of dielectronic
recombination), and for the heating and cooling rates within the
electron thermal balance \citep{Kubat99}. They also enter the Sobolev
line-rates for those lines with no information on the upper level (see
Appendix~\ref{no_upper}) from the selected background elements, as well as the
approximate NLTE calculations for the non-selected background. In both
latter cases, such rather smooth mean intensities should be used to
avoid contamination by narrow features within the detailed solution.
Since particularly the ionization and recombination integrals need to be
evaluated during each iteration step, and extend over a significant
frequency range, the calculation via re-mapped mean intensities is
computationally favorable, and introduces only small errors, because
of their specific conservation properties. 

In contrast, all scattering integrals, ALOs, and total line
acceleration are calculated by integrating over the highly resolved
CMF frequency grid, where the line acceleration is amended by results
for the ``outer'' ranges below and above the CMF regime (outside
$\lambda_{\rm min}$ {\ldots} $\lambda_{\rm max}$), via corresponding
integrals over the approximate pseudo-continuum fluxes. 

A second set of mean intensities is additionally calculated, to be
used when setting up the (non-coherent) electron scattering
emissivities. We note that using coherent scattering instead would
induce erroneous results within the line-cores, because of the large
electron thermal speeds, giving rise to a frequency redistribution
within a range of several thousand \kms. For the sake of simplicity
and computational performance, the corresponding terms
are approximated via convolving the CMF mean intensities by
electron thermal broadening. This procedure is at least qualitatively
similar to results when using the exact redistribution function as
provided by \citet[see also \citealt{RH94}]{Hummer67}.

\subsection{Total line list, and treatment of lines without information
on the upper level.} 
\label{no_upper}

Since for the explicit elements we use a flexible, {\sc detail-}
\citep{ButlerGiddings85} like input for atomic models and transitions,
while for the background elements we rely on the fixed-format, WM-{\sc
basic} \citep{pauldrach01} data base, we need to adapt our total line
list, to be used both in the CMF-transfer and in the final formal
integral (see below). The original line list, comprising roughly 2
million entries, is also taken from the WM-{\sc basic} data base,
whereas all transitions refering to explicit elements are replaced (at
the begin of the CMF-treatment) by corresponding ones from the {\sc
detail-}input. Because a number of transitions from the latter are
packed (for example, those for the \NV\ UV-resonance doublet), they need to
be de-packed within the line-list, which is done via one additional
data-file (LINES.dat) containing all necessary
information\footnote{The latter file is also used to provide the
line-broadening parameters required for the formal integral.}.

Whereas all transitions of the explicit elements have lower and upper
levels that are treated within our NLTE network, the line list for
the background elements also comprises numerous transitions (mostly
from Fe and Ni) where only the lower level is included into the
corresponding NLTE rate equations. Indeed, within the background
elements, there are roughly ``only'' 40,000 transitions where also the
upper level is explicitly considered. This, because the energy cut-off
for the corresponding atomic models has been chosen in such a way as
to allow for a numerically stable solution of the linear rate
equation system. For calculating the line-blocked radiation field, and
also the radiative acceleration, however, also the multitude of lines
that have a level beyond this cut-off need to be accounted for. For
all such lines (again: this only affects the background elements), we
use a two-level, Sobolev approach to estimate the corresponding source
functions. Here, the radiative rates are derived by considering the
re-mapped CMF mean intensities, and the collisional de-excitation rate
coefficients are either estimated from the \citet{vanregemorter62}
approximation for radiatively allowed transitions, or following the
semi-empirical expression by \citet[with collision strength $\Omega =
1$]{allen73} for the forbidden ones. The source functions estimated in
this way are then used within the subsequent CMF-transfer, such that
also these lines participate in the overall 
iteration scheme.  

Since such a two-level approach is justified only for strong
transitions connected to the ground or a meta-stable state, it needs
to be checked whether this approximation might also be used for
excited lower levels, as done per default in our current implementation. A
corresponding test has been performed by using an alternative
approach, namely by estimating the source-function from an approximate
occupation number of the (missing) upper level. This occupation number
(actually, divided by its statistical weight) was estimated from its
LTE-value (which can be calculated, since the energy of the lower
level, and the transition energy is known), times a NLTE departure
coefficient. As a crude approximation, the latter was assumed to be
identical to the departure coefficient of the highest-lying level (of
the considered ion) that is explicitly calculated, and which is
connected to the same lower state. Indeed, when comparing models
obtained with either approximation (two-level atom, or NLTE with
approximate departure coefficient for transitions with ``unknown''
upper level, and a lower one that is {\bf not} the ground or a
meta-stable state), it turned out that basically all results remain
unaffected, for all models within our complete grid (Table~\ref{tabgrid}). 

Though our two-level approach yields overall reasonable results, there
is one additional problem, which might introduce considerable errors
into important diagnostic lines. Namely, since there are numerous
lines treated in the above approximation (in particular, from Fe and
Ni), there is quite a chance of a coincidental line-overlap with lines
that are treated exactly. If, on the one side, the ``approximate''
line is stronger, the exact line can adapt to source-function
equality, since its upper level participates in the NLTE network. In
this situation, the final result is reasonable, and nothing needs to
be done. If, on the other side, the exact line is dominating, the
approximate one should develop a similar source-function as the 
exact one. This cannot happen though, since the run of its source
function depends, via the two-level Sobolev approach, mainly on the
escape probabilities, which are barely affected by the exact line.
And, since it is {\bf not} adapting, it might strongly influence the
radiation field and thus the source function of the exact line in
an erroneous way, at least if its opacity is not too low.
\begin{figure}
\resizebox{\hsize}{!}
  {\includegraphics[angle=90]{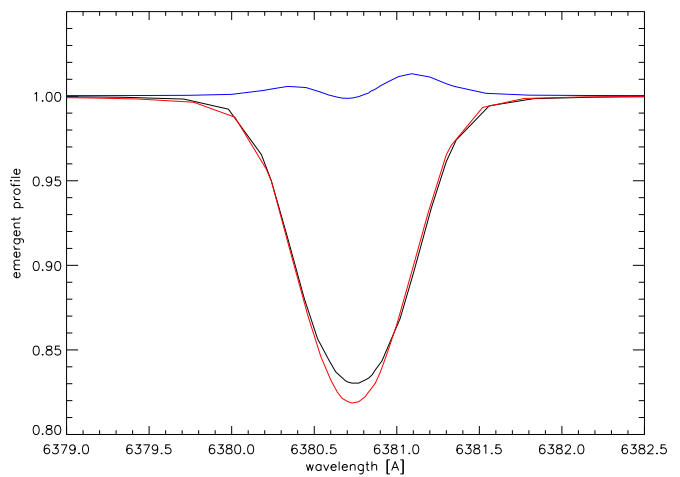}} 
\caption{\NIV~6380 for a hot dwarf model (d2v). Black: line
profile resulting from our current approach, when manipulating the
source functions of EUV background lines with no information on the upper
level, as described in the text. Blue: line profile when the EUV 
two-level-atom source functions are {\bf not} manipulated. Red: line
profile as calculated from our previous {\sc fastwind} version, v10.}
\label{plot_d2v_6380}
\end{figure}
An example for this situation is given by the (strategic)
\NIV~6380 line (see Fig.~\ref{plot_d2v_6380}), with a lower
level radiatively pumped by the third level of \NIV, at 
$\lambda \approx 387.36$~\AA. For hot (Galactic) O-stars, a strong \FeV\ line
at 387.37~\AA, with a lower meta-stable and unknown upper level
(at least in our database), is located just 11~\kms\ to the red, such
that it is overlapping for our standard value of $\vmic = 15~\kms$. If this
line is articifically neglected, there is no influence on the \NIV\ UV
line. In this case, the optical line appears in absorption, as, for example,
predicted by our previous {\sc fastwind} version (which does not
account for explicit line-overlaps; red profile in
Fig.~\ref{plot_d2v_6380}). If, however, the \FeV\, line is included, 
and treated by our two-level approach, line overlap-effects with the
\NIV\ UV line become significant. Because the \FeV\, line has only
slightly lower opacities, but a considerably lower source-function,
the radiation field at 387~\AA\, becomes weaker, and the lower level
of \NIV~6380 is less pumped. Indeed, it becomes even
depopulated compared to its upper level, and appears in emission (blue
profile in Fig.~\ref{plot_d2v_6380}), in stark contrast to our
previous results, and also to observations (\citealt{rivero12,
rivero122, grin17}).

To cure this problem, our code checks for potential overlaps between
exact lines and lines with unknown upper level. If the 
exact line is stronger, we reset the source function of the 
approximate one, to the value from the exact line. Otherwise,
nothing needs to be done. We have tested this approach by comparing between
models including or excluding this procedure. Indeed, in most cases
there is no effect at all, and only in few cases such as
\NIV~6380, we see considerable changes (black profile in
Fig.~\ref{plot_d2v_6380}).

\end{document}